\documentclass[prl,showpacs,twocolumn,floats,10pt,aps,citeautoscript,longbibliography,superscriptaddress]{revtex4-2}
\usepackage{microtype} 
\usepackage{graphicx}
\usepackage{enumerate}
\usepackage{bm}
\usepackage{amsmath}
\usepackage{color}
\usepackage{upgreek}
\usepackage[normalem]{ulem}

\allowdisplaybreaks 

\usepackage{xspace} 

\renewcommand{\vec}[1]{{\mathbf{#1}}} 
\newcommand{\bk}{{\vec{k}}}
\newcommand{\mi}{\mathrm{i}}
\usepackage{bbm}  

\newcommand{\fl}{{\scriptscriptstyle{\textrm{FL}}}}
\newcommand{\nfl}{{\scriptscriptstyle{\textrm{NFL}}}}

\newcommand{\vertex}{{\textrm{vtx}}}

\newcommand{\TFL}{T_\fl}
\newcommand{\TNF}{T_\nfl}
\newcommand{\epsilonf}{\epsilon_f}  
\newcommand{\epsilonck}{\epsilon_{c\bk}} 
\newcommand{\ahat}{\vec{a}} 

\newcommand{\YRS}{{YbRh${}_2$Si${}_2$}}

\newcommand{\CCI}{{CeCoIn${}_5$}}

\newcommand{\mc}[1]{\ensuremath{\mathcal{#1}}}
\newcommand{\mr}[1]{\ensuremath{\mathrm{#1}}}

\newcommand{\re}{\ensuremath{\mathrm{Re}}}
\newcommand{\im}{\ensuremath{\mathrm{Im}}}

\newcommand{\Eq}[1]{Eq.~\eqref{#1}}

\newcommand{\Fig}[1]{Fig.~\ref{#1}}

\newcommand{\Sec}[1]{Sec.~\ref{#1}}

\newcommand{\pdag}{{\vphantom{dagger}}}  

\newcommand{\inFL}[1]{#1  < \TFL}
\newcommand{\inNFL}[1]{\TFL \! < #1  < \! \TNF}

\RequirePackage[
  hyperindex,colorlinks,bookmarksnumbered,
  plainpages=true,pdfstartview=FitH,hypertexnames=false]{hyperref}
\hypersetup{linkcolor=blue,urlcolor=blue,citecolor=blue} 
\usepackage[hypertexnames=false]{hyperref}
\usepackage[all]{hypcap}

\makeatletter
\def\maketitle{
\@author@finish
\title@column\titleblock@produce
\suppressfloats[t]}
\makeatother

\begin{document} 

\title{Dynamical scaling and Planckian dissipation due to heavy-fermion quantum criticality}
\author{Andreas Gleis}
\email{andreas.gleis@lmu.de}
\affiliation{Arnold Sommerfeld Center for Theoretical Physics, 
Center for NanoScience,\looseness=-1\,  and 
Munich Center for \\ Quantum Science and Technology,\looseness=-2\, 
Ludwig-Maximilians-Universit\"at M\"unchen, 80333 Munich, Germany}
\author{Seung-Sup B.~Lee}
\email{sslee@snu.ac.kr}
\affiliation{Arnold Sommerfeld Center for Theoretical Physics, 
Center for NanoScience,\looseness=-1\,  and 
Munich Center for \\ Quantum Science and Technology,\looseness=-2\, 
Ludwig-Maximilians-Universit\"at M\"unchen, 80333 Munich, Germany}
\affiliation{Department of Physics and Astronomy, Seoul National University, Seoul 08826, Korea}
\affiliation{Center for Theoretical Physics, Seoul National University, Seoul 08826, Korea}
\author{Gabriel Kotliar}
\email{kotliar@physics.rutgers.edu}
\affiliation{Department of Physics and Astronomy, Rutgers University, Piscataway, NJ 08854, USA}
\affiliation{Condensed Matter Physics and Materials Science Department,\looseness=-1\,  
Brookhaven National Laboratory, Upton, NY 11973, USA}
\author{Jan von Delft}
\email{vondelft@lmu.de}
\affiliation{Arnold Sommerfeld Center for Theoretical Physics, 
Center for NanoScience,\looseness=-1\,  and 
Munich Center for \\ Quantum Science and Technology,\looseness=-2\, 
Ludwig-Maximilians-Universit\"at M\"unchen, 80333 Munich, Germany}

\date{\today}

\begin{abstract}\
We study dynamical scaling associated with
a Kondo-breakdown quantum critical point (KB-QCP) of the periodic Anderson model, treated by two-site cellular dynamical mean-field theory (2CDMFT).
In the quantum critical region, the staggered spin 
exhibits SYK-like slow dynamics and its dynamical susceptibility shows 
$\omega/T$ scaling. We propose a scaling Ansatz that describes this behavior.
It also implies Planckian dissipation for the longest-lived excitations.
The current susceptibility follows the same scaling ansatz, leading to 
strange-metal scaling.
This demonstrates that the KB-QCP described by 2CDMFT is an \textit{intrinsic} 
(i.e., disorder-free) strange-metal fixed point.
Surprisingly,
the SYK-like dynamics and scaling are driven by strong vertex contributions to the susceptibilities.  
Our results for the optical conductivity match experimental
observations on \YRS\ and \CCI.

\medskip

\noindent
DOI:
\end{abstract}

\maketitle

\textit{Introduction.---}%
Strange metals are
enigmatic states of matter which, despite extensive theoretical and experimental 
effort, still defy clear and unified understanding~\cite{Phillips2022,Hartnoll2022,Chowdhury2022,Zaanen2004,Checkelsky2024}. 
They are
found in the phase diagrams of a large number of strongly correlated materials,
such as cuprate superconductors~\cite{Keimer2015,Michon2019,Michon2023,Legros2019}, iron based superconductors, twisted bilayer graphene~\cite{Cao2018,Cao2020,Jaoui2022} or heavy fermion metals~\cite{Loehneysen1996a,Trovarelli2000,Paglione2003,Prochaska2020,Taupin2022,Li2023}. 

The phenomenology of strange metals is incompatible with our current understanding of conventional metals.
Most prominently, they show $T$-linear resistivity~\cite{Legros2019} down to temperatures too low to be of phononic origin (current record: $\sim15$ mK in {\YRS}~\cite{Nguyen2021})
and a $\sim T\ln T$ specific heat. Both are incompatible with the $\sim T^2$ resistivity and $\sim T$ specific heat expected in
normal Fermi liquids~\cite{Giuliani2005}. In many materials, $\omega/T$ scaling of dynamical susceptibilities~\cite{Aronson1995,Schroeder2000,Poudel2019} and more recently also of the optical conductivity~\cite{Prochaska2020,Yang2020,Michon2023,Li2023} is observed.
Dynamical scaling is incompatible with Fermi liquids, where quasiparticles with $\sim T^2$ decay rates lead to Lindhard-type susceptibilities and
to a Drude peak with width $\sim T^2$ in the optical conductivity.
A recent experiment on a strange-metal \YRS\ nanowire 
further found an almost complete suppression of the shot noise, 
indicating the absence of well-defined quasiparticles~\cite{Chen2022}.

Despite the ubiquity of materials and experiments showing strange metallicity, even basic questions are to date not fully settled~\cite{Phillips2022}.
Do strange metals arise due to quantum critical points and quantum critical phases, 
or are they intimately connected to quantum criticality at all? 
Do intrinsic strange metals, i.e., ones without disorder, exist~\cite{Else2021a,Else2021b}?
Recent work showed that many of the features of strange metals can arise from a critical boson coupled to fermions~\cite{Patel2023,Aldape2022}, provided
that the boson-fermion coupling is disordered. On the other hand, measurements on cuprates suggest that disorder only 
affects the residual resistivity, while the linear-in-$T$ slope is unaffected~\cite{Rullier-Albenque2000}. Further, there exist many stochiometric strange-metal compounds
with comparably small residual resistivities.

In this Letter, we show 
that intrinsic strange-metal scaling can arise due to a heavy-fermion quantum critical point~(QCP), described via cellular dynamical mean-field theory~(CDMFT)~\cite{Kotliar2001,Maier2005}.
We study the periodic Anderson model~(PAM). It exhibits a so-called Kondo breakdown~(KB) QCP~\cite{Si2001,Coleman2001,Coleman2002} arising as a continuous orbital-selective Mott transition~\cite{Tanaskovic2011,DeLeo2008,DeLeo2008a,Gleis2023}. 
Its hallmark 
is a partial localization of electrons, accompanied by a Fermi surface reconstruction, experimentally observable, e.g.,
via Hall effect or quantum oscillation measurements.
Experimental studies in the quantum critical region of KB--QCPs at $T>0$ 
often show strange-metal behavior.

In a long companion paper, Ref.~\cite{Gleis2023}, we showed that 
two-site CDMFT (2CDMFT)
combined with the numerical renormalization group~(NRG)~\cite{Bulla2008} 
describes many experimental features
of the KB--QCP. This includes a novel quantum critical point (stabilized by DMFT self-consistency) and strange-metal behavior, such as a $\sim T\ln T$ specific heat
in the non-Fermi liquid~(NFL) quantum critical region. 
Here, we focus on 
quantum critical dynamical scaling. 
We find (i) SYK-like slow dynamics;
(ii) $\omega/T$ scaling of dynamical susceptibilities; (iii) Planckian dissipation;
(iv) strange-metal-like $\omega/T$ scaling of the optical conductivity
$\sigma(\omega)$;
and (v) results for $\sigma(\omega)$
consistent with measurements 
on \YRS\ and \CCI.

\textit{Model and methods.---}%
We consider the PAM on a three-dimensional cubic lattice, consisting of an itinerant $c$ band and a localized $f$ band, 
described by the Hamiltonian 
\vspace{-0.2\baselineskip}
\begin{equation}
\label{eq:H_PAM}
\begin{aligned}
& H_{\mathrm{PAM}} =  \sum_{\bk\sigma} \bigl(\epsilonf - \mu\bigr)
f^\dagger_{\bk\sigma} f^\pdag_{\bk\sigma}  
+ U \sum_{i} f^\dagger_{i\uparrow} f^\pdag_{i\uparrow} f^\dagger_{i\downarrow} f^\pdag_{i\downarrow}
\\
& \quad
+ V \sum_{\bk\sigma} 
\bigl(c^\dagger_{\bk\sigma} f^\pdag_{\bk\sigma} + \mathrm{h.c.} \bigr)
 + \sum_{\bk\sigma} \left(\epsilonck - \mu\right)  c^\dagger_{\bk\sigma} c^\pdag_{\bk\sigma} .
\end{aligned}
\end{equation}
Here, $f_{\bk\sigma}^\dagger$ [$c_{\bk\sigma}^\dagger$] creates a spin-$\sigma$ $f$~[$c$] electron with momentum $\bk$, and
$\epsilonck = -2t \sum_{a = x,y,z} \cos (k_a)$ is the $c$-electron dispersion.
We set the $c$-electron hopping $t = 1/6$ as an energy unit (half bandwidth $=1$) and
fix the $f$-orbital level $\epsilonf=-5.5$, the interaction strength $U=10$, and the chemical potential $\mu=0.2$, as chosen in prior 2CDMFT studies~ \cite{DeLeo2008a,DeLeo2008,Gleis2023}. We tune the $c$-$f$ hybridization $V$ and the temperature $T$ as control parameters.

We study the PAM using 2CDMFT, 
which maps the lattice model to an effective two-impurity Anderson model (2IAM) with a self-consistent bath (cf.~Refs.~\cite{DeLeo2008a,DeLeo2008,Tanaskovic2011,Gleis2023} for more details).
The 2CDMFT approach allows us to study the competition between local Kondo correlations and nonlocal RKKY correlations,
which is believed to drive quantum criticality in heavy fermion systems~\cite{Doniach1977,Si2001,Senthil2003,Senthil2004}. 
We solve the effective 2IAM using NRG \cite{Bulla2008},   
enabling us to reach exponentially small frequency and energy scales.
We exploit and enforce 
$\mathrm{U}(1)$ charge and $\mathrm{SU}(2)$ spin symmetries
(using the QSpace tensor library~\cite{Weichselbaum2012a,Weichselbaum2020}), 
thereby excluding the possibility of symmetry-breaking order by hand.
We thus study KB quantum criticality in the
absence of symmetry breaking~\cite{Vojta2010,Yamamoto2010,Si2010,Coleman2010}, as can be observed, e.g., in experiments on {\YRS}~\cite{Friedemann2009}
and {\CCI}~\cite{Maksimovic2022}.
We do not find 
tendencies towards 
symmetry breaking (divergent susceptibilities) 
for $V > V_c$ or anywhere within the quantum critical region
emanating from the KB-QCP.

\begin{figure}
\includegraphics[width=\linewidth]{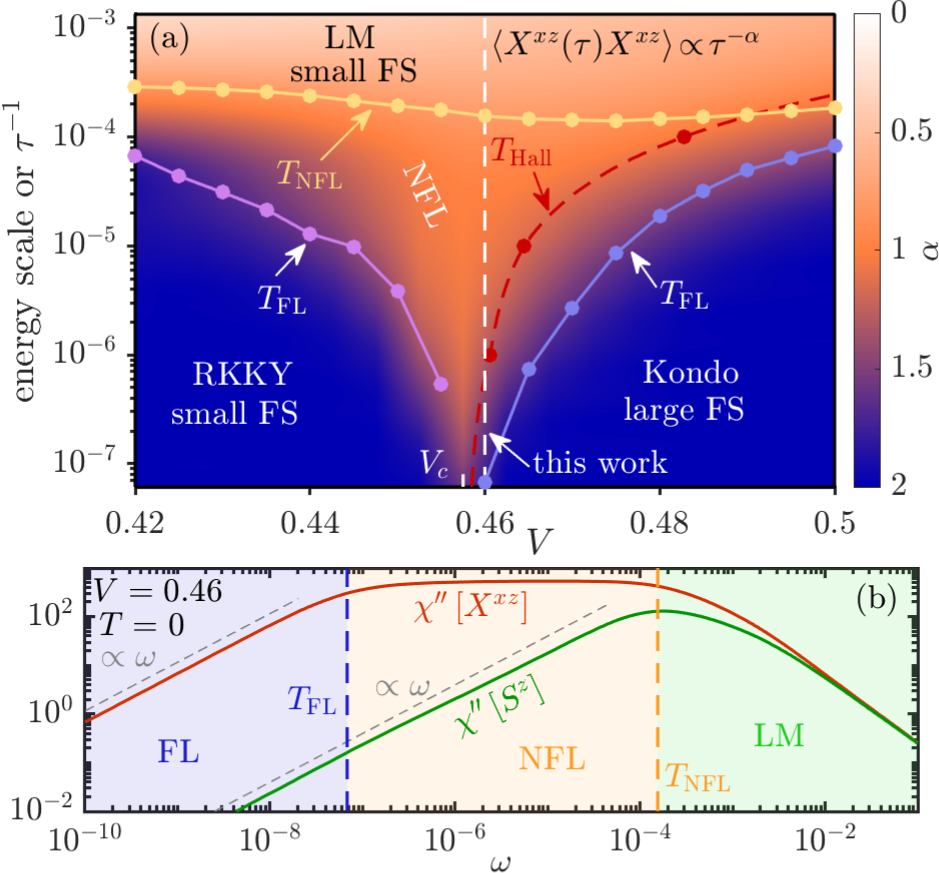}
\caption{%
(a) Phase diagram of the PAM obtained by 2CDMFT+NRG.
The dots (connected by lines as guides to the eye) denote 
relevant energy scales $\TFL$ and $\TNF$ below which we observe FL and NFL behavior, respectively, and $T_{\mr{Hall}}$,
the crossover scale between a large and small FS [see Ref.~\cite{Gleis2023} for details]. The color scale 
denotes the exponent $\alpha$ of the imaginary-time correlator 
$\left\langle X^{xz}(\tau) X^{xz} \right\rangle \propto \tau^{-\alpha}$.
The white dashed line denotes $V=0.46$, used for all subsequent plots 
in this work.
(b) Spectra of $X^{xz}$ and $S^{z}$ at $T=0$.
}
\label{fig:phase_diag}
\end{figure}

\textit{Phase diagram.---}%
Figure~\ref{fig:phase_diag}(a) shows our 2CDMFT+ NRG phase diagram in the $(V,T)$ plane close to the KB--QCP. At $T=0$, we find two Fermi liquid (FL) phases, separated by a KB--QCP located at $V_c = 0.4575(25)$, featuring a sudden Fermi surface~(FS) reconstruction~\cite{Gleis2023}. At finite excitation energies, we find two crossover scales, $\TFL(V)$ and $\TNF(V)$~\cite{Gleis2023}. 
FL behavior emerges below  $\TFL$, which decreases towards and vanishes at $V_c$. 
The high-energy region above $\TNF$ 
is characterized by thermally fluctuating $f$-electron local moments
decoupled from the $c$ electrons. $\TNF$ does not decrease for $V$ near $V_c$, hence strong scale separation between $\TNF$ and $\TFL$ occurs close to the QCP. 
For excitation energies 
between $\TFL$ and $\TNF$, 
we find NFL behavior---the main subject of this work.

\textit{Dynamical susceptibilities.---}%
The different regions can be most conveniently distinguished in terms of the dynamical behavior of response functions. 
For now, we focus on the staggered $f$-electron spin on a two-site cluster,
$
X^{xz} = S_1^z - S_2^z 
$, with
$
S_i^z = \tfrac{1}{2}\bigl[f^{\dag}_{i\uparrow}f^{\pdag}_{i\uparrow} - f^{\dag}_{i\downarrow}f^{\pdag}_{i\downarrow}\bigr]
$.
The color scale in Fig.~\ref{fig:phase_diag}(a) shows the exponent $\alpha$ of the imaginary-time autocorrelation function of $X^{xz}$,
$\langle X^{xz}(\tau)X^{xz} \rangle \propto \tau^{-\alpha}$, obtained via log-derivative. For long times, $\tau^{-1} < \TFL$, 
we find $\alpha = 2$, consistent with FL behavior and the presence of long-lived quasi-particles (QP)~\cite{Chowdhury2022} and thus quickly decaying, localized spin excitations. For short times, $\tau^{-1} > \TNF$, 
staggered spin excitations decay very slowly with an exponent $\alpha < 0.5$, consistent with local moment
behavior. For intermediate times, $\TFL < \tau^{-1} < \TNF$, 
we find an SYK-like exponent $\alpha \simeq 1$ in the NFL region, 
indicative of the absence of coherent QP~\cite{Chowdhury2022}. For at $V=0.46$, our data closest to $V_c$, this behavior 
extends over almost 4 orders of magnitude: in fact, our data suggests that it extends down to $\tau^{-1} \to 0$ at $V_c$, where $\TFL = 0$.
We note that we do \emph{not} find $\propto \tau^{-1/2}$ behavior of the single-electron Green's function $G(\tau)$, in contrast to the SYK model~\cite{Chowdhury2022}.
Thus, $\langle X^{xz}(\tau)X^{xz} \rangle$ is not $\propto G(\tau)^2$, i.e.,
the $\tau^{-1}$ behavior is governed by vertex contributions.

To understand the origin of the $\tau^{-1}$ dependence, we consider the spectral representation
of bosonic correlators,
\begin{equation}
\label{eq:imag_time_cor}
\langle \mc{A}^{\dag}(\tau) \mc{B} \rangle = \, \int_{-\infty}^{\infty} \mr{d}\omega \, \frac{\mr{e}^{-\tau \omega}}{1-\mr{e}^{-\beta \omega}} \, 
\chi'' [\mc{A},\mc{B}](\omega) \, .
\end{equation}
Here, the spectrum $\chi''(\omega)$ is obtained from the dynamical susceptibility $\chi(\omega) = \chi'(\omega) - \mr{i}\pi \chi''(\omega)$,
\begin{equation}
\label{eq:dynamical_susc}
\chi[\mc{A},\mc{B}](\omega) 
= -\mr{i} \int_{0}^{\infty} \mr{d}t \, \mr{e}^{\mr{i}(\omega+\mr{i}0^+)t} \left\langle\left[\mc{A}^{\dag}(t),\mc{B}\right]\right\rangle \, .
\end{equation}
We use the shorthand $\chi [\mc{A} ] (\omega) = \chi [ \mc{A}, \mc{A} ] (\omega)$.

The spectra for $X^{xz}$ and for the total spin $S^z = S^z_1 + S^z_2$ are shown 
in Fig.~\ref{fig:phase_diag}(b) at $V=0.46$ and $T=0$. The spectra 
$\chi''[X^{xz}]$ and $\chi''[S^z]$ both
show $\propto \omega$ behavior below 
$\TFL$, indicating that these fluctuations are screened in the FL, as expected. For 
long times, $\tau ^{-1} <\TFL$, the corresponding imaginary time correlation function Eq.~\eqref{eq:imag_time_cor} therefore decays as $\tau^{-2}$, as shown for $X^{xz}$ in Fig.~\ref{fig:phase_diag}(a). 

In the NFL region ($\TFL \! < \! \omega \! < \! \TNF$) 
 the spectra differ qualitatively: while $\chi''[S^z] \propto \omega$ still holds, 
$\chi''[X^{xz}]$ has 
an $\omega$-independent plateau; hence
$S^z$ fluctuations are screened, $X^{xz}$ fluctuations are
over-screened (reminiscent of the two-channel
 or two-impurity Kondo models \cite{Nozieres1974,Jones1988,Jones1989}). 
For intermediate times 
$\langle S^z(\tau) S^z \rangle$ 
thus decays as $\tau^{-2}$ (not shown) whereas 
$\langle X^{xz}(\tau) X^{xz} \rangle$ 
decays as $\tau^{-1}$ [cf.~Fig.~\ref{fig:phase_diag}(a)]. 
We note that besides $X^{xz}$, many other operators also have plateaus in their spectra, see Fig.~4 in Ref.~\cite{Gleis2023}.
Thus, the FL is reached via a two-stage screening process:
 as $\omega$ drops below $\TNF$, some excitations are screened, others over-screened; below $\TFL$, the latter are screened, too. 
\textit{Dynamical scaling.---}%
We now turn to the $T>0$ behavior, focusing on $V=0.46$ and $\chi[X^{xz}]$, whose spectrum shows a plateau in the NFL region at $T=0$. 
The $T$-dependent spectra $\chi''(\omega,T)$ and the corresponding real parts $\chi'(\omega,T)$ are shown in Figs.~\ref{fig:ChiS_scaling}(a) and (b), respectively.
As $T$ is decreased from around $\TNF$ to $\TFL$, the aforementioned plateau in $\chi''(\omega,T)$ emerges between $T < \omega < \TNF$, crossing over
to $\propto \omega$ behavior for $\omega < T$. For $T < \TFL$, the spectrum becomes $T$-independent, taking the same form as
already shown in Fig.~\ref{fig:phase_diag}(b) for $T=0$. 
$\chi'(\omega,T)$ is related to $\chi''(\omega,T)$ via a Kramers--Kronig relation. 
It thus shows a logarithmic~\cite{DefLogDep}
$\omega$-dependence 
for $\mr{max}(T,\TFL) < \omega < \TNF$ where $\chi''(\omega,T)$ has a plateau,  and is constant for $\omega < \mr{max}(T,\TFL)$ where $\chi''(\omega,T) \propto \omega$. 
As a result, $\chi'(0,T)$ [inset of Fig.~\ref{fig:ChiS_scaling}] has a $\propto \ln T$ dependence for 
$\inNFL{T}$ and is constant for 
$\inFL{T}$, 
where $X^{xz}$ fluctuations are screened.

\begin{figure}
\includegraphics[width=\linewidth]{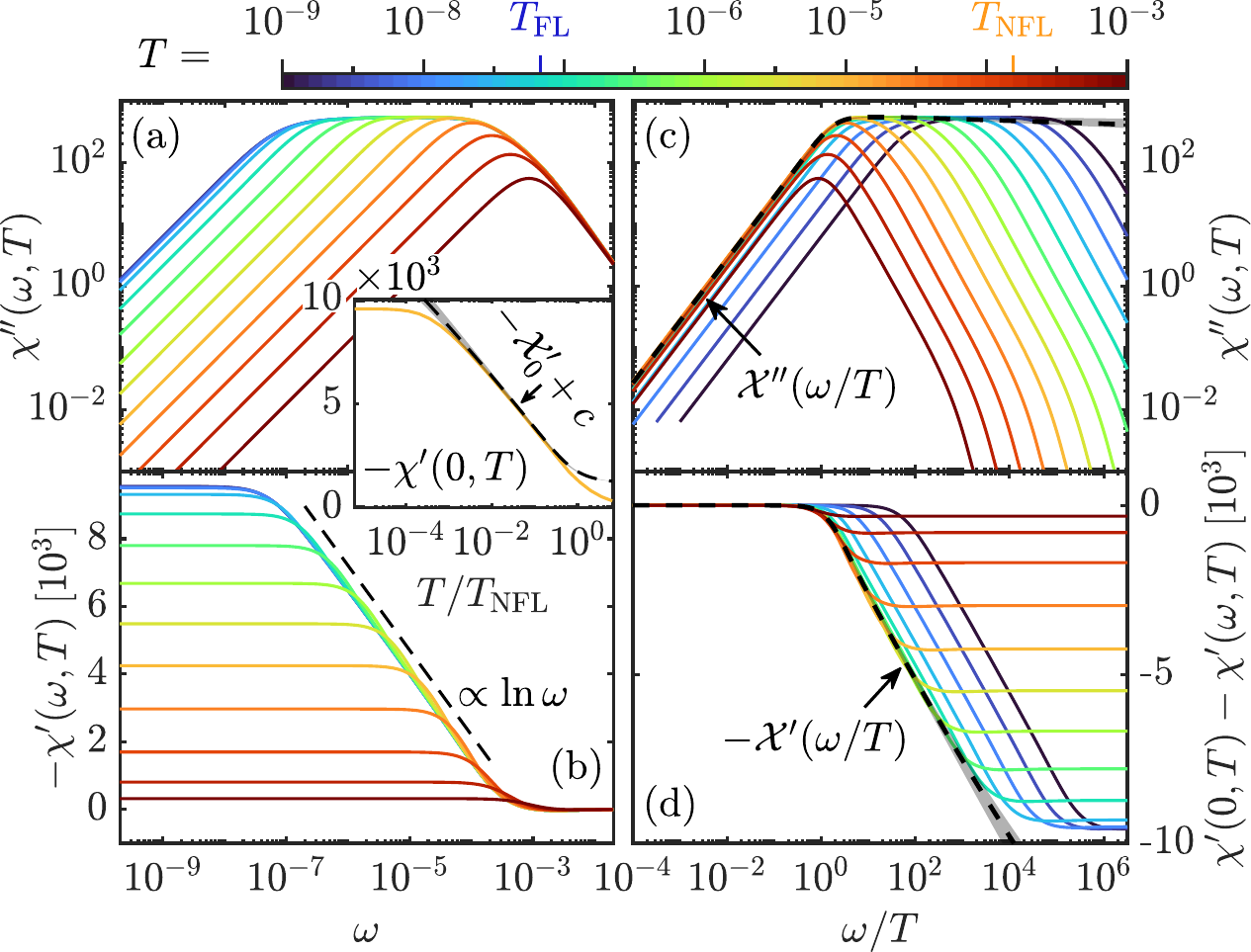}
\caption{Dynamical susceptibility $\chi[X^{xz}](\omega,T)$: (a) spectral part and (b) corresponding 
real part; (c,d) scaling collapse of spectral and real parts. Black dashed lines show the scaling functions $\mc{X}''(\omega/T)$ and $\mc{X}'(\omega/T)$,
respectively [cf.~Eq.~\eqref{eq:scaling_sus}]. Inset: $\chi'(0,T)$ (orange) and $\mc{X}'_0(T/\TNF)+c$ [black dashed, cf.~Eq.~\eqref{eq:scaling_sus}].
The constant shift $c$ accounts for spectral weight at $|\omega| > \TNF$. 
Grey areas indicate fitting uncertainties~\cite{supplement}.
}
\label{fig:ChiS_scaling}
\end{figure}

Figure~\ref{fig:ChiS_scaling}(c) shows $\chi''(\omega, T)$ vs.~$\omega/T$.
In the NFL region ($\inNFL{T}$, $|\omega| < \TNF$), the spectra all 
collapse onto a single curve. This demonstrates that the $T$-dependent spectra show dynamical scaling in the sense that in the NFL region,
$T^{\alpha} \chi''(\omega,T) = \mc{X}''(\omega/T)$ with $\alpha = 0$. Thus, $\chi''(\omega,T)$ depends on $\omega$ only via the ratio $\omega/T$, implying 
that $T$ is the only scale in this region. The scaling function $\mc{X}''(x)$ is flat for $x > 1$ and $\propto x$ for $x < 1$ (we discuss a phenomenological
fit below). The real part also shows $\omega/T$ scaling, $\chi'(\omega,T)  - \chi'(0,T) \simeq \mc{X}'(\omega/T)$. 
\textit{Scaling function and Planckian dissipation.---}%
In the NFL region ($\inNFL{T}$, $|\omega| < \TNF$), the spectra of dynamical susceptibilities showing plateaus (e.g., $\chi[X^{xz}]$)
can be fitted with a phenomenological ansatz for $\omega >  0$: 
\begin{align}
\label{eq:pheno_ansatz}
\widetilde{\chi}''(\omega,T) &= \chi_0  \! \int_{T}^{\TNF} \! \frac{\mr{d}\epsilon}{\pi}
\frac{(1-\mr{e}^{-\frac{\omega}{T}}) (\frac{\epsilon}{T})^{\nu} bT}{(\omega - a\epsilon)^2 + (b T)^2} \, .
\end{align}
$\omega < 0$ follows from anti-symmetry of $\widetilde{\chi}''$, the real part $\widetilde{\chi}'$ is
determined through a Kramers--Kronig relation. 
$\chi_0$, $a$, $b$, and $\nu$ are determined by fits to our spectra in the NFL region~\cite{supplement}. 
\nocite{Resta2018,Scalapino1993,Mitchell2014,Stadler2016,Kugler2022a,Lee2016}
We find $a\simeq 10^{-1}$, $b\simeq 1$ and $\nu \simeq 0$; $\chi_0$ determines the plateau value. (These parameters are $V$-independent within our fitting accuracy.)
When Eq.~\eqref{eq:pheno_ansatz} is evaluated for $|\omega|,T \ll \TNF$ 
one finds the scaling form
\begin{equation}
\label{eq:scaling_sus}
\widetilde{\chi}(\omega,T) \simeq 
\mc{X}'_0 \Big(\frac{T}{\TNF}\Big) + \mc{X}'\!\Big(\frac{\omega}{T}\Big)  - \mr{i}\pi \mc{X}''\!\Big(\frac{\omega}{T}\Big) \, .
\end{equation}
An explicit $T$-dependence, due to the high-energy cutoff $\TNF$, only enters via $\mc{X}'_0(T/\TNF) \simeq \widetilde{\chi}'(0,T)$; otherwise, $\widetilde{\chi}(\omega,T)$ only
depends on the ratio $\omega/T$ (for more information on the scaling functions $\mc{X}'_0$, $\mc{X}'$ and $\mc{X}''$, see Ref.~\cite{supplement}).
In Fig.~\ref{fig:ChiS_scaling}(c,d), we show that the scaling function $\mc{X}$ captures $\chi[X^{xz}]$
well in the NFL region (black dashed lines). 

The 
ansatz~\eqref{eq:pheno_ansatz} is motivated by a 
fit of $\langle X^{xz}(t) X^{xz} \rangle$ to a superposition of coherent excitations
with mean energy $a\epsilon$, decay rate $bT$ and density of states $(\epsilon/T)^{\nu}$~\cite{supplement}.
Since $b\simeq 1$, these
coherent excitations have a decay rate $\gamma \simeq T$ or correspondingly a lifetime $\tau \simeq 1/T$,
i.e., \emph{the longest-lived $X^{xz}$ excitations have a Planckian lifetime}. By contrast, we do \emph{not} observe a 
Planckian lifetime for single-particle excitations [cf.~Fig.~\ref{fig:optical}(d) and its discussion]. 
\textit{Optical conductivity.---}%
Our 2CDMFT approximation allows us to compute the \emph{local} current susceptibility $\chi[j_i^{a}](\omega,T)$ 
of the lattice model from the effective impurity model. Here, 
$
j^{a}_i = -\mr{i} t e \sum_{\sigma} \bigl( c^{\dag}_{i\sigma} c^{\pdag}_{i+\ahat\sigma} - c^{\dag}_{i+\ahat, \sigma} c^{\pdag}_{i\sigma} \bigr)
$
is the current operator in $a$-direction,
where $i$ and $i+\ahat$ are nearest neighbors on the lattice, 
chosen to also correspond to the two sites of the self-consistent impurity model.

For optical experiments and electronic transport, the \emph{uniform} current susceptibility $\chi[j^{a}_{\vec{q}=\vec{0}}](\omega,T)$
is relevant, where $j^{a}_{\vec{q}}$ is the $\vec{q}$-dependent current in $a$-direction, 
$j^{a}_{\vec{q}} = \frac{1}{N} \sum_{i\sigma} \mr{e}^{-\mr{i} \vec{q} \cdot \vec{r}_i} j^{a}_i$.
Assuming translation symmetry, 
$\chi[j^{a}_{\vec{0}}]$ 
can be expressed as a sum $\chi[j^{a}_i] + \chi_{\mr{nl}}[j]$ 
of local and nonlocal parts, with
$\chi_{\mr{nl}}[j] = \frac{1}{N} \sum_{\ell \neq i} \chi[j^{a}_{\ell},j^{a}_i] = \chi[j^{a}_{\vec{0}}] - \chi[j^{a}_i]$.
The computation of $\chi_{\mr{nl}}[j]$ would require four-point correlators~\cite{Lee2021,Kugler2021} for the self-consistent two-impurity model, 
which currently exceeds our computational resources. Hence 
we approximate it by its bubble contribution, $\chi_{\mr{nl,B}}[j] = \chi_{\mr{B}}[j^{a}_{\vec{0}}]-\chi_{\mr{B}}[j^{a}_i]$.
Thus, we use  
\begin{align}
\label{eq:chiJ_vertcorr}
\chi[j^{a}_{\vec{0}}] \approx 
\chi[j^{a}_i] + \chi_{\mr{nl,B}}[j] = \chi_{\mr{B}}[j^{a}_{\vec{0}}] + \chi_{\mr{vtx}}[j^{a}_i] \, ,
\end{align}
where $\chi_\vertex [j^{a}_i] = \chi[j^{a}_i] - \chi_{\mr{B}}[j^{a}_i]$ is the \emph{vertex} contribution to the local current susceptibility.
\begin{figure}
\includegraphics[width=\linewidth]{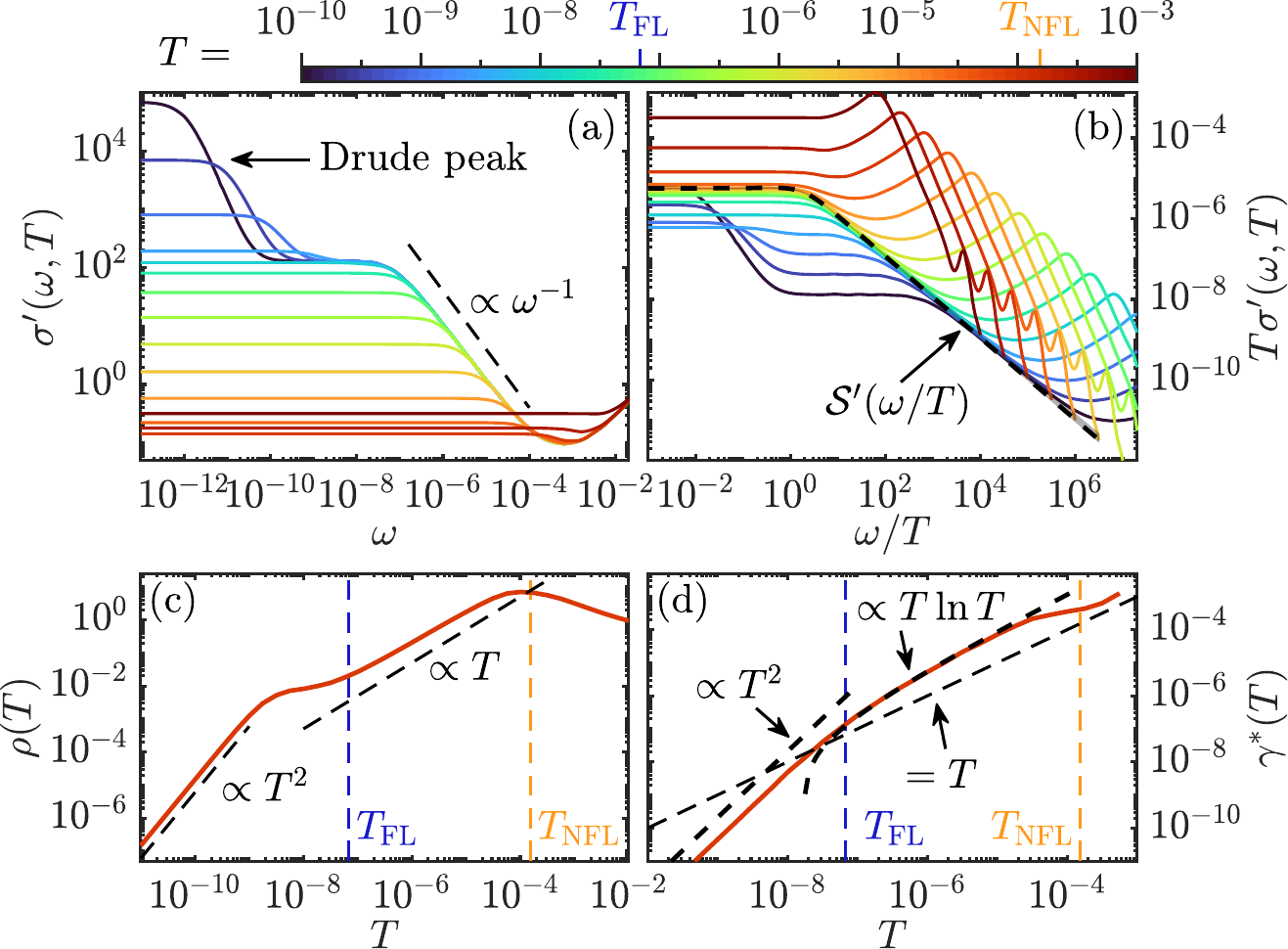}
\caption{
(a) Real part of the optical conductivity,
$\sigma'(\omega,T)$. 
(b) $\omega/T$ scaling of $T\sigma' (\omega,T)$; 
black dashed line: the scaling function $\mathcal{S}'$ of \Eq{eq:optical_scaling}.
(c) The resistivity $\rho(T)$. (d) The single-particle decay rate 
at the Fermi level, $\gamma^\ast (T)$.}
\label{fig:optical}
\end{figure}

The uniform current spectrum determines the real part of the optical conductivity, $\sigma'(\omega,T) = \frac{\pi}{\omega} \chi''[j^{a}_{\vec{0}}](\omega,T)$. It is shown in Fig.~\ref{fig:optical}(a). At $T \ll \TFL$ (blue/black), it features
a hybridization gap around $\omega \simeq \TNF$, $\omega^{-1}$ behavior for $\TFL < \omega < \TNF$, and a Drude peak at 
low frequencies below $\TFL$. These features emerge as the temperature is lowered from $T \gg \TNF$:
The hybridization gap forms around $T \simeq \TNF$ (red), the $\omega^{-1}$ feature emerges between $\TFL < T < \TNF$ (yellow/green)
and the Drude peak finally emerges for $T < \TFL$ (blue/black).

The most interesting feature is the $\omega^{-1}$ behavior in the NFL region. This feature is due to the fact 
 that $\chi''[j^{a}_i]$ (Fig.~S2 in Ref.~\cite{supplement}) exhibits a plateau similar to that of $\chi''[X^{xz}]$ [Fig.~\ref{fig:ChiS_scaling}(a,c)] for $|\omega|,T < \TNF$.
This plateau is entirely due to the vertex
contribution $\chi_\vertex''[j^{a}_i]$,
which in the NFL region completely
dominates the bubble contribution, 
$|\chi_\vertex''[j^{a}_i]| \gg |\chi''_{\mr{B}}[j^{a}_{\vec{0}}]|$~\cite{supplement}.
(The same is true  for $\chi[X^{xz}]$.)
Remarkably, $\chi''[j^{a}_i]$, just as $\chi''[X^{xz}]$, is well described by the ansatz~\eqref{eq:pheno_ansatz} (see Fig.~S6 of Ref.~\cite{supplement}),
implying $\omega/T$ scaling and Planckian dissipation of current fluctuations.
This 
implies that in the NFL region
$\TFL < T < \TNF$, 
$\sigma'(\omega,T)$ is governed by a scaling  function $\mc{S}'$: 
\begin{align}
\label{eq:optical_scaling}
T \sigma'(\omega,T) &= 
(T/\omega) \pi \mc{X}''(\omega/T) = \mc{S}'(\omega/T) \, .
\end{align}
Figure~\ref{fig:optical}(b) shows that $T\sigma'(\omega,T)$ 
is indeed well described by this scaling function (black dashed line).
We discuss scaling of the imaginary part $\sigma''(\omega,T)$ in 
Ref.~\cite{supplement}. 

The scaling behavior~\eqref{eq:optical_scaling} has 
two striking 
 implications for the NFL region $\TFL < T < \TNF$:
 First, a scaling collapse is achieved for 
$T^{\alpha} \sigma'(\omega,T)$ with $\alpha = 1$, an exponent which is also found experimentally~\cite{Prochaska2020,Yang2020,Li2023}. 
Second, the static conductivity $\sigma (T)= \sigma'( 0,T) = \mc{S}'(0)/T$ 
scales as $1/T$, implying $T$-linear behavior 
for the 
resistivity, $\rho (T)= 1/\sigma(T) \propto T$.  
This is born out in \Fig{fig:optical}(c):  $\rho(T)$ has a maximum around $\TNF$, where the hybridization gap forms, then
decreases with $T$ approximately $\propto T$ for $\TFL < T < \TNF$, before finally becoming $\propto T^2$ below $\TFL$.
In Ref.~\cite{supplement}, we analyze the complex optical conductivity, 
see Sec.~\ref{sec:compex_optical}, Fig.~\ref{fig:PAM_vs_CCI}. 
In the high-$T$ part of the NFL region $\TNF/10 \lesssim T \lesssim \TNF$, 
it shows qualitative similarities to data on {\CCI} of Ref.~\cite{Shi2023}: a dynamical transport scattering rate
$\propto \omega^2$, 
and a renormalized transport scattering rate $\propto T^2$.
In the FL region, on the other hand, the Drude peak and $\rho (T) \propto T^2$ behavior are due to the
nonlocal bubble part $\chi_{\mr{nl,B}}[j]$. These features can be understood from the single-particle decay rate~\cite{QPweightsPAM},
\begin{flalign}
& \gamma^{\ast} = Z \gamma \, , \quad \gamma =  \im \, G^{-1}_{\vec{k}_{\mr{F}}}(0) \, , \quad Z^{-1} =  \partial_{\omega} \re \, G^{-1}_{\vec{k}_{\mr{F}}}(0) \, , 
\hspace{-1cm} &
\end{flalign} 
shown in Fig.~\ref{fig:optical}(d).
In the FL region, $\gamma^{\ast} \propto T^{2}$ as expected, leading to a Drude peak of width $\propto T^2$ and $\rho(T) \propto T^2$, i.e., 
these features are due to long-lived coherent QP carrying the current. 
Since we neglect \textit{nonlocal} vertex contributions, the transport relaxation rate, and thus the $T^2$ prefactor of $\rho(T)$, is set purely by the QP decay rate and is therefore very likely overestimated~\cite{Coleman2015}.

In the NFL region, we find $Z \propto T$ and $\gamma \propto \ln T$, leading to $\gamma^{\ast} \propto T\ln T$. 
The latter is also found in 
the marginal FL~(MFL)~\cite{Varma1989} approach, but 
there, by contrast, one has  
$Z \propto \ln T$ and $\gamma \propto T$. 
Further, Fig.~\ref{fig:optical}(d) shows $\gamma^{\ast}(T) > T$ in the NFL region, i.e., single-particle excitations are \emph{not} Planckian
and decay faster than, for instance, current or $X^{xz}$ fluctuations. 
We emphasize that in the NFL region (in contrast to the FL region), the transport relaxation rate is \emph{not}
set by the single-particle decay rate: there, 
$\sigma (\omega, T)$ and thus $\rho (T)$
 are qualitatively influenced by the vertex contribution $\chi_\vertex''[j^{a}_i]$, as discussed above.

We conjecture that the following two features in Fig.~\ref{fig:optical}(c) are artifacts of neglecting \textit{nonlocal} vertex contributions:
First, $\rho(T)$ shows a slight deviation from 
perfect $T$-linear behavior. This deviation results from the fact that the \emph{bubble} part of the nonlocal current susceptibility,
$\chi_{\mr{B,nl}}[j]$, does not show $\omega/T$ scaling [not visible in Fig.~\ref{fig:optical}(b)].
In \Sec{sec:Jcorr_nonloc} of Ref.~\cite{supplement}, we provide indications that the full $\chi_{\mr{nl}}[j]$ does show scaling of the same type as $\chi''[j^{a}_i](\omega)$, which would imply perfect $T$-linear behavior. 
Second, $\rho(T)$ has a shoulder somewhat below $\TFL$. This likely reflects the above-mentioned overestimation of the $T^2$ prefactor of $\rho(T)$ in the FL region.

\textit{Discussion and Outlook.---}%
Our work provides a promising route towards an intrinsic strange metal. 
However, we have not yet achieved a full understanding of
the current decay mechanism. An inherent feature of (C)DMFT is 
that the interaction vertex does not ensure conservation of crystal momentum~\cite{Georges1996,Maier2005}.
Therefore, electron-electron scattering 
does not conserve crystal momentum, 
leading to current decay. 
This mechanism usually manifests as a dominant bubble contribution (in single-site DMFT, this is the only contribution).
A dominant bubble contribution is also
key to the Yukawa--SYK approach~\cite{Patel2023} to strange metals.
There,
a disordered Yukawa coupling leads to non-conserved momentum
in scattering processes. The result is a MFL 
where strange-metal scaling arises
in the bubble contribution and interaction disorder is needed to avoid its cancellation by
the vertex contribution. By contrast, in our 2CDMFT approach 
the strange-metal scaling in the NFL region arises 
\emph{entirely from the vertex contribution}, and not at all from the (much smaller) bubble contribution.
This strongly suggests that the current decay mechanism is not due to
the non-conservation of
crystal momentum at the interaction vertex. 
Our 2CDMFT approach also includes crystal momentum conserving Umklapp scattering
processes between momenta around $\vec{k} = (0,0,0)$ and $\vec{k} = (\pi,\pi,\pi)$ which flip the
current. We conjecture that these cause our observed strange-metal scaling.
A detailed analysis of the current decay mechanism is left for future work. 
This will involve studying (i) the frequency and temperature 
dependence of three- and four-point vertices in the NFL region and (ii) the relevance of Umklapp scattering in 2CDMFT.

\begin{acknowledgments}

We thank Assa Auerbach, Silke B\"uhler-Paschen, Andrey Chubukov, Piers Coleman, Dominic Else, Fabian Kugler, Antoine Georges, Sean Hartnoll, Andy Millis, Achim Rosch, Subir Sachdev, Qimiao Si, Katharina Stadler, Senthil Todadri, Matthias Vojta, Andreas Weichselbaum, and Krzysztof W\'ojcik for helpful discussions. This work was funded in part by the Deutsche Forschungsgemeinschaft under 
Germany's Excellence Strategy EXC-2111 (Project No.~390814868). It is part of the  Munich Quantum Valley, supported by the Bavarian state government with funds from the Hightech Agenda Bayern Plus. 
SSBL was supported by Samsung Electronics Co., Ltd (No.~IO220817-02066-01), and also by the National Research Foundation of Korea (NRF) grants funded by the Korean government (MEST) (No.~2019R1A6A1A10073437, No.~RS-2023-00214464).
The National Science Foundation supported GK under DMR-1733071,
and JvD in part under PHY-1748958.

\end{acknowledgments}
\bibliography{../../HeavyFermionCriticalityPRX/paper/references} 

\begin{thebibliography}{69}%
\makeatletter
\providecommand \@ifxundefined [1]{%
 \@ifx{#1\undefined}
}%
\providecommand \@ifnum [1]{%
 \ifnum #1\expandafter \@firstoftwo
 \else \expandafter \@secondoftwo
 \fi
}%
\providecommand \@ifx [1]{%
 \ifx #1\expandafter \@firstoftwo
 \else \expandafter \@secondoftwo
 \fi
}%
\providecommand \natexlab [1]{#1}%
\providecommand \enquote  [1]{``#1''}%
\providecommand \bibnamefont  [1]{#1}%
\providecommand \bibfnamefont [1]{#1}%
\providecommand \citenamefont [1]{#1}%
\providecommand \href@noop [0]{\@secondoftwo}%
\providecommand \href [0]{\begingroup \@sanitize@url \@href}%
\providecommand \@href[1]{\@@startlink{#1}\@@href}%
\providecommand \@@href[1]{\endgroup#1\@@endlink}%
\providecommand \@sanitize@url [0]{\catcode `\\12\catcode `\$12\catcode
  `\&12\catcode `\#12\catcode `\^12\catcode `\_12\catcode `\%12\relax}%
\providecommand \@@startlink[1]{}%
\providecommand \@@endlink[0]{}%
\providecommand \url  [0]{\begingroup\@sanitize@url \@url }%
\providecommand \@url [1]{\endgroup\@href {#1}{\urlprefix }}%
\providecommand \urlprefix  [0]{URL }%
\providecommand \Eprint [0]{\href }%
\providecommand \doibase [0]{https://doi.org/}%
\providecommand \selectlanguage [0]{\@gobble}%
\providecommand \bibinfo  [0]{\@secondoftwo}%
\providecommand \bibfield  [0]{\@secondoftwo}%
\providecommand \translation [1]{[#1]}%
\providecommand \BibitemOpen [0]{}%
\providecommand \bibitemStop [0]{}%
\providecommand \bibitemNoStop [0]{.\EOS\space}%
\providecommand \EOS [0]{\spacefactor3000\relax}%
\providecommand \BibitemShut  [1]{\csname bibitem#1\endcsname}%
\let\auto@bib@innerbib\@empty
\bibitem [{\citenamefont {Phillips}\ \emph {et~al.}(2022)\citenamefont
  {Phillips}, \citenamefont {Hussey},\ and\ \citenamefont
  {Abbamonte}}]{Phillips2022}%
  \BibitemOpen
  \bibfield  {author} {\bibinfo {author} {\bibfnamefont {P.~W.}\ \bibnamefont
  {Phillips}}, \bibinfo {author} {\bibfnamefont {N.~E.}\ \bibnamefont
  {Hussey}},\ and\ \bibinfo {author} {\bibfnamefont {P.}~\bibnamefont
  {Abbamonte}},\ }\bibfield  {title} {\bibinfo {title} {Stranger than metals},\
  }\href {https://doi.org/10.1126/science.abh4273} {\bibfield  {journal}
  {\bibinfo  {journal} {Science}\ }\textbf {\bibinfo {volume} {377}},\ \bibinfo
  {pages} {eabh4273} (\bibinfo {year} {2022})}\BibitemShut {NoStop}%
\bibitem [{\citenamefont {Hartnoll}\ and\ \citenamefont
  {Mackenzie}(2022)}]{Hartnoll2022}%
  \BibitemOpen
  \bibfield  {author} {\bibinfo {author} {\bibfnamefont {S.~A.}\ \bibnamefont
  {Hartnoll}}\ and\ \bibinfo {author} {\bibfnamefont {A.~P.}\ \bibnamefont
  {Mackenzie}},\ }\bibfield  {title} {\bibinfo {title} {Colloquium: Planckian
  dissipation in metals},\ }\href
  {https://doi.org/10.1103/RevModPhys.94.041002} {\bibfield  {journal}
  {\bibinfo  {journal} {Rev. Mod. Phys.}\ }\textbf {\bibinfo {volume} {94}},\
  \bibinfo {pages} {041002} (\bibinfo {year} {2022})}\BibitemShut {NoStop}%
\bibitem [{\citenamefont {Chowdhury}\ \emph {et~al.}(2022)\citenamefont
  {Chowdhury}, \citenamefont {Georges}, \citenamefont {Parcollet},\ and\
  \citenamefont {Sachdev}}]{Chowdhury2022}%
  \BibitemOpen
  \bibfield  {author} {\bibinfo {author} {\bibfnamefont {D.}~\bibnamefont
  {Chowdhury}}, \bibinfo {author} {\bibfnamefont {A.}~\bibnamefont {Georges}},
  \bibinfo {author} {\bibfnamefont {O.}~\bibnamefont {Parcollet}},\ and\
  \bibinfo {author} {\bibfnamefont {S.}~\bibnamefont {Sachdev}},\ }\bibfield
  {title} {\bibinfo {title} {{Sachdev-Ye-Kitaev} models and beyond: Window into
  non-{F}ermi liquids},\ }\href {https://doi.org/10.1103/RevModPhys.94.035004}
  {\bibfield  {journal} {\bibinfo  {journal} {Rev. Mod. Phys.}\ }\textbf
  {\bibinfo {volume} {94}},\ \bibinfo {pages} {035004} (\bibinfo {year}
  {2022})}\BibitemShut {NoStop}%
\bibitem [{\citenamefont {Zaanen}(2004)}]{Zaanen2004}%
  \BibitemOpen
  \bibfield  {author} {\bibinfo {author} {\bibfnamefont {J.}~\bibnamefont
  {Zaanen}},\ }\bibfield  {title} {\bibinfo {title} {Why the temperature is
  high},\ }\href {https://doi.org/10.1038/430512a} {\bibfield  {journal}
  {\bibinfo  {journal} {Nature}\ }\textbf {\bibinfo {volume} {430}},\ \bibinfo
  {pages} {512} (\bibinfo {year} {2004})}\BibitemShut {NoStop}%
\bibitem [{\citenamefont {Checkelsky}\ \emph {et~al.}(2024)\citenamefont
  {Checkelsky}, \citenamefont {Bernevig}, \citenamefont {Coleman},
  \citenamefont {Si},\ and\ \citenamefont {Paschen}}]{Checkelsky2024}%
  \BibitemOpen
  \bibfield  {author} {\bibinfo {author} {\bibfnamefont {J.~G.}\ \bibnamefont
  {Checkelsky}}, \bibinfo {author} {\bibfnamefont {B.~A.}\ \bibnamefont
  {Bernevig}}, \bibinfo {author} {\bibfnamefont {P.}~\bibnamefont {Coleman}},
  \bibinfo {author} {\bibfnamefont {Q.}~\bibnamefont {Si}},\ and\ \bibinfo
  {author} {\bibfnamefont {S.}~\bibnamefont {Paschen}},\ }\bibfield  {title}
  {\bibinfo {title} {Flat bands, strange metals and the {K}ondo effect},\
  }\bibfield  {journal} {\bibinfo  {journal} {Nat. Rev. Mater.}\ }\href
  {https://doi.org/10.1038/s41578-023-00644-z} {10.1038/s41578-023-00644-z}
  (\bibinfo {year} {2024})\BibitemShut {NoStop}%
\bibitem [{\citenamefont {Keimer}\ \emph {et~al.}(2015)\citenamefont {Keimer},
  \citenamefont {Kivelson}, \citenamefont {Norman}, \citenamefont {Uchida},\
  and\ \citenamefont {Zaanen}}]{Keimer2015}%
  \BibitemOpen
  \bibfield  {author} {\bibinfo {author} {\bibfnamefont {B.}~\bibnamefont
  {Keimer}}, \bibinfo {author} {\bibfnamefont {S.~A.}\ \bibnamefont
  {Kivelson}}, \bibinfo {author} {\bibfnamefont {M.~R.}\ \bibnamefont
  {Norman}}, \bibinfo {author} {\bibfnamefont {S.}~\bibnamefont {Uchida}},\
  and\ \bibinfo {author} {\bibfnamefont {J.}~\bibnamefont {Zaanen}},\
  }\bibfield  {title} {\bibinfo {title} {From quantum matter to
  high-temperature superconductivity in copper oxides},\ }\href
  {https://doi.org/10.1038/nature14165} {\bibfield  {journal} {\bibinfo
  {journal} {Nature}\ }\textbf {\bibinfo {volume} {518}},\ \bibinfo {pages}
  {179} (\bibinfo {year} {2015})}\BibitemShut {NoStop}%
\bibitem [{\citenamefont {Michon}\ \emph {et~al.}(2019)\citenamefont {Michon},
  \citenamefont {Girod}, \citenamefont {Badoux}, \citenamefont
  {Ka{\v{c}}mar{\v{c}}{\'i}k}, \citenamefont {Ma}, \citenamefont {Dragomir},
  \citenamefont {Dabkowska}, \citenamefont {Gaulin}, \citenamefont {Zhou},
  \citenamefont {Pyon}, \citenamefont {Takayama}, \citenamefont {Takagi},
  \citenamefont {Verret}, \citenamefont {Doiron-Leyraud}, \citenamefont
  {Marcenat}, \citenamefont {Taillefer},\ and\ \citenamefont
  {Klein}}]{Michon2019}%
  \BibitemOpen
  \bibfield  {author} {\bibinfo {author} {\bibfnamefont {B.}~\bibnamefont
  {Michon}}, \bibinfo {author} {\bibfnamefont {C.}~\bibnamefont {Girod}},
  \bibinfo {author} {\bibfnamefont {S.}~\bibnamefont {Badoux}}, \bibinfo
  {author} {\bibfnamefont {J.}~\bibnamefont {Ka{\v{c}}mar{\v{c}}{\'i}k}},
  \bibinfo {author} {\bibfnamefont {Q.}~\bibnamefont {Ma}}, \bibinfo {author}
  {\bibfnamefont {M.}~\bibnamefont {Dragomir}}, \bibinfo {author}
  {\bibfnamefont {H.~A.}\ \bibnamefont {Dabkowska}}, \bibinfo {author}
  {\bibfnamefont {B.~D.}\ \bibnamefont {Gaulin}}, \bibinfo {author}
  {\bibfnamefont {J.-S.}\ \bibnamefont {Zhou}}, \bibinfo {author}
  {\bibfnamefont {S.}~\bibnamefont {Pyon}}, \bibinfo {author} {\bibfnamefont
  {T.}~\bibnamefont {Takayama}}, \bibinfo {author} {\bibfnamefont
  {H.}~\bibnamefont {Takagi}}, \bibinfo {author} {\bibfnamefont
  {S.}~\bibnamefont {Verret}}, \bibinfo {author} {\bibfnamefont
  {N.}~\bibnamefont {Doiron-Leyraud}}, \bibinfo {author} {\bibfnamefont
  {C.}~\bibnamefont {Marcenat}}, \bibinfo {author} {\bibfnamefont
  {L.}~\bibnamefont {Taillefer}},\ and\ \bibinfo {author} {\bibfnamefont
  {T.}~\bibnamefont {Klein}},\ }\bibfield  {title} {\bibinfo {title}
  {Thermodynamic signatures of quantum criticality in cuprate
  superconductors},\ }\href {https://doi.org/10.1038/s41586-019-0932-x}
  {\bibfield  {journal} {\bibinfo  {journal} {Nature}\ }\textbf {\bibinfo
  {volume} {567}},\ \bibinfo {pages} {218} (\bibinfo {year}
  {2019})}\BibitemShut {NoStop}%
\bibitem [{\citenamefont {Michon}\ \emph {et~al.}(2023)\citenamefont {Michon},
  \citenamefont {Berthod}, \citenamefont {Rischau}, \citenamefont {Ataei},
  \citenamefont {Chen}, \citenamefont {Komiya}, \citenamefont {Ono},
  \citenamefont {Taillefer}, \citenamefont {van~der Marel},\ and\ \citenamefont
  {Georges}}]{Michon2023}%
  \BibitemOpen
  \bibfield  {author} {\bibinfo {author} {\bibfnamefont {B.}~\bibnamefont
  {Michon}}, \bibinfo {author} {\bibfnamefont {C.}~\bibnamefont {Berthod}},
  \bibinfo {author} {\bibfnamefont {C.~W.}\ \bibnamefont {Rischau}}, \bibinfo
  {author} {\bibfnamefont {A.}~\bibnamefont {Ataei}}, \bibinfo {author}
  {\bibfnamefont {L.}~\bibnamefont {Chen}}, \bibinfo {author} {\bibfnamefont
  {S.}~\bibnamefont {Komiya}}, \bibinfo {author} {\bibfnamefont
  {S.}~\bibnamefont {Ono}}, \bibinfo {author} {\bibfnamefont {L.}~\bibnamefont
  {Taillefer}}, \bibinfo {author} {\bibfnamefont {D.}~\bibnamefont {van~der
  Marel}},\ and\ \bibinfo {author} {\bibfnamefont {A.}~\bibnamefont
  {Georges}},\ }\bibfield  {title} {\bibinfo {title} {Reconciling scaling of
  the optical conductivity of cuprate superconductors with {P}lanckian
  resistivity and specific heat},\ }\href
  {https://doi.org/10.1038/s41467-023-38762-5} {\bibfield  {journal} {\bibinfo
  {journal} {Nat. Commun.}\ }\textbf {\bibinfo {volume} {14}},\ \bibinfo
  {pages} {3033} (\bibinfo {year} {2023})}\BibitemShut {NoStop}%
\bibitem [{\citenamefont {Legros}\ \emph {et~al.}(2019)\citenamefont {Legros},
  \citenamefont {Benhabib}, \citenamefont {Tabis}, \citenamefont
  {Lalibert{\'e}}, \citenamefont {Dion}, \citenamefont {Lizaire}, \citenamefont
  {Vignolle}, \citenamefont {Vignolles}, \citenamefont {Raffy}, \citenamefont
  {Li}, \citenamefont {Auban-Senzier}, \citenamefont {Doiron-Leyraud},
  \citenamefont {Fournier}, \citenamefont {Colson}, \citenamefont {Taillefer},\
  and\ \citenamefont {Proust}}]{Legros2019}%
  \BibitemOpen
  \bibfield  {author} {\bibinfo {author} {\bibfnamefont {A.}~\bibnamefont
  {Legros}}, \bibinfo {author} {\bibfnamefont {S.}~\bibnamefont {Benhabib}},
  \bibinfo {author} {\bibfnamefont {W.}~\bibnamefont {Tabis}}, \bibinfo
  {author} {\bibfnamefont {F.}~\bibnamefont {Lalibert{\'e}}}, \bibinfo {author}
  {\bibfnamefont {M.}~\bibnamefont {Dion}}, \bibinfo {author} {\bibfnamefont
  {M.}~\bibnamefont {Lizaire}}, \bibinfo {author} {\bibfnamefont
  {B.}~\bibnamefont {Vignolle}}, \bibinfo {author} {\bibfnamefont
  {D.}~\bibnamefont {Vignolles}}, \bibinfo {author} {\bibfnamefont
  {H.}~\bibnamefont {Raffy}}, \bibinfo {author} {\bibfnamefont {Z.~Z.}\
  \bibnamefont {Li}}, \bibinfo {author} {\bibfnamefont {P.}~\bibnamefont
  {Auban-Senzier}}, \bibinfo {author} {\bibfnamefont {N.}~\bibnamefont
  {Doiron-Leyraud}}, \bibinfo {author} {\bibfnamefont {P.}~\bibnamefont
  {Fournier}}, \bibinfo {author} {\bibfnamefont {D.}~\bibnamefont {Colson}},
  \bibinfo {author} {\bibfnamefont {L.}~\bibnamefont {Taillefer}},\ and\
  \bibinfo {author} {\bibfnamefont {C.}~\bibnamefont {Proust}},\ }\bibfield
  {title} {\bibinfo {title} {Universal {$T$}-linear resistivity and {P}lanckian
  dissipation in overdoped cuprates},\ }\href
  {https://doi.org/10.1038/s41567-018-0334-2} {\bibfield  {journal} {\bibinfo
  {journal} {Nat. Phys.}\ }\textbf {\bibinfo {volume} {15}},\ \bibinfo {pages}
  {142} (\bibinfo {year} {2019})}\BibitemShut {NoStop}%
\bibitem [{\citenamefont {Cao}\ \emph {et~al.}(2018)\citenamefont {Cao},
  \citenamefont {Fatemi}, \citenamefont {Fang}, \citenamefont {Watanabe},
  \citenamefont {Taniguchi}, \citenamefont {Kaxiras},\ and\ \citenamefont
  {Jarillo-Herrero}}]{Cao2018}%
  \BibitemOpen
  \bibfield  {author} {\bibinfo {author} {\bibfnamefont {Y.}~\bibnamefont
  {Cao}}, \bibinfo {author} {\bibfnamefont {V.}~\bibnamefont {Fatemi}},
  \bibinfo {author} {\bibfnamefont {S.}~\bibnamefont {Fang}}, \bibinfo {author}
  {\bibfnamefont {K.}~\bibnamefont {Watanabe}}, \bibinfo {author}
  {\bibfnamefont {T.}~\bibnamefont {Taniguchi}}, \bibinfo {author}
  {\bibfnamefont {E.}~\bibnamefont {Kaxiras}},\ and\ \bibinfo {author}
  {\bibfnamefont {P.}~\bibnamefont {Jarillo-Herrero}},\ }\bibfield  {title}
  {\bibinfo {title} {Unconventional superconductivity in magic-angle graphene
  superlattices},\ }\href {https://doi.org/10.1038/nature26160} {\bibfield
  {journal} {\bibinfo  {journal} {Nature}\ }\textbf {\bibinfo {volume} {556}},\
  \bibinfo {pages} {43} (\bibinfo {year} {2018})}\BibitemShut {NoStop}%
\bibitem [{\citenamefont {Cao}\ \emph {et~al.}(2020)\citenamefont {Cao},
  \citenamefont {Chowdhury}, \citenamefont {Rodan-Legrain}, \citenamefont
  {Rubies-Bigorda}, \citenamefont {Watanabe}, \citenamefont {Taniguchi},
  \citenamefont {Senthil},\ and\ \citenamefont {Jarillo-Herrero}}]{Cao2020}%
  \BibitemOpen
  \bibfield  {author} {\bibinfo {author} {\bibfnamefont {Y.}~\bibnamefont
  {Cao}}, \bibinfo {author} {\bibfnamefont {D.}~\bibnamefont {Chowdhury}},
  \bibinfo {author} {\bibfnamefont {D.}~\bibnamefont {Rodan-Legrain}}, \bibinfo
  {author} {\bibfnamefont {O.}~\bibnamefont {Rubies-Bigorda}}, \bibinfo
  {author} {\bibfnamefont {K.}~\bibnamefont {Watanabe}}, \bibinfo {author}
  {\bibfnamefont {T.}~\bibnamefont {Taniguchi}}, \bibinfo {author}
  {\bibfnamefont {T.}~\bibnamefont {Senthil}},\ and\ \bibinfo {author}
  {\bibfnamefont {P.}~\bibnamefont {Jarillo-Herrero}},\ }\bibfield  {title}
  {\bibinfo {title} {Strange metal in magic-angle graphene with near
  {P}lanckian dissipation},\ }\href
  {https://doi.org/10.1103/PhysRevLett.124.076801} {\bibfield  {journal}
  {\bibinfo  {journal} {Phys. Rev. Lett.}\ }\textbf {\bibinfo {volume} {124}},\
  \bibinfo {pages} {076801} (\bibinfo {year} {2020})}\BibitemShut {NoStop}%
\bibitem [{\citenamefont {Jaoui}\ \emph {et~al.}(2022)\citenamefont {Jaoui},
  \citenamefont {Das}, \citenamefont {Di~Battista}, \citenamefont
  {D{\'i}ez-M{\'e}rida}, \citenamefont {Lu}, \citenamefont {Watanabe},
  \citenamefont {Taniguchi}, \citenamefont {Ishizuka}, \citenamefont
  {Levitov},\ and\ \citenamefont {Efetov}}]{Jaoui2022}%
  \BibitemOpen
  \bibfield  {author} {\bibinfo {author} {\bibfnamefont {A.}~\bibnamefont
  {Jaoui}}, \bibinfo {author} {\bibfnamefont {I.}~\bibnamefont {Das}}, \bibinfo
  {author} {\bibfnamefont {G.}~\bibnamefont {Di~Battista}}, \bibinfo {author}
  {\bibfnamefont {J.}~\bibnamefont {D{\'i}ez-M{\'e}rida}}, \bibinfo {author}
  {\bibfnamefont {X.}~\bibnamefont {Lu}}, \bibinfo {author} {\bibfnamefont
  {K.}~\bibnamefont {Watanabe}}, \bibinfo {author} {\bibfnamefont
  {T.}~\bibnamefont {Taniguchi}}, \bibinfo {author} {\bibfnamefont
  {H.}~\bibnamefont {Ishizuka}}, \bibinfo {author} {\bibfnamefont
  {L.}~\bibnamefont {Levitov}},\ and\ \bibinfo {author} {\bibfnamefont {D.~K.}\
  \bibnamefont {Efetov}},\ }\bibfield  {title} {\bibinfo {title} {Quantum
  critical behaviour in magic-angle twisted bilayer graphene},\ }\href
  {https://doi.org/10.1038/s41567-022-01556-5} {\bibfield  {journal} {\bibinfo
  {journal} {Nat. Phys.}\ }\textbf {\bibinfo {volume} {18}},\ \bibinfo {pages}
  {633} (\bibinfo {year} {2022})}\BibitemShut {NoStop}%
\bibitem [{\citenamefont {von L\"ohneysen}(1996)}]{Loehneysen1996a}%
  \BibitemOpen
  \bibfield  {author} {\bibinfo {author} {\bibfnamefont {H.}~\bibnamefont {von
  L\"ohneysen}},\ }\bibfield  {title} {\bibinfo {title} {Non-{F}ermi-liquid
  behaviour in the heavy-fermion system {CeCu${}_{6-x}$Au${}_x$}},\ }\href
  {https://doi.org/10.1088/0953-8984/8/48/003} {\bibfield  {journal} {\bibinfo
  {journal} {J. Phys. Condens. Matter}\ }\textbf {\bibinfo {volume} {8}},\
  \bibinfo {pages} {9689} (\bibinfo {year} {1996})}\BibitemShut {NoStop}%
\bibitem [{\citenamefont {Trovarelli}\ \emph {et~al.}(2000)\citenamefont
  {Trovarelli}, \citenamefont {Geibel}, \citenamefont {Mederle}, \citenamefont
  {Langhammer}, \citenamefont {Grosche}, \citenamefont {Gegenwart},
  \citenamefont {Lang}, \citenamefont {Sparn},\ and\ \citenamefont
  {Steglich}}]{Trovarelli2000}%
  \BibitemOpen
  \bibfield  {author} {\bibinfo {author} {\bibfnamefont {O.}~\bibnamefont
  {Trovarelli}}, \bibinfo {author} {\bibfnamefont {C.}~\bibnamefont {Geibel}},
  \bibinfo {author} {\bibfnamefont {S.}~\bibnamefont {Mederle}}, \bibinfo
  {author} {\bibfnamefont {C.}~\bibnamefont {Langhammer}}, \bibinfo {author}
  {\bibfnamefont {F.~M.}\ \bibnamefont {Grosche}}, \bibinfo {author}
  {\bibfnamefont {P.}~\bibnamefont {Gegenwart}}, \bibinfo {author}
  {\bibfnamefont {M.}~\bibnamefont {Lang}}, \bibinfo {author} {\bibfnamefont
  {G.}~\bibnamefont {Sparn}},\ and\ \bibinfo {author} {\bibfnamefont
  {F.}~\bibnamefont {Steglich}},\ }\bibfield  {title} {\bibinfo {title}
  {{$\mathrm{YbRh}_2\mathrm{Si}_2$}: Pronounced non-{{F}ermi}-liquid effects
  above a low-lying magnetic phase transition},\ }\href
  {https://doi.org/10.1103/PhysRevLett.85.626} {\bibfield  {journal} {\bibinfo
  {journal} {Phys. Rev. Lett.}\ }\textbf {\bibinfo {volume} {85}},\ \bibinfo
  {pages} {626} (\bibinfo {year} {2000})}\BibitemShut {NoStop}%
\bibitem [{\citenamefont {Paglione}\ \emph {et~al.}(2003)\citenamefont
  {Paglione}, \citenamefont {Tanatar}, \citenamefont {Hawthorn}, \citenamefont
  {Boaknin}, \citenamefont {Hill}, \citenamefont {Ronning}, \citenamefont
  {Sutherland}, \citenamefont {Taillefer}, \citenamefont {Petrovic},\ and\
  \citenamefont {Canfield}}]{Paglione2003}%
  \BibitemOpen
  \bibfield  {author} {\bibinfo {author} {\bibfnamefont {J.}~\bibnamefont
  {Paglione}}, \bibinfo {author} {\bibfnamefont {M.~A.}\ \bibnamefont
  {Tanatar}}, \bibinfo {author} {\bibfnamefont {D.~G.}\ \bibnamefont
  {Hawthorn}}, \bibinfo {author} {\bibfnamefont {E.}~\bibnamefont {Boaknin}},
  \bibinfo {author} {\bibfnamefont {R.~W.}\ \bibnamefont {Hill}}, \bibinfo
  {author} {\bibfnamefont {F.}~\bibnamefont {Ronning}}, \bibinfo {author}
  {\bibfnamefont {M.}~\bibnamefont {Sutherland}}, \bibinfo {author}
  {\bibfnamefont {L.}~\bibnamefont {Taillefer}}, \bibinfo {author}
  {\bibfnamefont {C.}~\bibnamefont {Petrovic}},\ and\ \bibinfo {author}
  {\bibfnamefont {P.~C.}\ \bibnamefont {Canfield}},\ }\bibfield  {title}
  {\bibinfo {title} {Field-induced quantum critical point in
  {${\mathrm{C}\mathrm{e}\mathrm{C}\mathrm{o}\mathrm{I}\mathrm{n}}_{5}$}},\
  }\href {https://doi.org/10.1103/PhysRevLett.91.246405} {\bibfield  {journal}
  {\bibinfo  {journal} {Phys. Rev. Lett.}\ }\textbf {\bibinfo {volume} {91}},\
  \bibinfo {pages} {246405} (\bibinfo {year} {2003})}\BibitemShut {NoStop}%
\bibitem [{\citenamefont {Prochaska}\ \emph {et~al.}(2020)\citenamefont
  {Prochaska}, \citenamefont {Li}, \citenamefont {MacFarland}, \citenamefont
  {Andrews}, \citenamefont {Bonta}, \citenamefont {Bianco}, \citenamefont
  {Yazdi}, \citenamefont {Schrenk}, \citenamefont {Detz}, \citenamefont
  {Limbeck}, \citenamefont {Si}, \citenamefont {Ringe}, \citenamefont
  {Strasser}, \citenamefont {Kono},\ and\ \citenamefont
  {Paschen}}]{Prochaska2020}%
  \BibitemOpen
  \bibfield  {author} {\bibinfo {author} {\bibfnamefont {L.}~\bibnamefont
  {Prochaska}}, \bibinfo {author} {\bibfnamefont {X.}~\bibnamefont {Li}},
  \bibinfo {author} {\bibfnamefont {D.~C.}\ \bibnamefont {MacFarland}},
  \bibinfo {author} {\bibfnamefont {A.~M.}\ \bibnamefont {Andrews}}, \bibinfo
  {author} {\bibfnamefont {M.}~\bibnamefont {Bonta}}, \bibinfo {author}
  {\bibfnamefont {E.~F.}\ \bibnamefont {Bianco}}, \bibinfo {author}
  {\bibfnamefont {S.}~\bibnamefont {Yazdi}}, \bibinfo {author} {\bibfnamefont
  {W.}~\bibnamefont {Schrenk}}, \bibinfo {author} {\bibfnamefont
  {H.}~\bibnamefont {Detz}}, \bibinfo {author} {\bibfnamefont {A.}~\bibnamefont
  {Limbeck}}, \bibinfo {author} {\bibfnamefont {Q.}~\bibnamefont {Si}},
  \bibinfo {author} {\bibfnamefont {E.}~\bibnamefont {Ringe}}, \bibinfo
  {author} {\bibfnamefont {G.}~\bibnamefont {Strasser}}, \bibinfo {author}
  {\bibfnamefont {J.}~\bibnamefont {Kono}},\ and\ \bibinfo {author}
  {\bibfnamefont {S.}~\bibnamefont {Paschen}},\ }\bibfield  {title} {\bibinfo
  {title} {Singular charge fluctuations at a magnetic quantum critical point},\
  }\href {https://doi.org/10.1126/science.aag1595} {\bibfield  {journal}
  {\bibinfo  {journal} {Science}\ }\textbf {\bibinfo {volume} {367}},\ \bibinfo
  {pages} {285} (\bibinfo {year} {2020})}\BibitemShut {NoStop}%
\bibitem [{\citenamefont {Taupin}\ and\ \citenamefont
  {Paschen}(2022)}]{Taupin2022}%
  \BibitemOpen
  \bibfield  {author} {\bibinfo {author} {\bibfnamefont {M.}~\bibnamefont
  {Taupin}}\ and\ \bibinfo {author} {\bibfnamefont {S.}~\bibnamefont
  {Paschen}},\ }\bibfield  {title} {\bibinfo {title} {Are heavy fermion strange
  metals {Planckian}?},\ }\href {https://doi.org/10.3390/cryst12020251}
  {\bibfield  {journal} {\bibinfo  {journal} {Crystals}\ }\textbf {\bibinfo
  {volume} {12}},\ \bibinfo {pages} {251} (\bibinfo {year} {2022})}\BibitemShut
  {NoStop}%
\bibitem [{\citenamefont {Li}\ \emph {et~al.}(2023)\citenamefont {Li},
  \citenamefont {Kono}, \citenamefont {Si},\ and\ \citenamefont
  {Paschen}}]{Li2023}%
  \BibitemOpen
  \bibfield  {author} {\bibinfo {author} {\bibfnamefont {X.}~\bibnamefont
  {Li}}, \bibinfo {author} {\bibfnamefont {J.}~\bibnamefont {Kono}}, \bibinfo
  {author} {\bibfnamefont {Q.}~\bibnamefont {Si}},\ and\ \bibinfo {author}
  {\bibfnamefont {S.}~\bibnamefont {Paschen}},\ }\bibfield  {title} {\bibinfo
  {title} {Is the optical conductivity of heavy fermion strange metals
  {Planckian}?},\ }\href {https://doi.org/10.3389/femat.2022.934691} {\bibfield
   {journal} {\bibinfo  {journal} {Front. Electron. Mater}\ }\textbf {\bibinfo
  {volume} {2}},\ \bibinfo {pages} {934691} (\bibinfo {year}
  {2023})}\BibitemShut {NoStop}%
\bibitem [{\citenamefont {Nguyen}\ \emph {et~al.}(2021)\citenamefont {Nguyen},
  \citenamefont {Sidorenko}, \citenamefont {Taupin}, \citenamefont {Knebel},
  \citenamefont {Lapertot}, \citenamefont {Schuberth},\ and\ \citenamefont
  {Paschen}}]{Nguyen2021}%
  \BibitemOpen
  \bibfield  {author} {\bibinfo {author} {\bibfnamefont {D.~H.}\ \bibnamefont
  {Nguyen}}, \bibinfo {author} {\bibfnamefont {A.}~\bibnamefont {Sidorenko}},
  \bibinfo {author} {\bibfnamefont {M.}~\bibnamefont {Taupin}}, \bibinfo
  {author} {\bibfnamefont {G.}~\bibnamefont {Knebel}}, \bibinfo {author}
  {\bibfnamefont {G.}~\bibnamefont {Lapertot}}, \bibinfo {author}
  {\bibfnamefont {E.}~\bibnamefont {Schuberth}},\ and\ \bibinfo {author}
  {\bibfnamefont {S.}~\bibnamefont {Paschen}},\ }\bibfield  {title} {\bibinfo
  {title} {Superconductivity in an extreme strange metal},\ }\href
  {https://doi.org/10.1038/s41467-021-24670-z} {\bibfield  {journal} {\bibinfo
  {journal} {Nat. Commun.}\ }\textbf {\bibinfo {volume} {12}},\ \bibinfo
  {pages} {4341} (\bibinfo {year} {2021})}\BibitemShut {NoStop}%
\bibitem [{\citenamefont {Giuliani}\ and\ \citenamefont
  {Vignale}(2005)}]{Giuliani2005}%
  \BibitemOpen
  \bibfield  {author} {\bibinfo {author} {\bibfnamefont {G.}~\bibnamefont
  {Giuliani}}\ and\ \bibinfo {author} {\bibfnamefont {G.}~\bibnamefont
  {Vignale}},\ }\href {https://doi.org/10.1017/CBO9780511619915} {\emph
  {\bibinfo {title} {Quantum Theory of the Electron Liquid}}}\ (\bibinfo
  {publisher} {Cambridge University Press},\ \bibinfo {year}
  {2005})\BibitemShut {NoStop}%
\bibitem [{\citenamefont {Aronson}\ \emph {et~al.}(1995)\citenamefont
  {Aronson}, \citenamefont {Osborn}, \citenamefont {Robinson}, \citenamefont
  {Lynn}, \citenamefont {Chau}, \citenamefont {Seaman},\ and\ \citenamefont
  {Maple}}]{Aronson1995}%
  \BibitemOpen
  \bibfield  {author} {\bibinfo {author} {\bibfnamefont {M.~C.}\ \bibnamefont
  {Aronson}}, \bibinfo {author} {\bibfnamefont {R.}~\bibnamefont {Osborn}},
  \bibinfo {author} {\bibfnamefont {R.~A.}\ \bibnamefont {Robinson}}, \bibinfo
  {author} {\bibfnamefont {J.~W.}\ \bibnamefont {Lynn}}, \bibinfo {author}
  {\bibfnamefont {R.}~\bibnamefont {Chau}}, \bibinfo {author} {\bibfnamefont
  {C.~L.}\ \bibnamefont {Seaman}},\ and\ \bibinfo {author} {\bibfnamefont
  {M.~B.}\ \bibnamefont {Maple}},\ }\bibfield  {title} {\bibinfo {title}
  {Non-{F}ermi-liquid scaling of the magnetic response in {UCu
  ${}_{5-x}{\mathrm{Pd}}_{x}(x=1,1.5)$}},\ }\href
  {https://doi.org/10.1103/PhysRevLett.75.725} {\bibfield  {journal} {\bibinfo
  {journal} {Phys. Rev. Lett.}\ }\textbf {\bibinfo {volume} {75}},\ \bibinfo
  {pages} {725} (\bibinfo {year} {1995})}\BibitemShut {NoStop}%
\bibitem [{\citenamefont {Schr{\"o}der}\ \emph {et~al.}(2000)\citenamefont
  {Schr{\"o}der}, \citenamefont {Aeppli}, \citenamefont {Coldea}, \citenamefont
  {Adams}, \citenamefont {Stockert}, \citenamefont {L{\"o}hneysen},
  \citenamefont {Bucher}, \citenamefont {Ramazashvili},\ and\ \citenamefont
  {Coleman}}]{Schroeder2000}%
  \BibitemOpen
  \bibfield  {author} {\bibinfo {author} {\bibfnamefont {A.}~\bibnamefont
  {Schr{\"o}der}}, \bibinfo {author} {\bibfnamefont {G.}~\bibnamefont
  {Aeppli}}, \bibinfo {author} {\bibfnamefont {R.}~\bibnamefont {Coldea}},
  \bibinfo {author} {\bibfnamefont {M.}~\bibnamefont {Adams}}, \bibinfo
  {author} {\bibfnamefont {O.}~\bibnamefont {Stockert}}, \bibinfo {author}
  {\bibfnamefont {H.}~\bibnamefont {L{\"o}hneysen}}, \bibinfo {author}
  {\bibfnamefont {E.}~\bibnamefont {Bucher}}, \bibinfo {author} {\bibfnamefont
  {R.}~\bibnamefont {Ramazashvili}},\ and\ \bibinfo {author} {\bibfnamefont
  {P.}~\bibnamefont {Coleman}},\ }\bibfield  {title} {\bibinfo {title} {Onset
  of antiferromagnetism in heavy-fermion metals},\ }\href
  {https://doi.org/10.1038/35030039} {\bibfield  {journal} {\bibinfo  {journal}
  {Nature}\ }\textbf {\bibinfo {volume} {407}},\ \bibinfo {pages} {351}
  (\bibinfo {year} {2000})}\BibitemShut {NoStop}%
\bibitem [{\citenamefont {Poudel}\ \emph {et~al.}(2019)\citenamefont {Poudel},
  \citenamefont {Lawrence}, \citenamefont {Wu}, \citenamefont {Ehlers},
  \citenamefont {Qiu}, \citenamefont {May}, \citenamefont {Ronning},
  \citenamefont {Lumsden}, \citenamefont {Mandrus},\ and\ \citenamefont
  {Christianson}}]{Poudel2019}%
  \BibitemOpen
  \bibfield  {author} {\bibinfo {author} {\bibfnamefont {L.}~\bibnamefont
  {Poudel}}, \bibinfo {author} {\bibfnamefont {J.~M.}\ \bibnamefont
  {Lawrence}}, \bibinfo {author} {\bibfnamefont {L.~S.}\ \bibnamefont {Wu}},
  \bibinfo {author} {\bibfnamefont {G.}~\bibnamefont {Ehlers}}, \bibinfo
  {author} {\bibfnamefont {Y.}~\bibnamefont {Qiu}}, \bibinfo {author}
  {\bibfnamefont {A.~F.}\ \bibnamefont {May}}, \bibinfo {author} {\bibfnamefont
  {F.}~\bibnamefont {Ronning}}, \bibinfo {author} {\bibfnamefont {M.~D.}\
  \bibnamefont {Lumsden}}, \bibinfo {author} {\bibfnamefont {D.}~\bibnamefont
  {Mandrus}},\ and\ \bibinfo {author} {\bibfnamefont {A.~D.}\ \bibnamefont
  {Christianson}},\ }\bibfield  {title} {\bibinfo {title} {Multicomponent
  fluctuation spectrum at the quantum critical point in {CeCu$_{6-x}$Ag$_x$}},\
  }\href {https://doi.org/10.1038/s41535-019-0191-y} {\bibfield  {journal}
  {\bibinfo  {journal} {npj Quantum Mater.}\ }\textbf {\bibinfo {volume} {4}},\
  \bibinfo {pages} {52} (\bibinfo {year} {2019})}\BibitemShut {NoStop}%
\bibitem [{\citenamefont {Yang}\ \emph {et~al.}(2020)\citenamefont {Yang},
  \citenamefont {Pal}, \citenamefont {Zamani}, \citenamefont {Kliemt},
  \citenamefont {Krellner}, \citenamefont {Stockert}, \citenamefont
  {L\"ohneysen}, \citenamefont {Kroha},\ and\ \citenamefont
  {Fiebig}}]{Yang2020}%
  \BibitemOpen
  \bibfield  {author} {\bibinfo {author} {\bibfnamefont {C.-J.}\ \bibnamefont
  {Yang}}, \bibinfo {author} {\bibfnamefont {S.}~\bibnamefont {Pal}}, \bibinfo
  {author} {\bibfnamefont {F.}~\bibnamefont {Zamani}}, \bibinfo {author}
  {\bibfnamefont {K.}~\bibnamefont {Kliemt}}, \bibinfo {author} {\bibfnamefont
  {C.}~\bibnamefont {Krellner}}, \bibinfo {author} {\bibfnamefont
  {O.}~\bibnamefont {Stockert}}, \bibinfo {author} {\bibfnamefont {H.~v.}\
  \bibnamefont {L\"ohneysen}}, \bibinfo {author} {\bibfnamefont
  {J.}~\bibnamefont {Kroha}},\ and\ \bibinfo {author} {\bibfnamefont
  {M.}~\bibnamefont {Fiebig}},\ }\bibfield  {title} {\bibinfo {title}
  {Terahertz conductivity of heavy-fermion systems from time-resolved
  spectroscopy},\ }\href {https://doi.org/10.1103/PhysRevResearch.2.033296}
  {\bibfield  {journal} {\bibinfo  {journal} {Phys. Rev. Research}\ }\textbf
  {\bibinfo {volume} {2}},\ \bibinfo {pages} {033296} (\bibinfo {year}
  {2020})}\BibitemShut {NoStop}%
\bibitem [{\citenamefont {Chen}\ \emph {et~al.}()\citenamefont {Chen},
  \citenamefont {Lowder}, \citenamefont {Bakali}, \citenamefont {Andrews},
  \citenamefont {Schrenk}, \citenamefont {Waas}, \citenamefont {Svagera},
  \citenamefont {Eguchi}, \citenamefont {Prochaska}, \citenamefont {Wang},
  \citenamefont {Setty}, \citenamefont {Sur}, \citenamefont {Si}, \citenamefont
  {Paschen},\ and\ \citenamefont {Natelson}}]{Chen2022}%
  \BibitemOpen
  \bibfield  {author} {\bibinfo {author} {\bibfnamefont {L.}~\bibnamefont
  {Chen}}, \bibinfo {author} {\bibfnamefont {D.~T.}\ \bibnamefont {Lowder}},
  \bibinfo {author} {\bibfnamefont {E.}~\bibnamefont {Bakali}}, \bibinfo
  {author} {\bibfnamefont {A.~M.}\ \bibnamefont {Andrews}}, \bibinfo {author}
  {\bibfnamefont {W.}~\bibnamefont {Schrenk}}, \bibinfo {author} {\bibfnamefont
  {M.}~\bibnamefont {Waas}}, \bibinfo {author} {\bibfnamefont {R.}~\bibnamefont
  {Svagera}}, \bibinfo {author} {\bibfnamefont {G.}~\bibnamefont {Eguchi}},
  \bibinfo {author} {\bibfnamefont {L.}~\bibnamefont {Prochaska}}, \bibinfo
  {author} {\bibfnamefont {Y.}~\bibnamefont {Wang}}, \bibinfo {author}
  {\bibfnamefont {C.}~\bibnamefont {Setty}}, \bibinfo {author} {\bibfnamefont
  {S.}~\bibnamefont {Sur}}, \bibinfo {author} {\bibfnamefont {Q.}~\bibnamefont
  {Si}}, \bibinfo {author} {\bibfnamefont {S.}~\bibnamefont {Paschen}},\ and\
  \bibinfo {author} {\bibfnamefont {D.}~\bibnamefont {Natelson}},\ }\bibfield
  {title} {\bibinfo {title} {Shot noise in a strange metal},\ }\href
  {http://arxiv.org/abs/2206.00673} {\ }\Eprint
  {https://arxiv.org/abs/2206.00673} {arXiv:2206.00673} \BibitemShut {NoStop}%
\bibitem [{\citenamefont {Else}\ \emph {et~al.}(2021)\citenamefont {Else},
  \citenamefont {Thorngren},\ and\ \citenamefont {Senthil}}]{Else2021a}%
  \BibitemOpen
  \bibfield  {author} {\bibinfo {author} {\bibfnamefont {D.~V.}\ \bibnamefont
  {Else}}, \bibinfo {author} {\bibfnamefont {R.}~\bibnamefont {Thorngren}},\
  and\ \bibinfo {author} {\bibfnamefont {T.}~\bibnamefont {Senthil}},\
  }\bibfield  {title} {\bibinfo {title} {Non-{F}ermi liquids as ersatz {F}ermi
  liquids: General constraints on compressible metals},\ }\href
  {https://doi.org/10.1103/PhysRevX.11.021005} {\bibfield  {journal} {\bibinfo
  {journal} {Phys. Rev. X}\ }\textbf {\bibinfo {volume} {11}},\ \bibinfo
  {pages} {021005} (\bibinfo {year} {2021})}\BibitemShut {NoStop}%
\bibitem [{\citenamefont {Else}\ and\ \citenamefont
  {Senthil}(2021)}]{Else2021b}%
  \BibitemOpen
  \bibfield  {author} {\bibinfo {author} {\bibfnamefont {D.~V.}\ \bibnamefont
  {Else}}\ and\ \bibinfo {author} {\bibfnamefont {T.}~\bibnamefont {Senthil}},\
  }\bibfield  {title} {\bibinfo {title} {Strange metals as ersatz {F}ermi
  liquids},\ }\href {https://doi.org/10.1103/PhysRevLett.127.086601} {\bibfield
   {journal} {\bibinfo  {journal} {Phys. Rev. Lett.}\ }\textbf {\bibinfo
  {volume} {127}},\ \bibinfo {pages} {086601} (\bibinfo {year}
  {2021})}\BibitemShut {NoStop}%
\bibitem [{\citenamefont {Patel}\ \emph {et~al.}(2023)\citenamefont {Patel},
  \citenamefont {Guo}, \citenamefont {Esterlis},\ and\ \citenamefont
  {Sachdev}}]{Patel2023}%
  \BibitemOpen
  \bibfield  {author} {\bibinfo {author} {\bibfnamefont {A.~A.}\ \bibnamefont
  {Patel}}, \bibinfo {author} {\bibfnamefont {H.}~\bibnamefont {Guo}}, \bibinfo
  {author} {\bibfnamefont {I.}~\bibnamefont {Esterlis}},\ and\ \bibinfo
  {author} {\bibfnamefont {S.}~\bibnamefont {Sachdev}},\ }\bibfield  {title}
  {\bibinfo {title} {Universal theory of strange metals from spatially random
  interactions},\ }\href {https://doi.org/10.1126/science.abq6011} {\bibfield
  {journal} {\bibinfo  {journal} {Science}\ }\textbf {\bibinfo {volume}
  {381}},\ \bibinfo {pages} {790} (\bibinfo {year} {2023})}\BibitemShut
  {NoStop}%
\bibitem [{\citenamefont {Aldape}\ \emph {et~al.}(2022)\citenamefont {Aldape},
  \citenamefont {Cookmeyer}, \citenamefont {Patel},\ and\ \citenamefont
  {Altman}}]{Aldape2022}%
  \BibitemOpen
  \bibfield  {author} {\bibinfo {author} {\bibfnamefont {E.~E.}\ \bibnamefont
  {Aldape}}, \bibinfo {author} {\bibfnamefont {T.}~\bibnamefont {Cookmeyer}},
  \bibinfo {author} {\bibfnamefont {A.~A.}\ \bibnamefont {Patel}},\ and\
  \bibinfo {author} {\bibfnamefont {E.}~\bibnamefont {Altman}},\ }\bibfield
  {title} {\bibinfo {title} {Solvable theory of a strange metal at the
  breakdown of a heavy {F}ermi liquid},\ }\href
  {https://doi.org/10.1103/PhysRevB.105.235111} {\bibfield  {journal} {\bibinfo
   {journal} {Phys. Rev. B}\ }\textbf {\bibinfo {volume} {105}},\ \bibinfo
  {pages} {235111} (\bibinfo {year} {2022})}\BibitemShut {NoStop}%
\bibitem [{\citenamefont {Rullier-Albenque}\ \emph {et~al.}(2000)\citenamefont
  {Rullier-Albenque}, \citenamefont {Vieillefond}, \citenamefont {Alloul},
  \citenamefont {Tyler}, \citenamefont {Lejay},\ and\ \citenamefont
  {Marucco}}]{Rullier-Albenque2000}%
  \BibitemOpen
  \bibfield  {author} {\bibinfo {author} {\bibfnamefont {F.}~\bibnamefont
  {Rullier-Albenque}}, \bibinfo {author} {\bibfnamefont {P.~A.}\ \bibnamefont
  {Vieillefond}}, \bibinfo {author} {\bibfnamefont {H.}~\bibnamefont {Alloul}},
  \bibinfo {author} {\bibfnamefont {A.~W.}\ \bibnamefont {Tyler}}, \bibinfo
  {author} {\bibfnamefont {P.}~\bibnamefont {Lejay}},\ and\ \bibinfo {author}
  {\bibfnamefont {J.~F.}\ \bibnamefont {Marucco}},\ }\bibfield  {title}
  {\bibinfo {title} {Universal {$T_c$} depression by irradiation defects in
  underdoped and overdoped cuprates?},\ }\href
  {https://doi.org/10.1209/epl/i2000-00238-x} {\bibfield  {journal} {\bibinfo
  {journal} {Europhys. Lett.}\ }\textbf {\bibinfo {volume} {50}},\ \bibinfo
  {pages} {81} (\bibinfo {year} {2000})}\BibitemShut {NoStop}%
\bibitem [{\citenamefont {Kotliar}\ \emph {et~al.}(2001)\citenamefont
  {Kotliar}, \citenamefont {Savrasov}, \citenamefont {P{\'a}lsson},\ and\
  \citenamefont {Biroli}}]{Kotliar2001}%
  \BibitemOpen
  \bibfield  {author} {\bibinfo {author} {\bibfnamefont {G.}~\bibnamefont
  {Kotliar}}, \bibinfo {author} {\bibfnamefont {S.~Y.}\ \bibnamefont
  {Savrasov}}, \bibinfo {author} {\bibfnamefont {G.}~\bibnamefont
  {P{\'a}lsson}},\ and\ \bibinfo {author} {\bibfnamefont {G.}~\bibnamefont
  {Biroli}},\ }\bibfield  {title} {\bibinfo {title} {Cellular dynamical mean
  field approach to strongly correlated systems},\ }\href
  {https://doi.org/10.1103/PhysRevLett.87.186401} {\bibfield  {journal}
  {\bibinfo  {journal} {Phys. Rev. Lett.}\ }\textbf {\bibinfo {volume} {87}},\
  \bibinfo {pages} {186401} (\bibinfo {year} {2001})}\BibitemShut {NoStop}%
\bibitem [{\citenamefont {Maier}\ \emph {et~al.}(2005)\citenamefont {Maier},
  \citenamefont {Jarrell}, \citenamefont {Pruschke},\ and\ \citenamefont
  {Hettler}}]{Maier2005}%
  \BibitemOpen
  \bibfield  {author} {\bibinfo {author} {\bibfnamefont {T.}~\bibnamefont
  {Maier}}, \bibinfo {author} {\bibfnamefont {M.}~\bibnamefont {Jarrell}},
  \bibinfo {author} {\bibfnamefont {T.}~\bibnamefont {Pruschke}},\ and\
  \bibinfo {author} {\bibfnamefont {M.~H.}\ \bibnamefont {Hettler}},\
  }\bibfield  {title} {\bibinfo {title} {Quantum cluster theories},\ }\href
  {https://doi.org/10.1103/RevModPhys.77.1027} {\bibfield  {journal} {\bibinfo
  {journal} {Rev. Mod. Phys.}\ }\textbf {\bibinfo {volume} {77}},\ \bibinfo
  {pages} {1027} (\bibinfo {year} {2005})}\BibitemShut {NoStop}%
\bibitem [{\citenamefont {Si}\ \emph {et~al.}(2001)\citenamefont {Si},
  \citenamefont {Rabello}, \citenamefont {Ingersent},\ and\ \citenamefont
  {Smith}}]{Si2001}%
  \BibitemOpen
  \bibfield  {author} {\bibinfo {author} {\bibfnamefont {Q.}~\bibnamefont
  {Si}}, \bibinfo {author} {\bibfnamefont {S.}~\bibnamefont {Rabello}},
  \bibinfo {author} {\bibfnamefont {K.}~\bibnamefont {Ingersent}},\ and\
  \bibinfo {author} {\bibfnamefont {J.~L.}\ \bibnamefont {Smith}},\ }\bibfield
  {title} {\bibinfo {title} {Locally critical quantum phase transitions in
  strongly correlated metals},\ }\href {https://doi.org/10.1038/35101507}
  {\bibfield  {journal} {\bibinfo  {journal} {Nature}\ }\textbf {\bibinfo
  {volume} {413}},\ \bibinfo {pages} {804} (\bibinfo {year}
  {2001})}\BibitemShut {NoStop}%
\bibitem [{\citenamefont {Coleman}\ \emph {et~al.}(2001)\citenamefont
  {Coleman}, \citenamefont {P{\'e}pin}, \citenamefont {Si},\ and\ \citenamefont
  {Ramazashvili}}]{Coleman2001}%
  \BibitemOpen
  \bibfield  {author} {\bibinfo {author} {\bibfnamefont {P.}~\bibnamefont
  {Coleman}}, \bibinfo {author} {\bibfnamefont {C.}~\bibnamefont {P{\'e}pin}},
  \bibinfo {author} {\bibfnamefont {Q.}~\bibnamefont {Si}},\ and\ \bibinfo
  {author} {\bibfnamefont {R.}~\bibnamefont {Ramazashvili}},\ }\bibfield
  {title} {\bibinfo {title} {How do {{F}ermi} liquids get heavy and die?},\
  }\href {https://doi.org/10.1088/0953-8984/13/35/202} {\bibfield  {journal}
  {\bibinfo  {journal} {J. Phys. Condens. Matter}\ }\textbf {\bibinfo {volume}
  {13}},\ \bibinfo {pages} {R723} (\bibinfo {year} {2001})}\BibitemShut
  {NoStop}%
\bibitem [{\citenamefont {Coleman}\ and\ \citenamefont
  {P\'epin}(2002)}]{Coleman2002}%
  \BibitemOpen
  \bibfield  {author} {\bibinfo {author} {\bibfnamefont {P.}~\bibnamefont
  {Coleman}}\ and\ \bibinfo {author} {\bibfnamefont {C.}~\bibnamefont
  {P\'epin}},\ }\bibfield  {title} {\bibinfo {title} {What is the fate of the
  heavy electron at a quantum critical point?},\ }\href
  {https://doi.org/https://doi.org/10.1016/S0921-4526(01)01342-4} {\bibfield
  {journal} {\bibinfo  {journal} {Phys. B Condens. Matter}\ }\textbf {\bibinfo
  {volume} {312-313}},\ \bibinfo {pages} {383} (\bibinfo {year} {2002})},\
  \bibinfo {note} {the International Conference on Strongly Correlated Electron
  Systems}\BibitemShut {NoStop}%
\bibitem [{\citenamefont {Tanaskovi{\'c}}\ \emph {et~al.}(2011)\citenamefont
  {Tanaskovi{\'c}}, \citenamefont {Haule}, \citenamefont {Kotliar},\ and\
  \citenamefont {Dobrosavljevi{\'c}}}]{Tanaskovic2011}%
  \BibitemOpen
  \bibfield  {author} {\bibinfo {author} {\bibfnamefont {D.}~\bibnamefont
  {Tanaskovi{\'c}}}, \bibinfo {author} {\bibfnamefont {K.}~\bibnamefont
  {Haule}}, \bibinfo {author} {\bibfnamefont {G.}~\bibnamefont {Kotliar}},\
  and\ \bibinfo {author} {\bibfnamefont {V.}~\bibnamefont
  {Dobrosavljevi{\'c}}},\ }\bibfield  {title} {\bibinfo {title} {Phase diagram,
  energy scales, and nonlocal correlations in the {{A}nderson} lattice model},\
  }\href {https://doi.org/10.1103/PhysRevB.84.115105} {\bibfield  {journal}
  {\bibinfo  {journal} {Phys. Rev. B}\ }\textbf {\bibinfo {volume} {84}},\
  \bibinfo {pages} {115105} (\bibinfo {year} {2011})}\BibitemShut {NoStop}%
\bibitem [{\citenamefont {De~Leo}\ \emph
  {et~al.}(2008{\natexlab{a}})\citenamefont {De~Leo}, \citenamefont {Civelli},\
  and\ \citenamefont {Kotliar}}]{DeLeo2008}%
  \BibitemOpen
  \bibfield  {author} {\bibinfo {author} {\bibfnamefont {L.}~\bibnamefont
  {De~Leo}}, \bibinfo {author} {\bibfnamefont {M.}~\bibnamefont {Civelli}},\
  and\ \bibinfo {author} {\bibfnamefont {G.}~\bibnamefont {Kotliar}},\
  }\bibfield  {title} {\bibinfo {title} {Cellular dynamical mean-field theory
  of the periodic {{A}nderson} model},\ }\href
  {https://doi.org/10.1103/PhysRevB.77.075107} {\bibfield  {journal} {\bibinfo
  {journal} {Phys. Rev. B}\ }\textbf {\bibinfo {volume} {77}},\ \bibinfo
  {pages} {075107} (\bibinfo {year} {2008}{\natexlab{a}})}\BibitemShut
  {NoStop}%
\bibitem [{\citenamefont {De~Leo}\ \emph
  {et~al.}(2008{\natexlab{b}})\citenamefont {De~Leo}, \citenamefont {Civelli},\
  and\ \citenamefont {Kotliar}}]{DeLeo2008a}%
  \BibitemOpen
  \bibfield  {author} {\bibinfo {author} {\bibfnamefont {L.}~\bibnamefont
  {De~Leo}}, \bibinfo {author} {\bibfnamefont {M.}~\bibnamefont {Civelli}},\
  and\ \bibinfo {author} {\bibfnamefont {G.}~\bibnamefont {Kotliar}},\
  }\bibfield  {title} {\bibinfo {title} {{$T =0$} heavy-fermion quantum
  critical point as an orbital-selective {M}ott transition},\ }\href
  {https://doi.org/10.1103/PhysRevLett.101.256404} {\bibfield  {journal}
  {\bibinfo  {journal} {Phys. Rev. Lett.}\ }\textbf {\bibinfo {volume} {101}},\
  \bibinfo {pages} {256404} (\bibinfo {year} {2008}{\natexlab{b}})}\BibitemShut
  {NoStop}%
\bibitem [{\citenamefont {Gleis}\ \emph {et~al.}()\citenamefont {Gleis},
  \citenamefont {Lee}, \citenamefont {Kotliar},\ and\ \citenamefont {von
  Delft}}]{Gleis2023}%
  \BibitemOpen
  \bibfield  {author} {\bibinfo {author} {\bibfnamefont {A.}~\bibnamefont
  {Gleis}}, \bibinfo {author} {\bibfnamefont {S.-S.~B.}\ \bibnamefont {Lee}},
  \bibinfo {author} {\bibfnamefont {G.}~\bibnamefont {Kotliar}},\ and\ \bibinfo
  {author} {\bibfnamefont {J.}~\bibnamefont {von Delft}},\ }\bibfield  {title}
  {\bibinfo {title} {Emergent properties of the periodic {A}nderson model: a
  high-resolution, real-frequency study of heavy-fermion quantum criticality},\
  }\href {http://arxiv.org/abs/2310.12672} {\ }\Eprint
  {https://arxiv.org/abs/2310.12672} {arXiv:2310.12672} \BibitemShut {NoStop}%
\bibitem [{\citenamefont {Bulla}\ \emph {et~al.}(2008)\citenamefont {Bulla},
  \citenamefont {Costi},\ and\ \citenamefont {Pruschke}}]{Bulla2008}%
  \BibitemOpen
  \bibfield  {author} {\bibinfo {author} {\bibfnamefont {R.}~\bibnamefont
  {Bulla}}, \bibinfo {author} {\bibfnamefont {T.~A.}\ \bibnamefont {Costi}},\
  and\ \bibinfo {author} {\bibfnamefont {T.}~\bibnamefont {Pruschke}},\
  }\bibfield  {title} {\bibinfo {title} {Numerical renormalization group method
  for quantum impurity systems},\ }\href
  {https://doi.org/10.1103/RevModPhys.80.395} {\bibfield  {journal} {\bibinfo
  {journal} {Rev. Mod. Phys.}\ }\textbf {\bibinfo {volume} {80}},\ \bibinfo
  {pages} {395} (\bibinfo {year} {2008})}\BibitemShut {NoStop}%
\bibitem [{\citenamefont {Doniach}(1977)}]{Doniach1977}%
  \BibitemOpen
  \bibfield  {author} {\bibinfo {author} {\bibfnamefont {S.}~\bibnamefont
  {Doniach}},\ }\bibfield  {title} {\bibinfo {title} {The {{K}ondo} lattice and
  weak antiferromagnetism},\ }\href
  {https://doi.org/10.1016/0378-4363(77)90190-5} {\bibfield  {journal}
  {\bibinfo  {journal} {Physica}\ }\textbf {\bibinfo {volume} {91B+C}},\
  \bibinfo {pages} {231} (\bibinfo {year} {1977})}\BibitemShut {NoStop}%
\bibitem [{\citenamefont {Senthil}\ \emph {et~al.}(2003)\citenamefont
  {Senthil}, \citenamefont {Sachdev},\ and\ \citenamefont
  {Vojta}}]{Senthil2003}%
  \BibitemOpen
  \bibfield  {author} {\bibinfo {author} {\bibfnamefont {T.}~\bibnamefont
  {Senthil}}, \bibinfo {author} {\bibfnamefont {S.}~\bibnamefont {Sachdev}},\
  and\ \bibinfo {author} {\bibfnamefont {M.}~\bibnamefont {Vojta}},\ }\bibfield
   {title} {\bibinfo {title} {Fractionalized {F}ermi liquids},\ }\href
  {https://doi.org/10.1103/PhysRevLett.90.216403} {\bibfield  {journal}
  {\bibinfo  {journal} {Phys. Rev. Lett.}\ }\textbf {\bibinfo {volume} {90}},\
  \bibinfo {pages} {216403} (\bibinfo {year} {2003})}\BibitemShut {NoStop}%
\bibitem [{\citenamefont {Senthil}\ \emph {et~al.}(2004)\citenamefont
  {Senthil}, \citenamefont {Vojta},\ and\ \citenamefont
  {Sachdev}}]{Senthil2004}%
  \BibitemOpen
  \bibfield  {author} {\bibinfo {author} {\bibfnamefont {T.}~\bibnamefont
  {Senthil}}, \bibinfo {author} {\bibfnamefont {M.}~\bibnamefont {Vojta}},\
  and\ \bibinfo {author} {\bibfnamefont {S.}~\bibnamefont {Sachdev}},\
  }\bibfield  {title} {\bibinfo {title} {Weak magnetism and non-{{F}ermi}
  liquids near heavy-fermion critical points},\ }\href
  {https://doi.org/10.1103/PhysRevB.69.035111} {\bibfield  {journal} {\bibinfo
  {journal} {Phys. Rev. B}\ }\textbf {\bibinfo {volume} {69}},\ \bibinfo
  {pages} {035111} (\bibinfo {year} {2004})}\BibitemShut {NoStop}%
\bibitem [{\citenamefont {Weichselbaum}(2012)}]{Weichselbaum2012a}%
  \BibitemOpen
  \bibfield  {author} {\bibinfo {author} {\bibfnamefont {A.}~\bibnamefont
  {Weichselbaum}},\ }\bibfield  {title} {\bibinfo {title} {Non-{A}belian
  symmetries in tensor networks: A quantum symmetry space approach},\ }\href
  {https://doi.org/10.1016/j.aop.2012.07.009} {\bibfield  {journal} {\bibinfo
  {journal} {Ann. Phys. (N. Y.)}\ }\textbf {\bibinfo {volume} {327}},\ \bibinfo
  {pages} {2972} (\bibinfo {year} {2012})}\BibitemShut {NoStop}%
\bibitem [{\citenamefont {Weichselbaum}(2020)}]{Weichselbaum2020}%
  \BibitemOpen
  \bibfield  {author} {\bibinfo {author} {\bibfnamefont {A.}~\bibnamefont
  {Weichselbaum}},\ }\bibfield  {title} {\bibinfo {title} {X-symbols for
  non-{A}belian symmetries in tensor networks},\ }\href
  {https://doi.org/10.1103/PhysRevResearch.2.023385} {\bibfield  {journal}
  {\bibinfo  {journal} {Phys. Rev. Research}\ }\textbf {\bibinfo {volume}
  {2}},\ \bibinfo {pages} {023385} (\bibinfo {year} {2020})}\BibitemShut
  {NoStop}%
\bibitem [{\citenamefont {Vojta}(2010)}]{Vojta2010}%
  \BibitemOpen
  \bibfield  {author} {\bibinfo {author} {\bibfnamefont {M.}~\bibnamefont
  {Vojta}},\ }\bibfield  {title} {\bibinfo {title} {Orbital-selective {M}ott
  transitions: Heavy fermions and beyond},\ }\href
  {https://doi.org/10.1007/s10909-010-0206-3} {\bibfield  {journal} {\bibinfo
  {journal} {J. Low Temp. Phys.}\ }\textbf {\bibinfo {volume} {161}},\ \bibinfo
  {pages} {203} (\bibinfo {year} {2010})}\BibitemShut {NoStop}%
\bibitem [{\citenamefont {Yamamoto}\ and\ \citenamefont
  {Si}(2010)}]{Yamamoto2010}%
  \BibitemOpen
  \bibfield  {author} {\bibinfo {author} {\bibfnamefont {S.~J.}\ \bibnamefont
  {Yamamoto}}\ and\ \bibinfo {author} {\bibfnamefont {Q.}~\bibnamefont {Si}},\
  }\bibfield  {title} {\bibinfo {title} {Global phase diagram of the {K}ondo
  lattice: From heavy fermion metals to {K}ondo insulators},\ }\href
  {https://doi.org/10.1007/s10909-010-0221-4} {\bibfield  {journal} {\bibinfo
  {journal} {J. Low Temp. Phys.}\ }\textbf {\bibinfo {volume} {161}},\ \bibinfo
  {pages} {233} (\bibinfo {year} {2010})}\BibitemShut {NoStop}%
\bibitem [{\citenamefont {Si}(2010)}]{Si2010}%
  \BibitemOpen
  \bibfield  {author} {\bibinfo {author} {\bibfnamefont {Q.}~\bibnamefont
  {Si}},\ }\bibfield  {title} {\bibinfo {title} {Quantum criticality and global
  phase diagram of magnetic heavy fermions},\ }\href
  {https://doi.org/https://doi.org/10.1002/pssb.200983082} {\bibfield
  {journal} {\bibinfo  {journal} {Phys. Status Solidi B}\ }\textbf {\bibinfo
  {volume} {247}},\ \bibinfo {pages} {476} (\bibinfo {year}
  {2010})}\BibitemShut {NoStop}%
\bibitem [{\citenamefont {Coleman}\ and\ \citenamefont
  {Nevidomskyy}(2010)}]{Coleman2010}%
  \BibitemOpen
  \bibfield  {author} {\bibinfo {author} {\bibfnamefont {P.}~\bibnamefont
  {Coleman}}\ and\ \bibinfo {author} {\bibfnamefont {A.~H.}\ \bibnamefont
  {Nevidomskyy}},\ }\bibfield  {title} {\bibinfo {title} {Frustration and the
  {K}ondo effect in heavy fermion materials},\ }\href
  {https://doi.org/10.1007/s10909-010-0213-4} {\bibfield  {journal} {\bibinfo
  {journal} {J. Low Temp. Phys.}\ }\textbf {\bibinfo {volume} {161}},\ \bibinfo
  {pages} {182} (\bibinfo {year} {2010})}\BibitemShut {NoStop}%
\bibitem [{\citenamefont {Friedemann}\ \emph {et~al.}(2009)\citenamefont
  {Friedemann}, \citenamefont {Westerkamp}, \citenamefont {Brando},
  \citenamefont {Oeschler}, \citenamefont {Wirth}, \citenamefont {Gegenwart},
  \citenamefont {Krellner}, \citenamefont {Geibel},\ and\ \citenamefont
  {Steglich}}]{Friedemann2009}%
  \BibitemOpen
  \bibfield  {author} {\bibinfo {author} {\bibfnamefont {S.}~\bibnamefont
  {Friedemann}}, \bibinfo {author} {\bibfnamefont {T.}~\bibnamefont
  {Westerkamp}}, \bibinfo {author} {\bibfnamefont {M.}~\bibnamefont {Brando}},
  \bibinfo {author} {\bibfnamefont {N.}~\bibnamefont {Oeschler}}, \bibinfo
  {author} {\bibfnamefont {S.}~\bibnamefont {Wirth}}, \bibinfo {author}
  {\bibfnamefont {P.}~\bibnamefont {Gegenwart}}, \bibinfo {author}
  {\bibfnamefont {C.}~\bibnamefont {Krellner}}, \bibinfo {author}
  {\bibfnamefont {C.}~\bibnamefont {Geibel}},\ and\ \bibinfo {author}
  {\bibfnamefont {F.}~\bibnamefont {Steglich}},\ }\bibfield  {title} {\bibinfo
  {title} {Detaching the antiferromagnetic quantum critical point from the
  {{F}ermi}-surface reconstruction in {YbRh${}_2$Si${}_2$}},\ }\href
  {https://doi.org/10.1038/nphys1299} {\bibfield  {journal} {\bibinfo
  {journal} {Nat. Phys.}\ }\textbf {\bibinfo {volume} {5}},\ \bibinfo {pages}
  {465} (\bibinfo {year} {2009})}\BibitemShut {NoStop}%
\bibitem [{\citenamefont {Maksimovic}\ \emph {et~al.}(2022)\citenamefont
  {Maksimovic}, \citenamefont {Eilbott}, \citenamefont {Cookmeyer},
  \citenamefont {Wan}, \citenamefont {Rusz}, \citenamefont {Nagarajan},
  \citenamefont {Haley}, \citenamefont {Maniv}, \citenamefont {Gong},
  \citenamefont {Faubel}, \citenamefont {Hayes}, \citenamefont {Bangura},
  \citenamefont {Singleton}, \citenamefont {Palmstrom}, \citenamefont {Winter},
  \citenamefont {McDonald}, \citenamefont {Jang}, \citenamefont {Ai},
  \citenamefont {Lin}, \citenamefont {Ciocys}, \citenamefont {Gobbo},
  \citenamefont {Werman}, \citenamefont {Oppeneer}, \citenamefont {Altman},
  \citenamefont {Lanzara},\ and\ \citenamefont {Analytis}}]{Maksimovic2022}%
  \BibitemOpen
  \bibfield  {author} {\bibinfo {author} {\bibfnamefont {N.}~\bibnamefont
  {Maksimovic}}, \bibinfo {author} {\bibfnamefont {D.~H.}\ \bibnamefont
  {Eilbott}}, \bibinfo {author} {\bibfnamefont {T.}~\bibnamefont {Cookmeyer}},
  \bibinfo {author} {\bibfnamefont {F.}~\bibnamefont {Wan}}, \bibinfo {author}
  {\bibfnamefont {J.}~\bibnamefont {Rusz}}, \bibinfo {author} {\bibfnamefont
  {V.}~\bibnamefont {Nagarajan}}, \bibinfo {author} {\bibfnamefont {S.~C.}\
  \bibnamefont {Haley}}, \bibinfo {author} {\bibfnamefont {E.}~\bibnamefont
  {Maniv}}, \bibinfo {author} {\bibfnamefont {A.}~\bibnamefont {Gong}},
  \bibinfo {author} {\bibfnamefont {S.}~\bibnamefont {Faubel}}, \bibinfo
  {author} {\bibfnamefont {I.~M.}\ \bibnamefont {Hayes}}, \bibinfo {author}
  {\bibfnamefont {A.}~\bibnamefont {Bangura}}, \bibinfo {author} {\bibfnamefont
  {J.}~\bibnamefont {Singleton}}, \bibinfo {author} {\bibfnamefont {J.~C.}\
  \bibnamefont {Palmstrom}}, \bibinfo {author} {\bibfnamefont {L.}~\bibnamefont
  {Winter}}, \bibinfo {author} {\bibfnamefont {R.}~\bibnamefont {McDonald}},
  \bibinfo {author} {\bibfnamefont {S.}~\bibnamefont {Jang}}, \bibinfo {author}
  {\bibfnamefont {P.}~\bibnamefont {Ai}}, \bibinfo {author} {\bibfnamefont
  {Y.}~\bibnamefont {Lin}}, \bibinfo {author} {\bibfnamefont {S.}~\bibnamefont
  {Ciocys}}, \bibinfo {author} {\bibfnamefont {J.}~\bibnamefont {Gobbo}},
  \bibinfo {author} {\bibfnamefont {Y.}~\bibnamefont {Werman}}, \bibinfo
  {author} {\bibfnamefont {P.~M.}\ \bibnamefont {Oppeneer}}, \bibinfo {author}
  {\bibfnamefont {E.}~\bibnamefont {Altman}}, \bibinfo {author} {\bibfnamefont
  {A.}~\bibnamefont {Lanzara}},\ and\ \bibinfo {author} {\bibfnamefont {J.~G.}\
  \bibnamefont {Analytis}},\ }\bibfield  {title} {\bibinfo {title} {Evidence
  for a delocalization quantum phase transition without symmetry breaking in
  {$\mathrm{CeCoIn}_5$}},\ }\href {https://doi.org/10.1126/science.aaz4566}
  {\bibfield  {journal} {\bibinfo  {journal} {Science}\ }\textbf {\bibinfo
  {volume} {375}},\ \bibinfo {pages} {76} (\bibinfo {year} {2022})}\BibitemShut
  {NoStop}%
\bibitem [{\citenamefont {Nozi{\`e}res}(1974)}]{Nozieres1974}%
  \BibitemOpen
  \bibfield  {author} {\bibinfo {author} {\bibfnamefont {P.}~\bibnamefont
  {Nozi{\`e}res}},\ }\bibfield  {title} {\bibinfo {title} {A ``{F}ermi-liquid''
  description of the {{K}ondo} problem at low temperatures},\ }\href
  {https://doi.org/10.1007/BF00654541} {\bibfield  {journal} {\bibinfo
  {journal} {J. Low Temp. Phys.}\ }\textbf {\bibinfo {volume} {17}},\ \bibinfo
  {pages} {31} (\bibinfo {year} {1974})}\BibitemShut {NoStop}%
\bibitem [{\citenamefont {Jones}\ \emph {et~al.}(1988)\citenamefont {Jones},
  \citenamefont {Varma},\ and\ \citenamefont {Wilkins}}]{Jones1988}%
  \BibitemOpen
  \bibfield  {author} {\bibinfo {author} {\bibfnamefont {B.~A.}\ \bibnamefont
  {Jones}}, \bibinfo {author} {\bibfnamefont {C.~M.}\ \bibnamefont {Varma}},\
  and\ \bibinfo {author} {\bibfnamefont {J.~W.}\ \bibnamefont {Wilkins}},\
  }\bibfield  {title} {\bibinfo {title} {Low-temperature properties of the
  two-impurity {K}ondo {H}amiltonian},\ }\href
  {https://doi.org/10.1103/PhysRevLett.61.125} {\bibfield  {journal} {\bibinfo
  {journal} {Phys. Rev. Lett.}\ }\textbf {\bibinfo {volume} {61}},\ \bibinfo
  {pages} {125} (\bibinfo {year} {1988})}\BibitemShut {NoStop}%
\bibitem [{\citenamefont {Jones}\ and\ \citenamefont
  {Varma}(1989)}]{Jones1989}%
  \BibitemOpen
  \bibfield  {author} {\bibinfo {author} {\bibfnamefont {B.~A.}\ \bibnamefont
  {Jones}}\ and\ \bibinfo {author} {\bibfnamefont {C.~M.}\ \bibnamefont
  {Varma}},\ }\bibfield  {title} {\bibinfo {title} {Critical point in the
  solution of the two magnetic impurity problem},\ }\href
  {https://doi.org/10.1103/PhysRevB.40.324} {\bibfield  {journal} {\bibinfo
  {journal} {Phys. Rev. B}\ }\textbf {\bibinfo {volume} {40}},\ \bibinfo
  {pages} {324} (\bibinfo {year} {1989})}\BibitemShut {NoStop}%
\bibitem [{Def()}]{DefLogDep}%
  \BibitemOpen
  \href@noop {} {\bibinfo  {journal} {We call frequency dependencies
  $(\omega^{a}-1)/a$ with $a \simeq 0$ logarithmic if the exponent $a$ cannot
  be reasonably distinguished from $a \to 0$ based on our numerical data. Ditto
  for temperature dependencies. Similarly, ``plateau'' means $\omega^{a}$ with
  $a \simeq 0$ and ``linear'' $\omega^{a}$ with $a \simeq 1$}\ }\BibitemShut
  {NoStop}%
\bibitem [{sup()}]{supplement}%
  \BibitemOpen
\bibfield  {journal} {  }\href@noop {} {\bibinfo  {journal} {See Supplemental
  Material at [url] for additional information on basic formulas involving the
  optical conductivity; the numerical computation of the optical conductivity;
  the role of vertex contributions; the Drude weight; the scaling functions;
  scaling of the imaginary part of the optical conductivity and an analysis on
  the complex optical conductivity, including a comparison to data on {\CCI}.
  The Supplemental Material includes
  Refs.~\cite{Resta2018,Scalapino1993,Mitchell2014,Stadler2016,Kugler2022a,Lee2016}}\
  }\BibitemShut {NoStop}%
\bibitem [{\citenamefont {Resta}(2018)}]{Resta2018}%
  \BibitemOpen
\bibfield  {journal} {  }\bibfield  {author} {\bibinfo {author} {\bibfnamefont
  {R.}~\bibnamefont {Resta}},\ }\bibfield  {title} {\bibinfo {title} {Drude
  weight and superconducting weight},\ }\href
  {https://doi.org/10.1088/1361-648X/aade19} {\bibfield  {journal} {\bibinfo
  {journal} {J. Phys. Condens. Matter}\ }\textbf {\bibinfo {volume} {30}},\
  \bibinfo {pages} {414001} (\bibinfo {year} {2018})}\BibitemShut {NoStop}%
\bibitem [{\citenamefont {Scalapino}\ \emph {et~al.}(1993)\citenamefont
  {Scalapino}, \citenamefont {White},\ and\ \citenamefont
  {Zhang}}]{Scalapino1993}%
  \BibitemOpen
  \bibfield  {author} {\bibinfo {author} {\bibfnamefont {D.~J.}\ \bibnamefont
  {Scalapino}}, \bibinfo {author} {\bibfnamefont {S.~R.}\ \bibnamefont
  {White}},\ and\ \bibinfo {author} {\bibfnamefont {S.}~\bibnamefont {Zhang}},\
  }\bibfield  {title} {\bibinfo {title} {Insulator, metal, or superconductor:
  The criteria},\ }\href {https://doi.org/10.1103/PhysRevB.47.7995} {\bibfield
  {journal} {\bibinfo  {journal} {Phys. Rev. B}\ }\textbf {\bibinfo {volume}
  {47}},\ \bibinfo {pages} {7995} (\bibinfo {year} {1993})}\BibitemShut
  {NoStop}%
\bibitem [{\citenamefont {Mitchell}\ \emph {et~al.}(2014)\citenamefont
  {Mitchell}, \citenamefont {Galpin}, \citenamefont {Wilson-Fletcher},
  \citenamefont {Logan},\ and\ \citenamefont {Bulla}}]{Mitchell2014}%
  \BibitemOpen
  \bibfield  {author} {\bibinfo {author} {\bibfnamefont {A.~K.}\ \bibnamefont
  {Mitchell}}, \bibinfo {author} {\bibfnamefont {M.~R.}\ \bibnamefont
  {Galpin}}, \bibinfo {author} {\bibfnamefont {S.}~\bibnamefont
  {Wilson-Fletcher}}, \bibinfo {author} {\bibfnamefont {D.~E.}\ \bibnamefont
  {Logan}},\ and\ \bibinfo {author} {\bibfnamefont {R.}~\bibnamefont {Bulla}},\
  }\bibfield  {title} {\bibinfo {title} {Generalized {Wilson} chain for solving
  multichannel quantum impurity problems},\ }\href
  {https://doi.org/10.1103/PhysRevB.89.121105} {\bibfield  {journal} {\bibinfo
  {journal} {Phys. Rev. B}\ }\textbf {\bibinfo {volume} {89}},\ \bibinfo
  {pages} {121105} (\bibinfo {year} {2014})}\BibitemShut {NoStop}%
\bibitem [{\citenamefont {Stadler}\ \emph {et~al.}(2016)\citenamefont
  {Stadler}, \citenamefont {Mitchell}, \citenamefont {von Delft},\ and\
  \citenamefont {Weichselbaum}}]{Stadler2016}%
  \BibitemOpen
  \bibfield  {author} {\bibinfo {author} {\bibfnamefont {K.~M.}\ \bibnamefont
  {Stadler}}, \bibinfo {author} {\bibfnamefont {A.~K.}\ \bibnamefont
  {Mitchell}}, \bibinfo {author} {\bibfnamefont {J.}~\bibnamefont {von
  Delft}},\ and\ \bibinfo {author} {\bibfnamefont {A.}~\bibnamefont
  {Weichselbaum}},\ }\bibfield  {title} {\bibinfo {title} {Interleaved
  numerical renormalization group as an efficient multiband impurity solver},\
  }\href {https://doi.org/10.1103/PhysRevB.93.235101} {\bibfield  {journal}
  {\bibinfo  {journal} {Phys. Rev. B}\ }\textbf {\bibinfo {volume} {93}},\
  \bibinfo {pages} {235101} (\bibinfo {year} {2016})}\BibitemShut {NoStop}%
\bibitem [{\citenamefont {Kugler}(2022)}]{Kugler2022a}%
  \BibitemOpen
  \bibfield  {author} {\bibinfo {author} {\bibfnamefont {F.~B.}\ \bibnamefont
  {Kugler}},\ }\bibfield  {title} {\bibinfo {title} {Improved estimator for
  numerical renormalization group calculations of the self-energy},\ }\href
  {https://doi.org/10.1103/PhysRevB.105.245132} {\bibfield  {journal} {\bibinfo
   {journal} {Phys. Rev. B}\ }\textbf {\bibinfo {volume} {105}},\ \bibinfo
  {pages} {245132} (\bibinfo {year} {2022})}\BibitemShut {NoStop}%
\bibitem [{\citenamefont {Lee}\ and\ \citenamefont
  {Weichselbaum}(2016)}]{Lee2016}%
  \BibitemOpen
  \bibfield  {author} {\bibinfo {author} {\bibfnamefont {S.-S.~B.}\
  \bibnamefont {Lee}}\ and\ \bibinfo {author} {\bibfnamefont {A.}~\bibnamefont
  {Weichselbaum}},\ }\bibfield  {title} {\bibinfo {title} {Adaptive broadening
  to improve spectral resolution in the numerical renormalization group},\
  }\href {https://doi.org/10.1103/PhysRevB.94.235127} {\bibfield  {journal}
  {\bibinfo  {journal} {Phys. Rev. B}\ }\textbf {\bibinfo {volume} {94}},\
  \bibinfo {pages} {235127} (\bibinfo {year} {2016})}\BibitemShut {NoStop}%
\bibitem [{\citenamefont {Lee}\ \emph {et~al.}(2021)\citenamefont {Lee},
  \citenamefont {Kugler},\ and\ \citenamefont {von Delft}}]{Lee2021}%
  \BibitemOpen
  \bibfield  {author} {\bibinfo {author} {\bibfnamefont {S.-S.~B.}\
  \bibnamefont {Lee}}, \bibinfo {author} {\bibfnamefont {F.~B.}\ \bibnamefont
  {Kugler}},\ and\ \bibinfo {author} {\bibfnamefont {J.}~\bibnamefont {von
  Delft}},\ }\bibfield  {title} {\bibinfo {title} {Computing local multipoint
  correlators using the numerical renormalization group},\ }\href
  {https://doi.org/10.1103/PhysRevX.11.041007} {\bibfield  {journal} {\bibinfo
  {journal} {Phys. Rev. X}\ }\textbf {\bibinfo {volume} {11}},\ \bibinfo
  {pages} {041007} (\bibinfo {year} {2021})}\BibitemShut {NoStop}%
\bibitem [{\citenamefont {Kugler}\ \emph {et~al.}(2021)\citenamefont {Kugler},
  \citenamefont {Lee},\ and\ \citenamefont {von Delft}}]{Kugler2021}%
  \BibitemOpen
  \bibfield  {author} {\bibinfo {author} {\bibfnamefont {F.~B.}\ \bibnamefont
  {Kugler}}, \bibinfo {author} {\bibfnamefont {S.-S.~B.}\ \bibnamefont {Lee}},\
  and\ \bibinfo {author} {\bibfnamefont {J.}~\bibnamefont {von Delft}},\
  }\bibfield  {title} {\bibinfo {title} {Multipoint correlation functions:
  Spectral representation and numerical evaluation},\ }\href
  {https://doi.org/10.1103/PhysRevX.11.041006} {\bibfield  {journal} {\bibinfo
  {journal} {Phys. Rev. X}\ }\textbf {\bibinfo {volume} {11}},\ \bibinfo
  {pages} {041006} (\bibinfo {year} {2021})}\BibitemShut {NoStop}%
\bibitem [{\citenamefont {Shi}\ \emph {et~al.}()\citenamefont {Shi},
  \citenamefont {Tagay}, \citenamefont {Liang}, \citenamefont {Duong},
  \citenamefont {Wu}, \citenamefont {Ronning}, \citenamefont {Schlom},
  \citenamefont {Shen},\ and\ \citenamefont {Armitage}}]{Shi2023}%
  \BibitemOpen
  \bibfield  {author} {\bibinfo {author} {\bibfnamefont {L.~Y.}\ \bibnamefont
  {Shi}}, \bibinfo {author} {\bibfnamefont {Z.}~\bibnamefont {Tagay}}, \bibinfo
  {author} {\bibfnamefont {J.}~\bibnamefont {Liang}}, \bibinfo {author}
  {\bibfnamefont {K.}~\bibnamefont {Duong}}, \bibinfo {author} {\bibfnamefont
  {Y.}~\bibnamefont {Wu}}, \bibinfo {author} {\bibfnamefont {F.}~\bibnamefont
  {Ronning}}, \bibinfo {author} {\bibfnamefont {D.~G.}\ \bibnamefont {Schlom}},
  \bibinfo {author} {\bibfnamefont {K.}~\bibnamefont {Shen}},\ and\ \bibinfo
  {author} {\bibfnamefont {N.~P.}\ \bibnamefont {Armitage}},\ }\bibfield
  {title} {\bibinfo {title} {Low energy electrodynamics and a hidden {F}ermi
  liquid in the heavy-fermion {CeCoIn$_5$}},\ }\href
  {http://arxiv.org/abs/2310.10916} {\ }\Eprint
  {https://arxiv.org/abs/2310.10916} {arXiv:2310.10916} \BibitemShut {NoStop}%
\bibitem [{QPw()}]{QPweightsPAM}%
  \BibitemOpen
  \href@noop {} {\bibinfo  {journal} {In the PAM, $\vec{G}_{\vec{k}}(\omega)$
  is matrix-valued in the orbital space, with orbital index $\alpha \in\{f,
  c\}$. $G^{-1}_{\vec{k}_{\mr{F}}}(\omega)$ is defined in terms of the
  eigenvalue which fulfils $\re \, G^{-1}_{\vec{k}_{\mr{F}}}(0) = 0$. To
  determine the single-particle decay rate $\gamma^{\ast}$, the
  $\omega$-dependence of the corresponding eigenvalue is fitted to a Lorentzian
  lineshape around $\omega=0$, $G_{\vec{k}_{\mr{F}}}(\omega) \simeq Z/(\omega +
  \mr{i} \gamma^{\ast})$, as usual}\ }\BibitemShut {NoStop}%
\bibitem [{\citenamefont {Coleman}(2015)}]{Coleman2015}%
  \BibitemOpen
\bibfield  {journal} {  }\bibfield  {author} {\bibinfo {author} {\bibfnamefont
  {P.}~\bibnamefont {Coleman}},\ }\href
  {https://doi.org/10.1017/CBO9781139020916} {\emph {\bibinfo {title}
  {Introduction to Many-Body Physics}}}\ (\bibinfo  {publisher} {Cambridge
  University Press},\ \bibinfo {address} {Cambridge},\ \bibinfo {year}
  {2015})\BibitemShut {NoStop}%
\bibitem [{\citenamefont {Varma}\ \emph {et~al.}(1989)\citenamefont {Varma},
  \citenamefont {Littlewood}, \citenamefont {Schmitt-Rink}, \citenamefont
  {Abrahams},\ and\ \citenamefont {Ruckenstein}}]{Varma1989}%
  \BibitemOpen
  \bibfield  {author} {\bibinfo {author} {\bibfnamefont {C.~M.}\ \bibnamefont
  {Varma}}, \bibinfo {author} {\bibfnamefont {P.~B.}\ \bibnamefont
  {Littlewood}}, \bibinfo {author} {\bibfnamefont {S.}~\bibnamefont
  {Schmitt-Rink}}, \bibinfo {author} {\bibfnamefont {E.}~\bibnamefont
  {Abrahams}},\ and\ \bibinfo {author} {\bibfnamefont {A.~E.}\ \bibnamefont
  {Ruckenstein}},\ }\bibfield  {title} {\bibinfo {title} {Phenomenology of the
  normal state of {Cu-O} high-temperature superconductors},\ }\href
  {https://doi.org/10.1103/PhysRevLett.63.1996} {\bibfield  {journal} {\bibinfo
   {journal} {Phys. Rev. Lett.}\ }\textbf {\bibinfo {volume} {63}},\ \bibinfo
  {pages} {1996} (\bibinfo {year} {1989})}\BibitemShut {NoStop}%
\bibitem [{\citenamefont {Georges}\ \emph {et~al.}(1996)\citenamefont
  {Georges}, \citenamefont {Kotliar}, \citenamefont {Krauth},\ and\
  \citenamefont {Rozenberg}}]{Georges1996}%
  \BibitemOpen
  \bibfield  {author} {\bibinfo {author} {\bibfnamefont {A.}~\bibnamefont
  {Georges}}, \bibinfo {author} {\bibfnamefont {G.}~\bibnamefont {Kotliar}},
  \bibinfo {author} {\bibfnamefont {W.}~\bibnamefont {Krauth}},\ and\ \bibinfo
  {author} {\bibfnamefont {M.~J.}\ \bibnamefont {Rozenberg}},\ }\bibfield
  {title} {\bibinfo {title} {Dynamical mean-field theory of strongly correlated
  fermion systems and the limit of infinite dimensions},\ }\href
  {https://doi.org/10.1103/RevModPhys.68.13} {\bibfield  {journal} {\bibinfo
  {journal} {Rev. Mod. Phys.}\ }\textbf {\bibinfo {volume} {68}},\ \bibinfo
  {pages} {13} (\bibinfo {year} {1996})}\BibitemShut {NoStop}%
\end{thebibliography}%
%



\clearpage

\thispagestyle{empty}

\setcounter{equation}{0}
\setcounter{figure}{0}
\setcounter{page}{1}

\renewcommand{\theequation}{S\arabic{equation}}
\renewcommand{\thefigure}{S\arabic{figure}}
\renewcommand{\thepage}{S\arabic{page}}

\setcounter{secnumdepth}{2} 
\renewcommand{\thefigure}{S\arabic{figure}}
\setcounter{figure}{0}
\setcounter{section}{0}
\setcounter{equation}{0}
\renewcommand{\thesection}{S-\Roman{section}}
\renewcommand{\theequation}{S\arabic{equation}}
%

%
\title{Supplemental Material for \\
``Dynamical scaling and Planckian dissipation due to heavy-fermion quantum criticality''}

\date{\today}
\maketitle

In Sec.~\ref{sec:optical},
we provide basic definitions and expressions regarding the Fourier transforms of operators and regarding the optical conductivity.
Section~\ref{sec:optical_numerics} provides additional information on the numerical computation of the optical conductivity, the role of vertex contributions, and to what extent the Drude term vanishes.
Section~\ref{sec:scaling_function} provides more information on the scaling functions $\mc{X}$ and $\mc{S}$.
In Sec.~\ref{sec:compex_optical}, we discuss scaling of the imaginary part of the optical conductivity and provide an analysis 
of the complex optical conductivity similar in spirit to the analysis of experimental data on \CCI\ of Ref.~\cite{Shi2023}.

\section{Optical conductivity}
\label{sec:optical}
In this section, we state some textbook~\cite{Coleman2015} formulas that are important in the context of the optical conductivity for the PAM. 
\subsection{Fourier transforms of operators}
\label{sec:FT}
We define the Fourier transform of fermionic creation and annihilation operators in a unitary fashion,
\begin{align}
c^{\pdag}_{\vec{k}\sigma} &= \frac{1}{\sqrt{N}} \sum_{i} \mr{e}^{-\mi \vec{k} \cdot \vec{r}_i} c^{\pdag}_{i\sigma} \, ,
\end{align}
ensuring
$
\{c^{\dag}_{\vec{k}\sigma}, c^{\pdag}_{\vec{k}'\sigma'} \} = \delta_{\sigma\sigma'} \delta_{\vec{k}\vec{k}'} 
$.
For bosonic observables $\mc{O}_i$ like the current density, on the other hand, we define it as an orthogonal but non-unitary transformation,
\begin{align}
\mc{O}_{\vec{q}} &= \frac{1}{N} \sum_{i} \mr{e}^{-\mi \vec{q} \cdot \vec{r}_i} \mc{O}_{i} \, .
\end{align}
This ensures that the expectation values $\langle\mc{O}_{\vec{q}}\rangle$ and $\langle\mc{O}_{i}\rangle$ scale the same way with $N$ in the thermodynamic limit.
(if we had used a unitary
Fourier transforms for bosonic observables, $\langle\mc{O}_{\vec{q}}\rangle \sim \sqrt{N}$ would not be well-defined in the thermodynamic limit). 
The same goes for source fields like the vector potential.
\subsection{Current and conductivity}

In presence of a vector potential $\vec{A}$, the Hamiltonian~\eqref{eq:H_PAM} is modified by replacing the hopping between site $i$ and $i+\ahat$ by $t \to t\exp\left(-\mi e A^{a}_i\right)$,
where $\ahat$ is some unit lattice vector. The current density is
\begin{align}
\label{eq:current}
j^{a}_i = -\frac{\partial H}{\partial A^{a}_i} = -\mi t e \sum_{\sigma} \left(\mr{e}^{-\mi e A^a_i} c^{\dag}_{i\sigma} c^{\pdag}_{i+\ahat\sigma} - \mr{h.c.}\right) \, .
\end{align}
If no lattice symmetry is broken,
the current response to a $\vec{q}$- and $\omega$-dependent electric field $\vec{E}_{\vec{q}}(\omega) = \mr{i}\omega^{+}\vec{A}_{\vec{q}}(\omega)$ (where $\omega^{+} = \omega + \mr{i}0^{+}$) 
takes the form 
$\langle j_{\vec{q}}^{a} \rangle(\omega) = \sigma_{\vec{q}}(\omega) E^{a}_{\vec{q}}(\omega)$,
where the dynamical conductivity is given by
\begin{align}
\sigma_{\vec{q}}(\omega) &= \frac{1}{\mr{i}\omega^{+}} \bigl[\langle \hat{K} \rangle - \chi[j^a_{\vec{q}}](\omega) \bigr] \, , \\
\hat{K} &= -\frac{te^2}{N} \sum_{i\sigma} \bigl( c^{\dag}_{i\sigma} c^{\pdag}_{i+\ahat\sigma} + \mr{h.c.} \bigr) \, , \nonumber
\end{align}
and $j^{a}_{\vec{q}} = \frac{1}{N} \sum_{i} \mr{e}^{-\mr{i} \vec{q} \cdot \vec{r}_i} j^{a}_i$. 
In a $d$-dimensional hypercubic lattice, $\langle \hat{K} \rangle$ is proportional to the kinetic energy density $\epsilon_{\mathrm{kin}} = \frac{d}{e^2} \langle \hat{K} \rangle$.

The optical conductivity $\sigma(\omega) = \sigma_{\vec{q} = \vec{0}}(\omega)$ is the response to a uniform electric field.
It can be decomposed as~\cite{Resta2018,Scalapino1993} $\sigma(\omega) = \sigma^{\mr{D}}(\omega) + \sigma^{\mr{reg}}(\omega)$, with
\begin{align}
\label{eq:sigma_inf}
\sigma^{\mr{D}}(\omega) &= D\left[\delta(\omega) + \mc{P}\frac{\mr{i}}{\pi \omega}\right] \, ,\\
D &= \pi \bigl[\chi'[j^a_{\vec{0}}](0) - \langle \hat{K} \rangle \bigr] 
 \, , \label{eq:DrudeWeight} \\
\label{eq:sigma_norm}
 \sigma^{\mr{reg}}(\omega) &= \mc{P}\frac{1}{\mr{i}\omega} \bigl[\chi'[j^a_{\vec{0}}](0) - \chi[j^a_{\vec{0}}](\omega) \bigr] \, ,
\end{align}
where $\mc{P}$ denotes the principal part. 
The regular term $\sigma^{\mr{reg}}(\omega)$ describes currents that decay at long times; the Drude term $\sigma^{\mr{D}}(\omega)$ with Drude weight $D$ describes 
persistent currents. For a non-superconducting, thermodynamically large lattice model
at non-zero temperature, one expects $D=0$.

The optical conductivity fulfills the $f$-sum rule,
\begin{align}
\int_{-\infty}^{\infty} \frac{\mr{d}\omega}{\pi} \sigma'(\omega) = -\langle \hat{K} \rangle \, ,
\end{align}
which follows when evaluating 
$\chi'[j^{a}_{\vec{0}}](0)$ using the Kramers--Kronig relation for 
general susceptibilities,  
\begin{align}
\label{eq:KramersKronig}
\chi'[\mathcal{O}](\omega') = -
\mathcal{P} \hspace{-1.5mm}
 \int_{-\infty}^\infty \mathrm{d} \omega \, 
 \chi''[\mathcal{O}](\omega)/(\omega- \omega').
\end{align} 

\subsection{Bubble contribution}
The bubble contribution to the current susceptibility is defined as the susceptibility of a free system but with the Green's functions replaced
by the Green's function of the interacting system. 
We shortly outline the corresponding formulas for the bubble contribution to the local current 
susceptibility, $\chi_{\mathrm{B}}[j^{a}_i]$ and to the uniform $\vec{q}=0$ susceptibility, $\chi_{\mathrm{B}}[j^{a}_\vec{0}]$.
Since the current operator in \Eq{eq:current} consists only of $c$-electron operators, the formulas for the bubble contribution only involve $c$-electron Green's functions.
For brevity, we suppress the $c$ labels on all Green's functions, spectral functions and self-energies in this section and in Sec.~\ref{sec:Bubble_comp}.
The current operators can be written in terms of the bare current vertex $\mc{J}^{a}$,
\begin{subequations}
\begin{align}
j^{a}_i &= \sum_{\ell\ell'\sigma} \mc{J}^{a}_{i\ell\ell'} c^{\dag}_{\ell\sigma} c^{\pdag}_{\ell'\sigma} \, , \\
\mc{J}^{a}_{i\ell\ell'} &= -\mi t e \left(\delta_{i\ell} \delta_{i+\ahat\ell'} - \delta_{i+\ahat\ell} \delta_{i\ell'}\right) \, , \\
j^{a}_{\vec{q}} &= \frac{1}{N} \sum_i \mr{e}^{-\mi \vec{q} \cdot \vec{r}_i} j^{a}_i = \sum_{\vec{k}\vec{k}'\sigma} \mc{J}^{a}_{\vec{q}\vec{k}\vec{k}'} 
c^{\dag}_{\vec{k}\sigma} c^{\pdag}_{\vec{k}'\sigma} \\
 \mc{J}^{a}_{\vec{q}\vec{k}\vec{k}'}  &= \frac{-2te}{N} \delta_{\vec{q},\vec{k}-\vec{k}'} \mr{e}^{\mi \frac{\vec{q} \cdot \ahat}{2}} 
\sin\left[\left(\vec{k} - \frac{\vec{q}}{2}\right)\!\cdot \ahat\right] \, .
\end{align}
\end{subequations}

We define the polarization bubble (with $\im\, z > 0$),
\begin{align}
\mc{P}_{g,g'}(z) =& \, T\sum_{m} G_{g}(\mi\omega_m) G_{g'}(\mi \omega_m + z) \\
 =& \int_{-\infty}^{\infty}\!\!\! \mr{d}\omega \, f(\omega) [A_{g}(\omega) G_{g'}(\omega+z)  \\
 &+ A_{g'}(\omega) G_{g}(\omega - z) ] \, , \nonumber
\end{align}
where $G(z)$ is the Masubara Green's function, $A(\omega)$ the corresponding spectral function, $f(\omega)$ the Fermi-Dirac distribution function and $g$ and $g'$ are quantum numbers like momentum, spin or spatial distance, $\vec{r}_{ij} = \vec{r}_i - \vec{r}_j$ and we assume $G$ depends on $|\vec{r}_i - \vec{r}_j|$ only.

The bubble contribution to the $\vec{q} = 0$ current susceptibility is
\begin{flalign}
\chi_{\mr{B}}[j^{a}_{\vec{0}}](z) =& \frac{8t^2e^2}{N} \sum_{\vec{k}} \sin^2(\vec{k} \cdot \ahat) \mc{P}_{\vec{k},\vec{k}}(z) \nonumber \\
=& \frac{8t^2e^2}{N} \sum_{\vec{k}} \sin^2(\vec{k} \cdot \ahat) \int_{-\infty}^{\infty} \mr{d}\omega \, f(\omega) \times \\
&[A_{\vec{k}}(\omega) G_{\vec{k}}(\omega+z) + A_{\vec{k}}(\omega) G_{\vec{k}}(\omega - z) ] \, .  \nonumber
\end{flalign}
The corresponding spectral function is ($\nu^{\pm} = \nu \pm \mr{i} 0^{+}$) 
\begin{subequations}
\begin{align}
\label{eq:chiJ_q0_bubble}
\chi''_{\mr{B}}[j^{a}_\vec{0}](\nu) &= \frac{\mi}{2\pi} \left[\chi_{\mr{B}}[j^{a}_\vec{0}](\nu^+) - \chi_{\mr{B}}[j^{a}_\vec{0}](\nu^-)\right] \nonumber \\
&= \frac{8t^2e^2}{N} \sum_{\vec{k}} \sin^2(\vec{k}\cdot\ahat) I_{\vec{k}}(\nu) \\
\label{eq:Ik_nu}
I_{\vec{k}}(\nu) &= \!\! \int_{-\infty}^{\infty}\!\!\! \mr{d}\omega  \left[f(\omega) \! - \! f(\omega \! +\! \nu)\right] A_{\vec{k}}(\omega) A_{\vec{k}}(\omega \! + \! \nu) \, . 
\end{align}
\end{subequations}

The bubble contribution to the local current-current susceptibility (involving 
one link in the lattice, i.e., two sites) is then 
\begin{align}
\chi_{\mr{B}}&[j^{a}_i](z) = 2\sum_{mm'} \sum_{nn'} \mc{J}^{a}_{imm'} \mc{J}^{a}_{inn'} \mc{P}_{\vec{r}_{mn'},\vec{r}_{m'n}}(z) \nonumber \\
=&  -4 t^2e^2 \int_{-\infty}^{\infty}\!\!\! \mr{d}\omega \, f(\omega) \times \\
&\big[ [A_{\vec{r}_{i,i+\ahat}}(\omega) G_{\vec{r}_{i+\ahat,i}}(\omega+z) - A_{\vec{r}_{i,i}}(\omega) G_{\vec{r}_{i,i}}(\omega+z)] \nonumber \\
 +&[A_{\vec{r}_{i+\ahat,i}}(\omega) G_{\vec{r}_{i,i+\ahat}}(\omega - z) - A_{\vec{r}_{i,i}}(\omega) G_{\vec{r}_{i,i}}(\omega - z)] \big] \, .\nonumber 
\end{align}
The local current-current spectral function is
\begin{align}
\chi''_{\mr{B}}&[j^{a}_i](\nu) = \frac{\mi}{2\pi} \left[\chi_{\mr{B}}[j^{a}_i](\nu^+) - \chi_{\mr{B}}[j^{a}_i](\nu^-)\right] \\
\label{eq:jjcorr_bubble}
&=  4 t^2e^2 \int_{-\infty}^{\infty}\!\!\! \mr{d}\omega \, \left[f(\omega) - f(\omega+\nu)\right] \times \\
&[A_{\vec{r}_{i,i}}(\omega) A_{\vec{r}_{i,i}}(\omega+\nu) - A_{\vec{r}_{i,i+\ahat}}(\omega) A_{\vec{r}_{i+\ahat,i}}(\omega+\nu)] \, . \nonumber
\end{align}
\section{Optical conductivity: numerical computation}
\label{sec:optical_numerics}
In this section, we describe how we compute the bubble contribution $\chi''_{\mr{B}}[j^{a}_\vec{0}](\nu)$ [Eq.~\eqref{eq:chiJ_q0_bubble}] in a numerically 
efficient way, how we treat the electronic self-energy close to zero frequency and temperature, and how we deal with vertex contributions and fulfillment of the $f$ sum rule.
We further discuss the potential role of vertex contributions for short-ranged nonlocal current fluctuations.
\subsection{Bubble contribution \label{sec:Bubble_comp}}
Computing the bubble contribution to the optical conductivity requires numerical evaluation of Eq.~\eqref{eq:chiJ_q0_bubble}.
This is challenging, especially close to $\nu=0$ or $T=0$, due to the close-to-singular behavior of $A_{\vec{k}}(\omega) A_{\vec{k}}(\omega \! + \! \nu)$
in the integrand. 

To deal with this, we exploit our knowledge of $G_{\vec{k}}^{-1}(\omega^+) = \omega^+ + \mu - \epsilon_{\vec{k}} - \Sigma_{\vec{k}}(\omega^+)$.
It is a smooth function of $\omega$ and known on a predetermined frequency grid $\omega \in \{\omega_{i}\}$. 
Since $G_{\vec{k}}^{-1}(\omega^+)$ is a smooth function, we represent it by linear interpolation, $G_{\vec{k}}^{-1}(\omega^+) = a_i+b_i\omega$, for $I_i = [\omega_i,\omega_{i+1}]$.
Due to the logarithmic resolution of NRG, we use a logarithmic frequency grid with $10^{-12} \leq |\omega_i| \leq 10^{4}$ and 200 grid points per decade.

By writing 
\begin{align}
A_{\vec{k}}(\omega) &A_{\vec{k}}(\omega \! + \! \nu) \! = \nonumber \\ 
 \tfrac{1}{\pi}\im \! &\left[ G_{\vec{k}}(\omega^+)\frac{G_{\vec{k}}(\omega \! + \! \nu^+)  -  G_{\vec{k}}(\omega \! + \! \nu^-)}{2\pi\mr{i}} \right] \, , 
\end{align}
the frequency integral in Eq.~\eqref{eq:Ik_nu} can be computed by evaluating the integrals,
\begin{align}
I^{\pm}_{\vec{k}}(\nu)  &=
\int_{-\infty}^{\infty}\!\!\! \mr{d}\omega  \left[f(\omega) \! - \! f(\omega \! +\! \nu)\right] G_{\vec{k}}(\omega^+) 
G_{\vec{k}}(\omega \! + \! \nu^{\pm}) \nonumber \\
&= \sum_{i} \int_{I_i} \mr{d}\omega \, \frac{\alpha_i + \beta_i \omega}{(a_i + b_i \omega) (c_{i,\pm} + d_{i,\pm} \omega)}
\label{eq:I_1_numerical}
\end{align}
where $\alpha_i + \beta_i \omega$ is a linear interpolation of $f(\omega) \! - \! f(\omega \! + \! \nu)$ 
on the interval $I_i$
and $a_i + b_i\omega$ and $c_{i,\pm} + d_{i,\pm} \omega$ are the linear interpolations of $G^{-1}_{\vec{k}}(\omega^+)$ and $G^{-1}_{\vec{k}}(\omega+\nu^{\pm})$, respectively.
The integral over every interval $I_i$ in Eq.~\eqref{eq:I_1_numerical} is very simple to evaluate exactly,
summing up the contributions from all intervals gives $I^{\pm}_{\vec{k}}(\nu)$.

The $\vec{k}$ sum/integral in Eq.~\eqref{eq:chiJ_q0_bubble} is finally computed using a standard integrator. (We use MATLAB's \texttt{integral} function.)
We use the periodized self-energy when computing Eq.~\eqref{eq:chiJ_q0_bubble}, cf.~App.~A.3 of Ref.~\cite{Gleis2023}.
In our case, this allows us to reduce the three-dimensional 
$\vec{k}$ integral in Eq.~\eqref{eq:chiJ_q0_bubble} 
to a one-dimensional one,
cf.~Eq.~(A10) of Ref.~\cite{Gleis2023}.

\begin{figure}
\includegraphics[width=\linewidth]{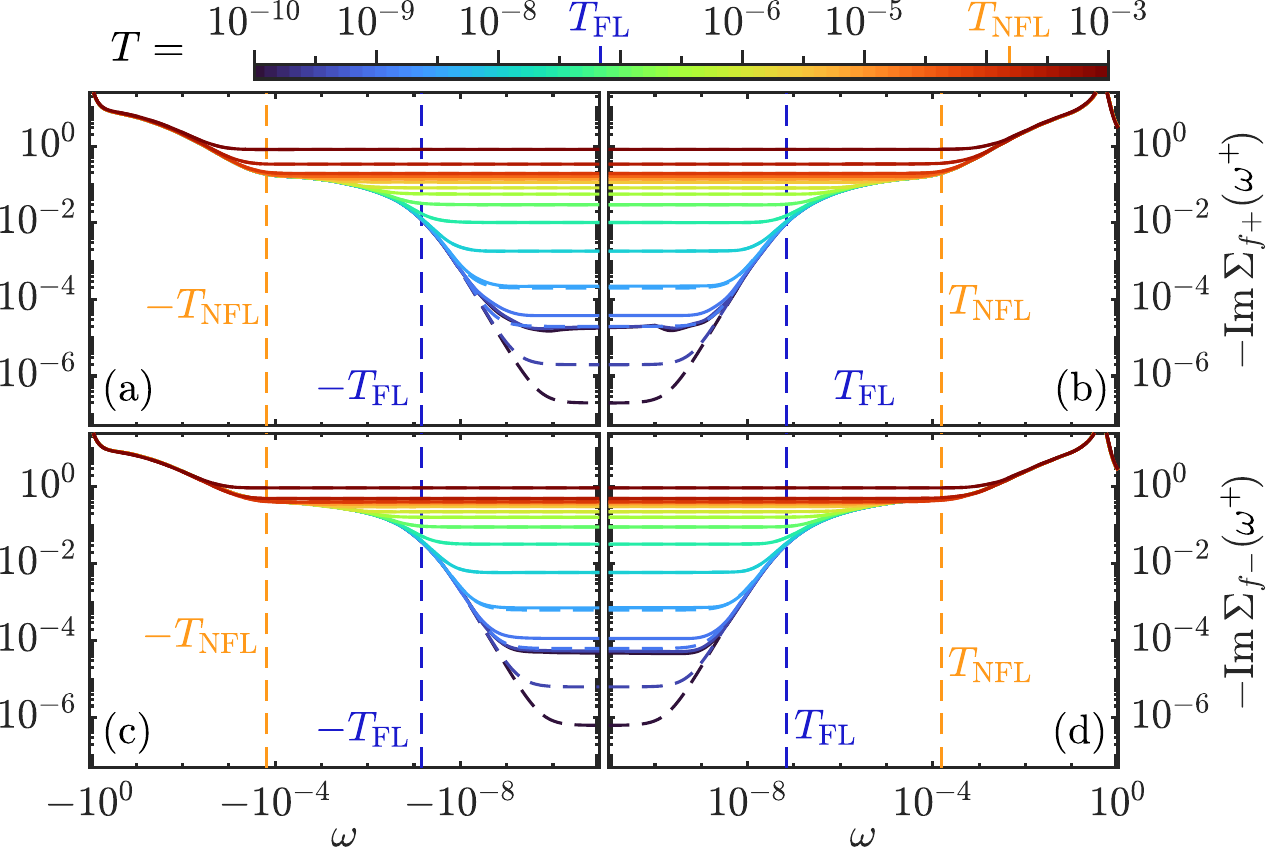}
\caption{%
Self-energy of the $f$ electrons for the self-consistent 2IAM at different temperatures. (a,b) Bonding orbital ($+$) and (c,d) anti-bonding orbital ($-$).
Solid lines denote the numerical data, dashed lines (not visible whenever they coincide with solid lines) denote the extrapolated self-energy.
Visible differences occur only for $|\omega|,T < 10^{-1} \TFL$, i.e., well below the FL scale $\TFL$. 
}
\label{fig:SE_PAM_extrapFL}
\end{figure}

\subsection{Self-energy at $\omega,T \simeq 0$ \label{sec:SE_extrap}}
The Drude peak which emerges in the optical conductivity at $T<\TFL$ for small frequencies
arises due to $-\im \Sigma(\omega^+) = a\omega^2 + b T^2$ behavior 
for $|\omega|, T  < \TFL$.
Capturing this $\omega, T$-dependence for very small $\TFL$ ($\ll \TNF$), as is the current case close to the QCP, is highly challenging.
To achieve this, we keep a large number of states---up to 40,000 $\mr{U}(1) \times \mr{SU}(2)$ symmetry multiplets---in iterative diagonalization
and use an interleaved Wilson chain~\cite{Mitchell2014,Stadler2016} to keep the computational cost manageable.
We compute the $f$-electon self-energy by using the symmetric improved estimator of Ref.~\cite{Kugler2022a}
which significantly reduces numerical artifacts and leads to state-of-the-art accuracy.
This accuracy allows us to obtain
 $-\im \Sigma_{f\pm}(\omega^+) = a\omega^2 + b T^2$ behavior 
  for $|\omega|, T \in (\TFL / 10, \TFL)$ (but not  
 for $|\omega|, T \in  (0, \TFL/10)$, because there $-\im \Sigma_{f\pm}(\omega^+)$  becomes smaller than $10^{-4}$, and numerical inaccuracies become
 significant). 
Therefore, we fit the coefficients $a$ and $b$ with the data for $(\TFL / 10, \TFL)$ then extrapolate $-\im \Sigma_{f\pm}(\omega^+)$ to $ (0, \TFL/10)$ based on the fitting.
Figure~\ref{fig:SE_PAM_extrapFL} shows the low $T$ and $\omega$ behavior of $-\im \Sigma_{f\pm}(\omega^+)$ before (solid) and after (dashed) extrapolation.
The $c$-electron self-energy $\Sigma_{c\pm}(\omega^+) = V^2/\bigl(\omega^+ -\epsilon_f - \Sigma_{f\pm}(\omega^+)\bigr)$ (which is not one-particle irreducible) follows from $\Sigma_{f\pm}(\omega^+)$.

Note that in an FL, $a \pi^2/b = 1$ should hold. 
On the other hand,
our fits yield $a \pi^2/b = \mc{O}(2\text{--}3)$ due to the broadening used in NRG, which overestimates $a$. We have checked that $a \to b/\pi^2$ when we 
lower the broadening width. This however comes at the expense of severe discretization artifacts. Since the exact value of $a$ is irrelevant to the present work, we preferred to adopt the procedure described above. 
\subsection{Local vertex contributions}
We stated several times in the main text that vertex contributions are crucial for the current-current correlation functions to capture the strange metallicity
and Planckian dissipation. 
As described in the main text, we have included vertex contributions only for the local contribution, $\chi[j^{a}_i]$, to the uniform current susceptibility, $\chi[j^{a}_{\vec{0}}] = \chi[j^{a}_i] + \chi_{\mr{nl}}[j^{a}_\vec{0}]$, 
where $j^{a}_i = -\mr{i} t e \sum_{\sigma} (c^{\dag}_{i\sigma} c^{\pdag}_{i+\ahat\sigma} - \mr{h.c.})$
is the current between lattice sites $i$ and $i+\ahat$, the neighbor of $i$ in $a$-direction.
The main reason is that we do not currently have access to three- or four-point correlation functions.
By choosing sites $i$ and $i+\ahat$ as the two sites of our self-consistent two-impurity model, 
we can compute $\chi[j^{a}_i]$ directly as a two-point correlation function using NRG. 
Here, we provide supplemental data that shows to what extent 
the full local susceptibility
$\chi[j^{a}_i]$ is influenced by its vertex contribution $\chi_\vertex[j^{a}_i] = \chi[j^{a}_i] - \chi_{\mr{B}}[j^{a}_i]$. To this end, we compare
$\chi[j^{a}_i]$ to its bubble contribution
$\chi_{\mr{B}}[j^{a}_i]$, computed via Eq.~\eqref{eq:jjcorr_bubble}. 
The integrand of the latter is \emph{not} close-to-singular [in contrast to
that of~\Eq{eq:Ik_nu}] and can therefore be efficiently evaluated via a standard integrator.

The bare output of NRG are discrete spectra for $\chi''[j^{a}_i]$, which are subsequently broadened through log-Gaussian broadening kernels,
see Ref.~\cite{Lee2016} for more details. The spectral functions used in Eq.~\eqref{eq:jjcorr_bubble} to compute $\chi''_{\mr{B}}[j^{a}_i]$ on the other hand 
are obtained by computing the \emph{self-energy} via the symmetric improved estimators of Ref.~\cite{Kugler2022a}; 
$\chi''_{\mr{B}}[j^{a}_i]$ therefore contains finer high-frequency details than achievable with NRG for $\chi''[j^{a}_i]$.
To compare the full $\chi''[j^{a}_i]$ and its bubble contribution $\chi''_{\mr{B}}[j^{a}_i]$ [computed from Eq.~\eqref{eq:jjcorr_bubble}], we, therefore, smear out the continuous curve of $\chi''_{\mr{B}}[j^{a}_i]$ by further applying the log-Gaussian kernel used to broaden the discrete data for $\chi''[j^{a}_i]$, to match their resolution levels.
We emphasize here that this broadening of $\chi''_{\mr{B}}[j^{a}_i]$ only affects high-frequency details at $|\omega| > \TNF$, the basic features remain the same. 

\begin{figure}
\includegraphics[width=\linewidth]{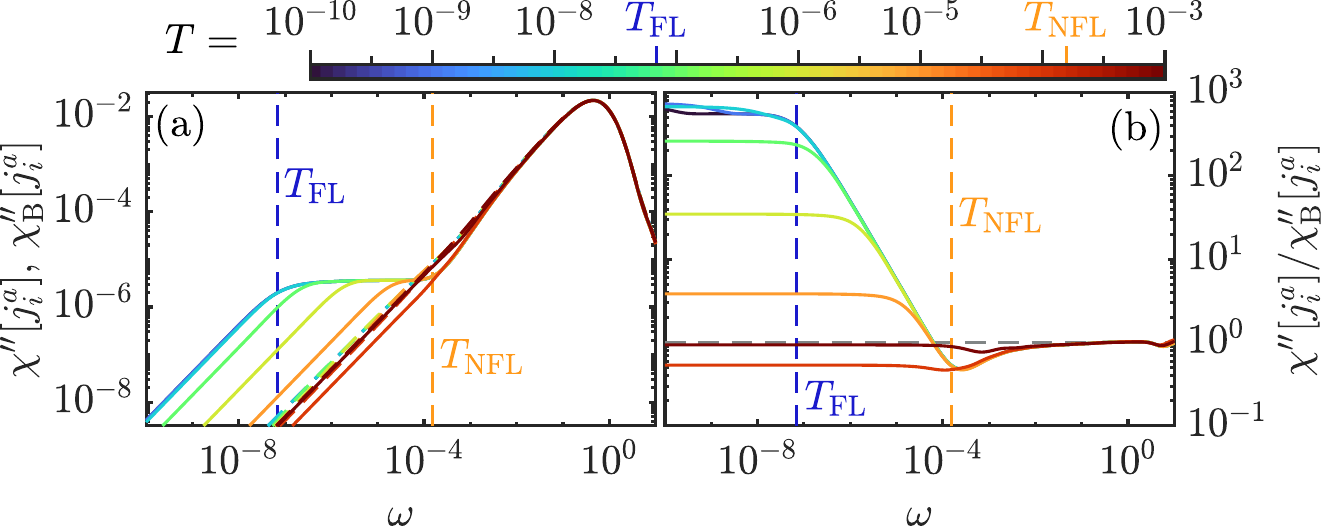}
\caption{%
(a) Spectrum of the local current susceptibility $\chi''[j^{a}_i](\omega,T)$. Solid lines are full susceptibilities $\chi''[j^{a}_i]$, dashed lines 
are the bubble contributions 
$\chi''_{\mr{B}}[j^{a}_i]$. $\chi''_{\mr{B}}[j]$ 
is almost temperature independent, which is why 
the $\chi''_{\mr{B}}[j^{a}_i]$ curves for $T<10^{-3}$ are covered by the $T=10^{-3}$ curve.  $\chi''_{\mr{B}}[j^{a}_i]$ and
$\chi''[j^{a}_i]$ are almost identical at $T=10^{-3}$. (b) The ratio between full susceptibility and bubble contribution.
}
\label{fig:ChiJ_Bubble_vs_full}
\end{figure}

Figure~\ref{fig:ChiJ_Bubble_vs_full}(a) shows the spectrum of the full local current susceptibility $\chi''[j^{a}_i]$ and of the corresponding bubble contribution $\chi''_{\mr{B}}[j^{a}_i]$, while
Fig.~\ref{fig:ChiJ_Bubble_vs_full}(b) shows their ratio.
The bubble contribution captures only the high-frequency behavior at $|\omega|,T > \TNF$ well: the spectra in  Fig.~\ref{fig:ChiJ_Bubble_vs_full}(a) almost 
coincide and the ratios in Fig.~\ref{fig:ChiJ_Bubble_vs_full}(b) are close to $1$. 

On the other hand, the plateau emerging below $|\omega|,T < \TNF$ is not captured at all by the bubble contribution,
i.e., both the $\omega/T$ scaling and the Planckian dissipation discussed in the main text and in Sec.~\ref{sec:scaling_function} result from vertex contributions.
 The ratio shown in Fig.~\ref{fig:ChiJ_Bubble_vs_full}(b) increases dramatically in the NFL region ($\TFL < |\omega|,T < \TNF$)
 by several orders of magnitude and saturates close to $10^{3}$ in the FL region ($|\omega|,T < \TFL$).
\subsection{Estimate of nonlocal vertex contributions \label{sec:non_loc_vert}}
\label{sec:Jcorr_nonloc}

To estimate what to expect for nonlocal current fluctuations in terms of scaling and vertex contributions, 
we define ``current'' operators that lie across the cluster boundaries,
\begin{equation}
j_{i} = (-1)^{i} \frac{\mr{i}te}{\sqrt{5}}\left(c^{\dagger}_{i\sigma}a^\pdag_{i\sigma} - \mr{h.c.}\right), 
\label{eq:j_i}
\end{equation}
where $a_{i\sigma}$ annihilates a spin-$\sigma$ electron in the first bath orbital (within the Wilson chain) that directly couples to the $c$ orbital of the cluster site $i = 1,2$.
According to the effective medium construction of DMFT (which defines bath sites by replacing the interaction on the original lattice sites by the self-energy, cf.~Sec.~III D of Ref.~\cite{Georges1996}),
the Green's function of $a_{i\sigma}$ is the same as that of
a symmetric superposition of the five nearest neighbors (on the lattice) of site $i$ which are not located on the same cluster.
Due to that, we can interpret these orbitals as a proxy for the aforementioned symmetric superposition. The current operators in Eq.~\eqref{eq:j_i}
can therefore be interpreted as a proxy for the average (hence normalization by $\sqrt{5}$) current between these five nearest-neighbor sites and the corresponding cluster site.
Since there is no specific direction in the lattice associated with these currents, we did not specify a superscript $a$ in Eq.~\eqref{eq:j_i}.
We emphasize that this correspondence is \emph{not} exact since the first bath sites are non-interacting orbitals that belong to the dynamical mean field.
Correlators involving $j_1$ or $j_2$ do not enter the results shown in the main text.

\begin{figure}
\includegraphics[width=\linewidth]{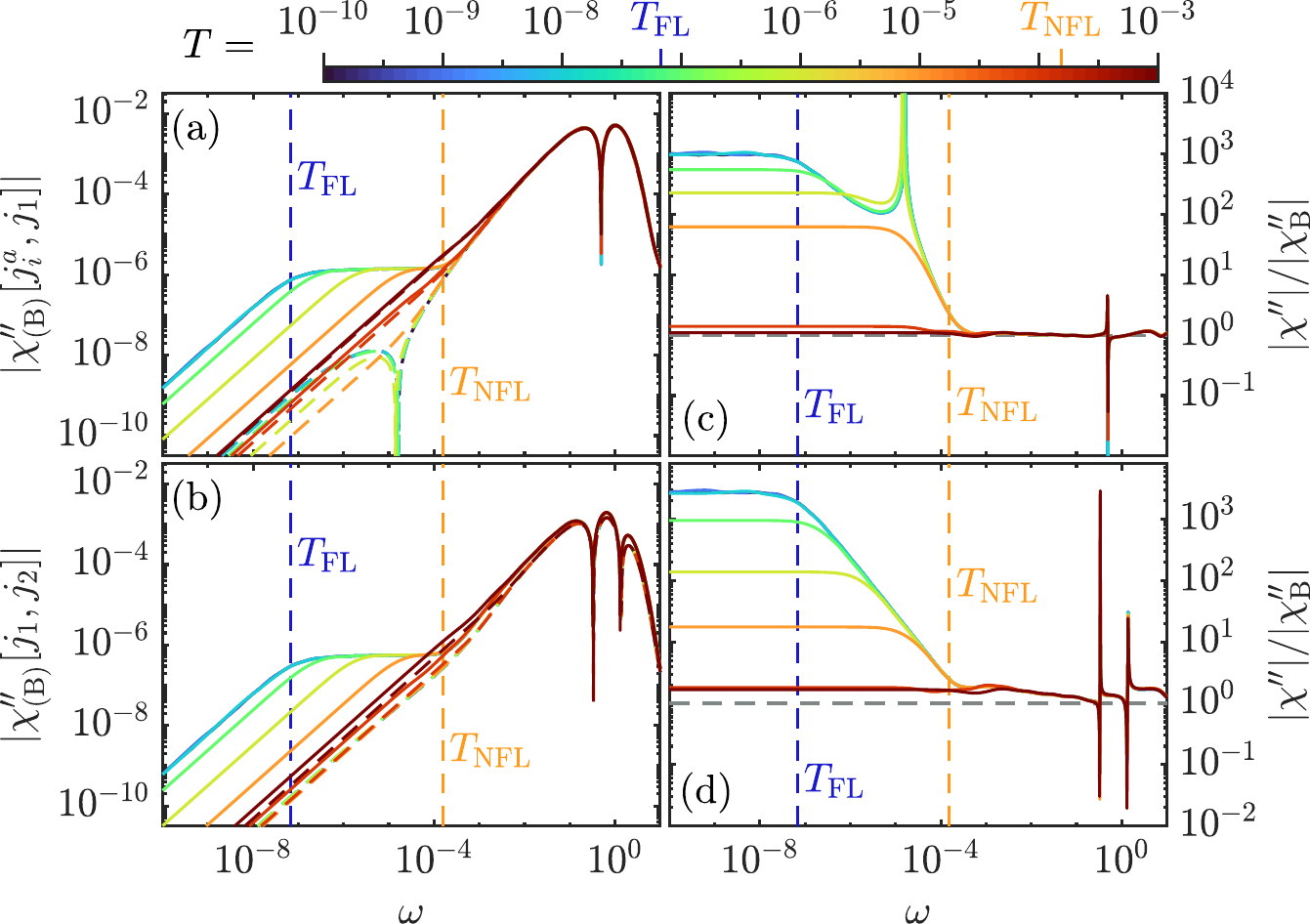}
\caption{%
(a,b)
Absolute values of the spectra of different nonlocal current susceptibilities, $\chi''[j^{a}_i,j_1](\omega,T)$ and $\chi''[j_1,j_2](\omega,T)$. Solid lines are full susceptibilities, dashed lines 
are the bubble contributions. 
Cusps indicate sign changes in the spectra.
(c,d) 
Ratios between the spectra of the full susceptibility and the bubble contribution.
The cusps at $|\omega| > 10^{-1}$ arise due to a slight misalignment between the sign changes in $\chi''$ and $\chi''_{\mr{B}}$.
}
\label{fig:ChiJ_nonloc}
\end{figure}

We compute $\chi[j^{a}_i,j_1]$ and $\chi[j_1,j_2]$ to estimate
the behavior of nearest-neighbor and next-nearest-neighbor 
current susceptibilities, respectively.
Their spectra, including the corresponding bubble contribution, are shown in Fig.~\ref{fig:ChiJ_nonloc}(a,b). 
The spectra of the full susceptibilities
again show a similar plateau as observed for the local current susceptibility. 
Figure~\ref{fig:ChiJ_nonloc}(c,d) shows the ratio between full susceptibility and bubble contribution.
Similarly to the local current susceptibility, the ratio is somewhat close to 1 for 
$|\omega|, T > \TNF$ and becomes large for $|\omega|, T < \TNF$, suggesting that vertex contributions are important also on the nonlocal level in this region. 

\begin{figure}
\includegraphics[width=\linewidth]{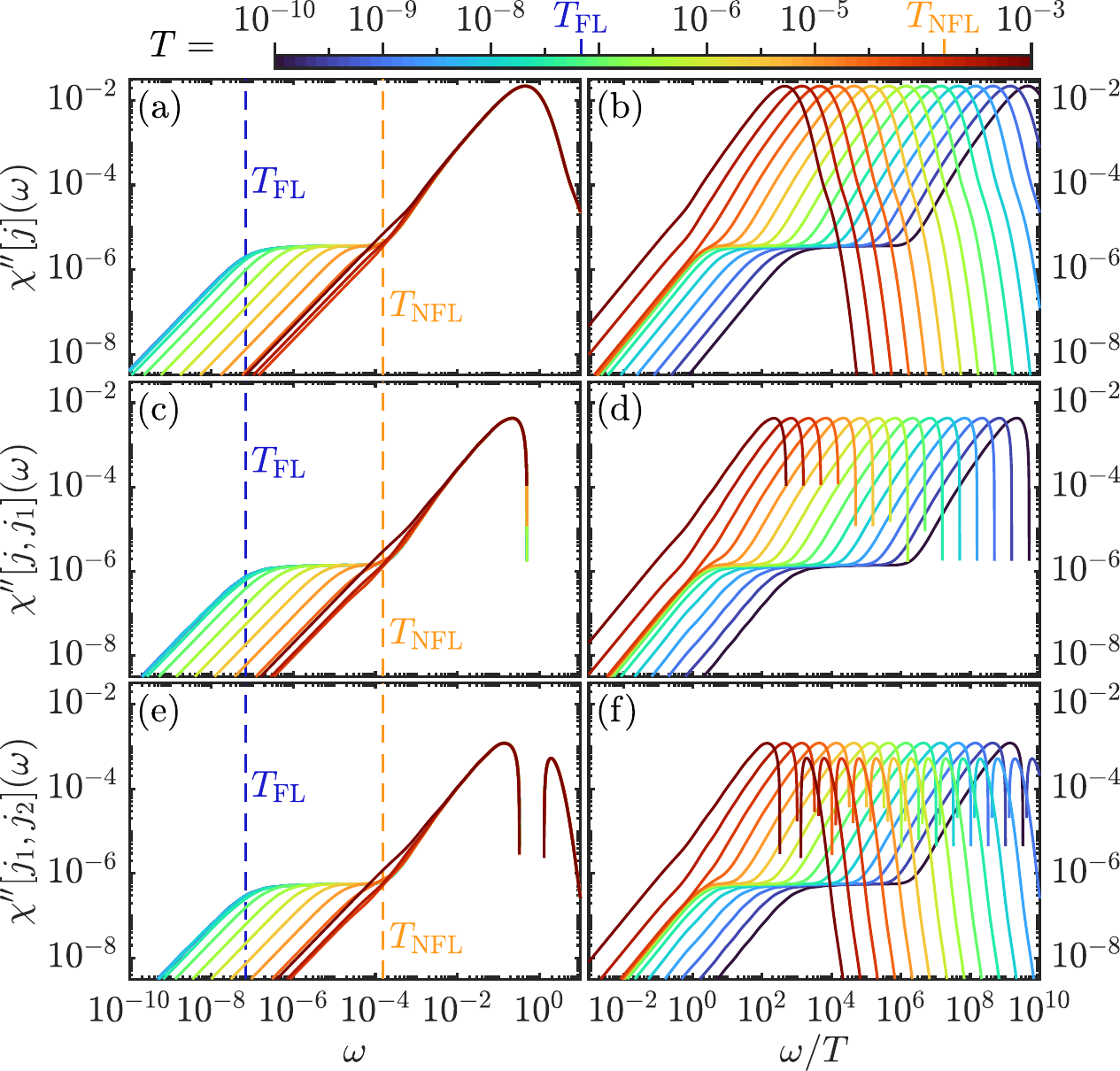}
\caption{%
Current spectra versus frequency (left columns) and versus $\omega/T$ (right columns).
(a,b) Local current spectrum. (c,d) Proxy to the nearest-neighbor current spectrum.
(e,f) Proxy to the next-nearest-neighbor current spectrum.
}
\label{fig:ChiJ_all_scaling}
\end{figure}

In Fig.~\ref{fig:ChiJ_all_scaling}, we further illustrate that $\chi''[j^{a}_i,j_1]$ and $\chi''[j_1,j_2]$ show $\omega/T$ scaling very similar to $\chi''[j^{a}_i]$.
Since the behavior of the nonlocal susceptibilities is qualitatively similar to that of the local susceptibility,
we expect that the full nonlocal current susceptibility $\chi''_{\mr{nl}}[j]$, in contrast to its bubble contribution $\chi''_{\mr{B,nl}}[j]$, will show similar $\omega/T$ scaling as $\chi''[j^{a}_i]$.
As discussed in the main text, we expect that the full inclusion of vertex contributions in $\chi''_{\mr{nl}}[j]$ will ameliorate or
fully avoid the artifacts seen in Fig.~\ref{fig:optical}(c) for 
the resistivity $\rho(T)$: 
(i) in the NFL region, the nearly-$T$-linear behavior will become fully-$T$-linear; 
and (ii) in the FL-to-NFL crossover region, the shoulder will become less prominent or disappear.

\subsection{Drude weight}

In this section, we discuss the Drude weight of \Eq{eq:DrudeWeight},
 $D/\pi = \chi'[j^{a}_{\vec{0}}](0) - \langle \hat{K} \rangle$.
According to \Eq{eq:sigma_inf},  
if $D \neq 0$ that would imply (i) a $\delta(\omega)$ contribution to $\sigma'(\omega)$ and therefore zero resistivity (i.e., persistent currents), and (ii) a $1/\omega$ contribution in $\sigma''(\omega)$.
Since our study of $\sigma(\omega)$ considers only 
non-superconducting solutions 
at $T>0$, we expect that our system does not support persistent currents and 
$D=0$. Accordingly, we have set  $D=0$ for all results shown in this manuscript. 

\begin{figure}
\includegraphics[width=\linewidth]{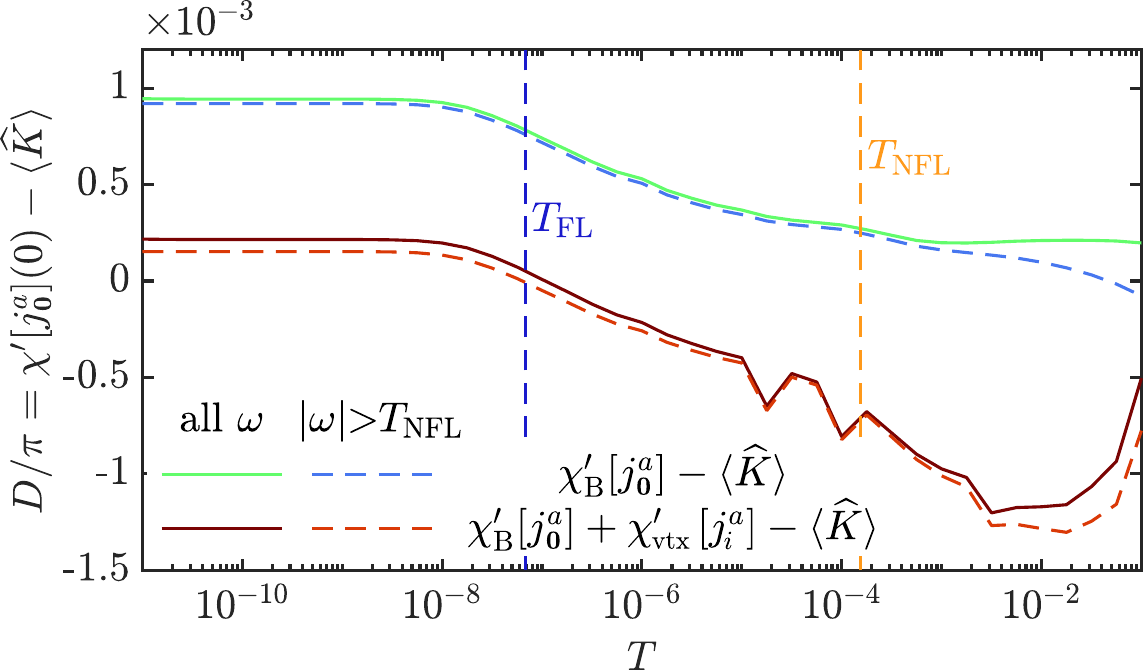}
\caption{%
The Drude weight $D/\pi = \chi'[j^{a}_{\vec{0}}](0) - \bigl\langle \hat{K} \bigr\rangle$ vs.~temperature. 
For the upper (or lower) row of the legend, 
$\chi'[j^{a}_{\vec{0}}](0)$ was approximated 
as $\chi'_{\mr{B}}[j^{a}_{\vec{0}}](0)$ (or $\chi'_{\mr{B}}[j^{a}_{\vec{0}}](0) + \chi'_\vertex[j^{a}_{i}](0)$), i.e., using only 
the bubble contribution (or including also the local vertex contribution).
When computing these $\chi'[j](0)$ terms 
via the Kramers--Kronig transformation \eqref{eq:KramersKronig}, we either 
integrated over all $\omega \in \mathbbm{R}$ (solid lines) 
or only high frequencies $|\omega| > \TNF$ (dashed lines). 
Since solid and dashed lines almost match,  
$\chi'[j](0)$ is governed by high-frequency contributions, where NRG has poorer frequency resolution.
From that perspective, the values for the Drude weight found here,
$D/\pi \lesssim 10^{-3}$, are remarkably close to the expected value of zero. 
}
\label{fig:f_sum_rule}
\end{figure}

As a consistency check, we have also computed the Drude weight directly. This is a difficult task, since the numerical challenges involved in computing  $\chi'[j^a_{\vec{0}}](0)$,  a \emph{uniform, zero-frequency} susceptibility, and $\langle \hat{K} \rangle$, a \emph{local, equal-time} expectation value, are quite different.
Moreover, our computation of $\chi'[j^a_{\vec{0}}]$ involves
a rather crude approximation [\Eq{eq:chiJ_vertcorr} of the main text]. 
Nevertheless,
we find $|D|/\pi$ to be remarkably small, $\lesssim 10^{-3}$,
with an estimated numerical uncertainty that is likewise of the order of $10^{-3}$.
This justifies our choice to set $D=0$.  Below, we describe how we obtained these values.

Figure~\ref{fig:f_sum_rule} shows the Drude weight 
$D/\pi$, with the static current response $\chi'[j^{a}_{\vec{0}}](0)$ computed via Eq.~\eqref{eq:chiJ_vertcorr} in the main text. 
Both the bubble contribution $\chi'_{\mr{B}}[j^{a}_{\vec{0}}](0)$ 
and our locally vertex-corrected result $\chi'_{\mr{B}}[j^{a}_{\vec{0}}](0) + \chi'_\vertex[j^{a}_{i}](0)$ show a deviation from $\langle \hat{K} \rangle$ of the
order of $10^{-3}$. The inclusion of $\chi'_\vertex[j^{a}_{i}](0)$ slightly reduces this deviation at low $T$ but slightly increases it at high $T$.
The solid and dashed lines in Fig.~\ref{fig:f_sum_rule} compare 
results obtained by computing the $\chi'[j^{a}_{\vec{0}}](0)$ contributions via the Kramers--Kronig transform \eqref{eq:KramersKronig} in two ways, either including the spectral weight from all frequencies, $\omega \in \mathbbm{R}$ (solid), or only from large frequencies, $|\omega|>\TNF$ (dashed). Since the solid and 
dashed lines almost match, the contribution to $D$ from low frequencies $|\omega|<\TNF$ (including the contribution from the plateau in $\chi''_\vertex[j^{a}_{i}](\omega)$) is negligible. Therefore, the non-fulfillment of $D = 0$ is mainly due to inaccuracies
at high frequencies.  

High-frequency inaccuracies are to be expected in NRG spectra,
due to the use of logarithmic discretization and an asymmetric log-Gaussian broadening
kernel (cf. Eqs.~(17) and (21) from Ref.~\cite{Lee2016}), which can lead
to slight shifts in spectral weight.
The broadened spectral function is evaluated on a logarithmic frequency grid and approximated by linear interpolation between grid points.
In practice, this means that if a discrete spectrum of the form $\chi''(\omega) = \sum_{j} \chi''_j \delta(\omega - E_j)$ is broadened, the integral of the broadened spectrum can differ slightly from the actual weight, $\sum_j \chi_j''$, typically by an amount $\sim \mc{O}(10^{-3})$. 
As a result, the Kramers--Kronig transformation used to compute $\chi'(0) = - \mathcal{P} \int \chi''(\omega)/\omega$ 
usually induces an error $\sim \mc{O}(10^{-3})$, compared to
the result directly computed from the discrete data, $\chi'(0) = -\sum_{j} \chi''_j /E_j$.
Since our approximation of $\chi''[j^{a}_{\vec{0}}](\omega)$ involves the bubble contributions $\chi_{\mr{B}}''[j^{a}_{\vec{0}}](\omega)$
and $\chi_{\mr{B}}''[j^{a}_{i}](\omega)$ which are only available as broadened spectral functions, direct computation of 
$\chi_{\mr{B}}'[j^{a}_{\vec{0}}](0)$ from discrete data is not possible.
All of the aforementioned issues, on top of the approximation \eqref{eq:chiJ_vertcorr}, can lead to inaccuracies in the spectral weights and their corresponding frequencies.
We have checked that shifting spectral positions by $\mc{O}(1\%)$,
i.e., $\omega \to  (1\pm10^{-2})\omega$ and 
normalizing the spectra accordingly, i.e.,
$\chi''(\omega) \to (1\pm10^{-2})^{-1} \chi''(\omega)$, is sufficient to
change $\chi'[j^{a}_{\vec{0}}](0)$ 
by $\mc{O}(10^{-3})$.
For all these reasons, we estimate the numerical uncertainty 
of our determination of the Drude weight $D$ to be at least of the order of $10^{-3}$.

\section{Scaling function}
\label{sec:scaling_function}
In the main text, we proposed the phenomenological ansatz $\widetilde{\chi}''(\omega,T)$
to capture the $X^{xz}$ and current spectra in the NFL region.
In the limit of $|\omega| \ll \TNF$, the ansatz is governed by the scaling function~\eqref{eq:scaling_sus}, $\mc{X}(x) = \mc{X}'(x) - \mr{i}\pi \mc{X}''(x)$.
To ease future referencing, we duplicate the ansatz~\eqref{eq:pheno_ansatz} and its relation~\eqref{eq:scaling_sus} to the scaling function here:
\begin{align}
\widetilde{\chi}''(\omega,T) &= \chi_0  \! \int_{T}^{\TNF} \! \frac{\mr{d}\epsilon}{\pi}
\frac{(1-\mr{e}^{-\frac{\omega}{T}}) (\frac{\epsilon}{T})^{\nu} bT}{(\omega - a\epsilon)^2 + (b T)^2} \, , \label{eq:pheno_ansatz_supp} \\
\widetilde{\chi}(\omega,T) &\simeq 
\mc{X}'_0\Big(\frac{T}{\TNF}\Big) + \mc{X}'\Big(\frac{\omega}{T}\Bigr) - \mr{i}\pi \mc{X}''\Big(\frac{\omega}{T}\Big) \, ,
\label{eq:scaling_sus_supp} 
\end{align}
In this Section, we motivate the ansatz, derive the scaling function, and provide the detail of the fitting, for the $X^{xz}$ susceptibility.
The discussion for other susceptibilities showing a plateau and scaling in the NFL region (e.g., the current susceptibility) is analogous.

We start with representing the greater correlation function $\chi_{>}[X^{xz}]$ in terms of a superposition of coherent excitations,
\begin{align}
\label{eq:t_ansatz}
\chi_{>}[X^{xz}](t) &= -\mr{i}\theta(t) \langle X^{xz}(t) X^{xz} \rangle \\
\simeq \widetilde{\chi}_{>}[X^{xz}](t) &= -\mr{i}\int_{T}^{\TNF} \mr{d}\epsilon \, \left(\frac{\epsilon}{T}\right)^{\nu} \mr{e}^{-\mr{i}(a\epsilon - \mr{i} bT)t} \, . \nonumber
\end{align}
These coherent excitations have mean energy $a\epsilon$, decay rate $bT$, and a power-law density of states with exponent $\nu$. 
We assume that the spectrum of this ansatz,
\begin{align}
\label{eq:gtr_spectrum}
\widetilde{\chi}_{>}[X^{xz}](\omega) &= -\mr{i} \int_{0}^{\infty} \mr{d}t \, \langle X^{xz}(t) X^{xz} \rangle \mr{e}^{\mr{i}\omega^{+} t} \, , \\
\widetilde{\chi}''_{>}[X^{xz}](\omega) &= -\frac{1}{\pi} \im \, \widetilde{\chi}_{>}[X^{xz}](\omega) \, , \nonumber
\end{align}
captures the low-frequency behavior, $|\omega| < \TNF$. 
High frequencies $|\omega| > \TNF$ are not governed
by the quantum critical point and contain information on the local-moment behavior which is not of interest here.
The spectrum should also fulfill the fluctuation-dissipation theorem,
\begin{align}
\label{eq:constraint_spectra}
\widetilde{\chi}''_{>}[X^{xz}](-\omega) &= -\frac{1-\mr{e}^{-\omega/T}}{1-\mr{e}^{\,\omega/T}} \widetilde{\chi}''_{>}[X^{xz}](\omega) \, ,
\end{align}
which mainly affects and constrains the very low-frequency spectrum, $|\omega| \lesssim T$.
We therefore use our ansatz~\eqref{eq:t_ansatz} to compute the $\omega>0$
part of the spectrum~\eqref{eq:gtr_spectrum} and we then determine the $\omega<0$ part via Eq.~\eqref{eq:constraint_spectra}, i.e., we enforce Eq.~\eqref{eq:constraint_spectra}.

The spectrum of the corresponding retarded correlator
is given by
\begin{align}
\widetilde{\chi}''[X^{xz}](\omega) = (1-\mr{e}^{-\omega/T}) \widetilde{\chi}''_{>}[X^{xz}](\omega) \, ,
\end{align}
which leads to the ansatz \eqref{eq:pheno_ansatz_supp} for $\omega > 0$.
The $\omega < 0$ side is given by the oddity, $\widetilde{\chi}''[X^{xz}](-\omega) = -\widetilde{\chi}''[X^{xz}](\omega)$.
The real part is obtained via the Kramers--Kronig relation,
\begin{align}
\label{eq:Kramers_Kronig_scalingansatz}
\widetilde{\chi}'[X^{xz}](\omega) = \mc{P} \int_{-\infty}^{\infty} \mr{d}\omega' \frac{\widetilde{\chi}''[X^{xz}](\omega')}{\omega-\omega'} \, .
\end{align}

To get the scaling function $\mc{X}''$, we take the limit of $\TNF \to \infty$ in \Eq{eq:pheno_ansatz_supp}.
(This limit exists for $\nu < 1$, while our data shows $\nu \simeq 0$.)
Equation~\eqref{eq:pheno_ansatz_supp} is then a function of $x = \omega/T$,
\begin{align}
\label{eq:X_x_scaling}
 \mc{X}''(x) &= \chi_0  \! \int_{1}^{\infty} \! \frac{\mr{d}y}{\pi}
\frac{(1-\mr{e}^{-x}) y^{\nu} b}{(x - a y)^2 + b^2} \; , \, x>0 \, , \\
\mc{X}''(-x) &= - \mc{X}''(x) \, . \nonumber
\end{align}

In Eq.~\eqref{eq:Kramers_Kronig_scalingansatz}, $\widetilde{\chi}'[X^{xz}](\omega)$ is singular in $\TNF/T \to \infty$ if $\nu \geq 0 $.
Therefore, we split the real part into a potentially singular static part, $\widetilde{\chi}'[X^{xz}](0)$, and a non-singular part,
$\widetilde{\chi}'[X^{xz}](\omega) - \widetilde{\chi}'[X^{xz}](0)$.
Using 
\begin{align}
\frac{1}{\omega-\omega'} - \frac{1}{-\omega'} = \frac{\omega}{(\omega-\omega')\omega'} \, , \nonumber
\end{align}
we can take the $\TNF/T \to \infty$ limit of the non-singular $\widetilde{\chi}'[X^{xz}](\omega) - \widetilde{\chi}'[X^{xz}](0)$ part,
\begin{align}
\label{eq:X_x_scale_real}
\mc{X}'(x) = \mc{P} \int_{-\infty}^{\infty} \mr{d}x' \frac{x\mc{X}''(x')}{(x-x')x'} \, .
\end{align}
This defines the scaling function $\mc{X}(x) = \mc{X}'(x) - \mr{i}\pi \mc{X}''(x)$.

For the potentially singular static contribution $\widetilde{\chi}'[X^{xz}](0)$, we cannot safely take the $\TNF \to \infty$ limit.
In the $\TNF/T \gg 1$ limit, the spectral part $\widetilde{\chi}''[X^{xz}](\omega)$ sharply drops to zero for $|\omega| > \TNF$,
so that we can approximate $\widetilde{\chi}'[X^{xz}](0) \simeq  \mc{X}_0'(T/\TNF)$, with
\begin{align}
\label{eq:X_T_scale}
\mc{X}_0'(y) = -\mc{P} \int_{-y}^{y} \mr{d}x' \, \frac{\mc{X}''(x')}{x'} \, .
\end{align}
$\mc{X}_0'(T)$ describes the contribution of the excitations within the NFL region to the static response,
\begin{align}
\chi_\nfl'(0) = -\mc{P} \int_{-\TNF}^{\TNF} \mr{d}\omega' \,  \frac{\chi''(\omega')}{\omega'} \, .
\end{align}
The remaining contribution from high-energy excitations,
\begin{align}
\chi_{\mr{high}}'(0) = \chi'(0) - \chi_\nfl'(0) \, ,
\end{align}
may dominate the temperature dependence of $\chi'(0)$. In that case, $\mc{X}_0'(T/\TNF)$ only governs $\chi_\nfl'(0)$ 
but \emph{not} $\chi'(0)$. This is for instance the case for the static current susceptibility, 
where only $\chi_\nfl[j]'(0)$ follows $\mc{X}_0'(T)$. On the other hand, $\chi'[X^{xz}](0)$ is well
described by $\mc{X}_0'(T)$ up to an additive constant.

We determine the parameters $a$, $b$, $\nu$ and $\chi_0$ in Eq.~\eqref{eq:X_x_scaling} by fitting logarithms of $\chi''_{>}(\omega)$ to the logarithm of our scaling ansatz~\eqref{eq:X_x_scaling}.
We employ a least-square fit on a logarithmic frequency grid with 20 grid points per decade and frequencies between $\omega_{\mr{min}} = 10^{-9}$ and $\omega_{\mr{max}} = \TNF/4$,
i.e., we stay well below the crossover temperature $\TNF$. 
Our fits are done for seven logarithmically spaced temperatures between $(\TFL \ll) 10^{-6.5}$ and $10^{-5} (\ll \TNF)$, i.e., for temperatures well separated from the crossover temperatures $\TFL$
and $\TNF$. We then determine a scaling curve by the geometric average over the fitted curves at different temperatures. The largest deviations from the geometric average serve as an error bar. 
$\mc{X}'(x)$ and $\mc{X}'_0(T)$ are determined via Eqs.~\eqref{eq:X_x_scale_real} and~\eqref{eq:X_T_scale}, respectively.

\begin{figure}
\includegraphics[width=\linewidth]{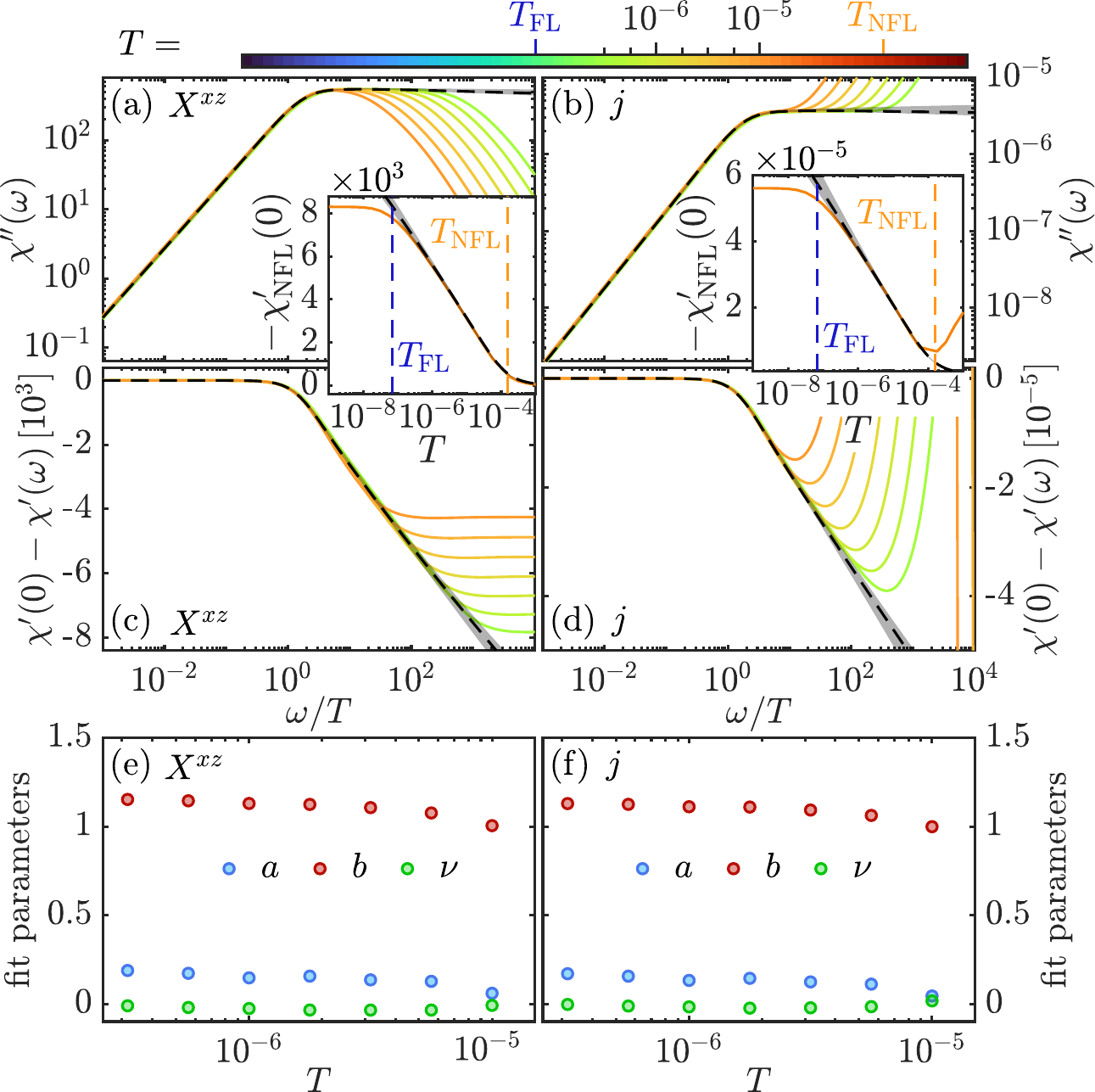}
\caption{%
(a) $\chi''[X^{xz}](\omega)$ and (b) $\chi''[j](\omega)$ (solid lines) versus scaling function $\mc{X}''(x)$ (black dashed line). The
grey shaded area indicates the deviation when fitting at different temperatures. Only curves used in the fitting process are shown,
the ticks on the color bar at the top indicate the temperature, and the color range is the same as in Fig.~\ref{fig:optical} of the main text.
(c,d) Corresponding real parts. Insets: NFL contribution to the static susceptibility. 
(e,f) Fit parameters at different 
temperatures. The 95\% confidence interval is smaller than the symbol size. 
}
\label{fig:scaling_fit}
\end{figure}

Figure~\ref{fig:scaling_fit}(a--d) shows the fitting result for $\chi''[X^{xz}](\omega)$ and $\chi''[j](\omega)$. In both cases, our ansatz fits our data very well, with all temperatures yielding 
very similar curves (the grey area, indicating the largest deviations from the geometric mean, is relatively small). Fig.~\ref{fig:scaling_fit}(e,f) shows the results for the fit parameters
$a$, $b$ and $\nu$. The fitting parameters for both $\chi''[X^{xz}](\omega)$ and $\chi''[j](\omega)$ are very similar and the variation with temperature is small. 
We note that the fits for the highest temperatures are a little less reliable because the plateau in $\chi''(\omega)$ is not that well developed yet.
Most important to us is the result for $b$, which varies between $1.153$ at $T=10^{-6.5}$ and $1.005$ at $T=10^{-5}$ for $\chi''[X^{xz}](\omega)$
and between $1.130$ at $T=10^{-6.5}$ and $0.999$ at $T=10^{-5}$ for $\chi''[j](\omega)$. Thus, our results are consistent with Planckian dissipation,
i.e., the lifetime of $X^{xz}$ or current excitations is $\tau \simeq 1/T$, up to a prefactor close to $1$.

\begin{figure}
\includegraphics[width=\linewidth]{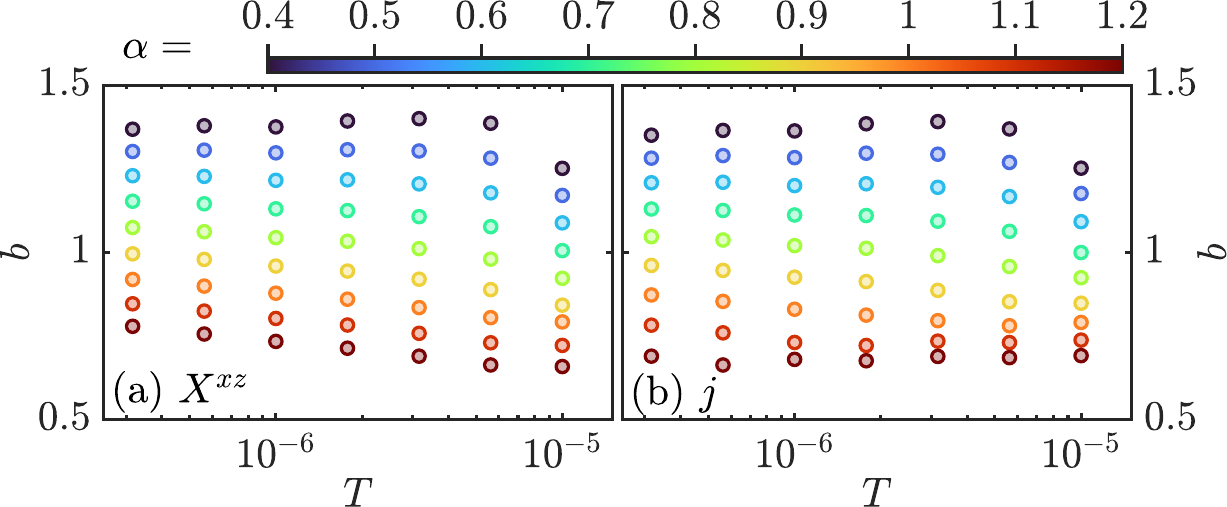}
\caption{%
Effect of log-Gaussian broadening width $\sigma = \alpha \ln \Lambda$ on the fit parameter $b$ for (a) $\chi''[X^{xz}](\omega)$ and (b) $\chi''[j](\omega)$.
}
\label{fig:scaling_fit_broadening_dep}
\end{figure}

The fit parameters also depend on how the discrete spectral data from NRG is broadened. For our scaling analysis, we used both a log-Gaussian broadening
kernel (cf.~Eq.~(17) of Ref.~\cite{Lee2016}) with width $\sigma = 0.7 \ln \Lambda$ ($\Lambda = 3$) and the derivative of the Fermi-Dirac distribution with width $\gamma = T/10$ (cf.~Eq.~(21) of Ref.~\cite{Lee2016}) as linear broadening kernel. The broadening parameters are chosen such that the data is almost underbroadened (i.e., discretization artifacts become visible for smaller broadening width). 
In Fig.~\ref{fig:scaling_fit_broadening_dep}, we show the effect on $b$ of varying the width $\sigma = \alpha \ln \Lambda$ of the log-Gaussian broadening kernel. Most importantly, $b$ remains of order 1 and changes from $b \simeq 1.4$ for  $\alpha = 0.4$ (underbroadened) to $b \simeq 0.66$ for $\alpha = 1.2$ (overbroadened). Interestingly, the parameter $b$ which determines the decay rate \emph{decreases} with \emph{increasing}
broadening width. The linear broadening parameter $\gamma$ (not shown) appears to have the converse effect, i.e., lower $\gamma$ leads to lower $b$ and vice versa.

\begin{figure}
\includegraphics[width=\linewidth]{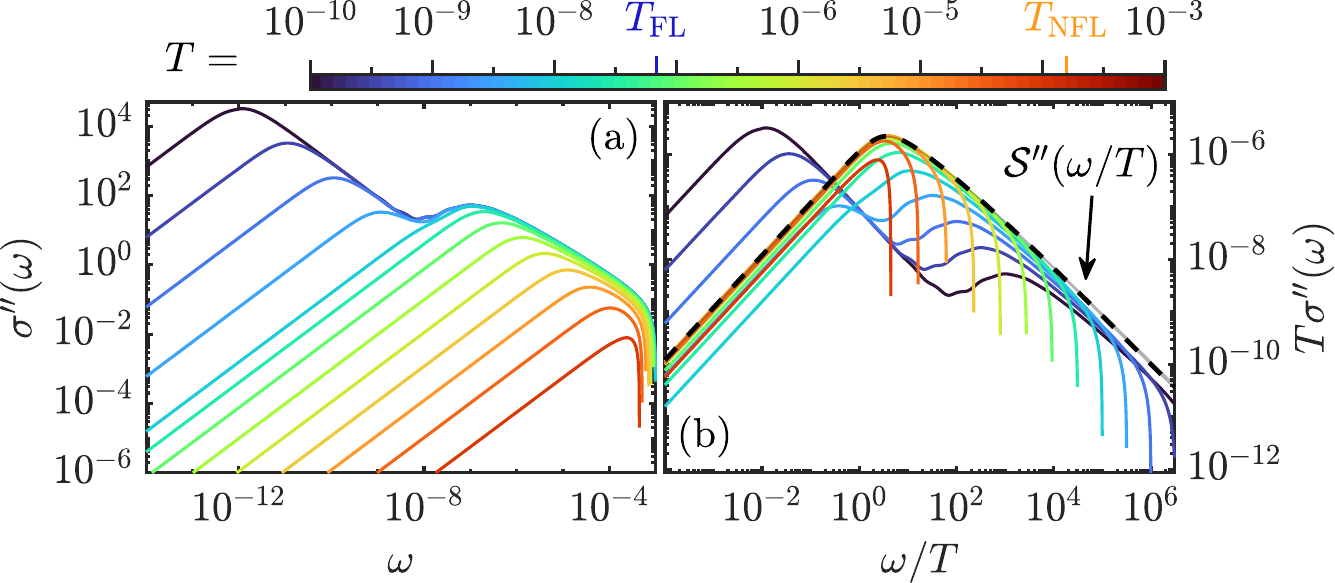}
\caption{%
(a) Imaginary part of the optical conductivity at different temperatures. $\sigma''(\omega)$ becomes negative around $\omega \lesssim 10^{-3}$. 
(b) Dynamical scaling of the imaginary part. In the NFL region, 
all curves fall onto the scaling curve $\mc{S}''(\omega/T)$. Data at $\omega > 10^{-3}$ has been omitted for clarity.}
\label{fig:sigma_imag_scaling}
\end{figure}

The scaling function $\mc{S}(x)$ for the optical conductivity follows from the scaling function $\mc{X}(x)$
for the current susceptibility,
\begin{align}
T\sigma(\omega) &= \mc{S}(x) =  -\frac{1}{\mr{i}x}\mc{X}(x) = \mc{S}'(x) + \mr{i} \mc{S}''(x) \, , \nonumber \\
\mc{S}'(x) &= \frac{\pi}{x} \mc{X}''(x) \, ,\\
\mc{S}''(x) &= \frac{1}{x} \mc{X}'(x) = \mc{P} \int_{-\infty}^{\infty} \mr{d}x' \frac{\mc{X}''(x')}{(x-x')x'} \, . \nonumber
\end{align}
We show and discuss the scaling of the real part of $\sigma$ in the main text; the scaling of the imaginary part is shown in Fig.~\ref{fig:sigma_imag_scaling}(b).

\section{Complex optical conductivity}
\label{sec:compex_optical}

In this section, we provide data on the imaginary part of the optical conductivity. Further, we make contact with recent experimental data on \CCI\ from Ref.~\cite{Shi2023},
where the dynamical scattering rate $\tau^{-1}(\omega)$ and the dynamical effective mass $m^{\ast}(\omega)$ were studied. 
Even though \CCI\ has a strange-metal-like $\tau^{-1}(0) \propto \rho(T) \propto T$, the authors of Ref.~\cite{Shi2023} found that both the freqnency dependence of
$\tau^{-1}(\omega)$ and the temperature dependence of the renormalized scattering rate $\tau^{\ast -1} = \tau^{-1}(0)/m^{\ast}(0)$ show FL-like $\omega^2$ and $T^2$ behavior, respectively.
We make contact with these surprising experimental findings by showing that in our CDMFT+NRG data, (i) $\tau^{-1}(\omega) \propto \omega^2$ holds at low frequencies ($|\omega|\lesssim T$) throughout the strange-metal region $\TFL < T < \TNF$;
and (ii) $\tau^{\ast -1} \propto T^{2}$ holds for $\TNF/10 \lesssim T < \TNF$ while deep in the strange metal, $\tau^{\ast -1} \propto T$ holds.

The imaginary part $\sigma''(\omega) = \im \, \sigma(\omega)$ is shown in Fig.~\ref{fig:sigma_imag_scaling}. As expected from our discussion on $\sigma'$, also $\sigma''$ exhibits $\omega/T$ scaling
in the NFL region, where it is well described by the scaling curve $\mc{S}''(x)$.

Following Ref.~\cite{Shi2023}, we define the $\omega$-dependent transport scattering rate $\tau^{-1}(\omega)$ and effective mass $m^{\ast}(\omega)$,
\begin{flalign}
\tau^{-1}(\omega) &= \re \left[\frac{1}{\sigma(\omega)}\right] \, , \;
m^{\ast}(\omega) = -\frac{1}{\omega}  \im\left[ \frac{1}{\sigma(\omega)} \right] \, , \\
\sigma(\omega) &= \frac{1}{\tau^{-1}(\omega) - \mr{i}\omega \, m^{\ast}(\omega)} \, .
\end{flalign}
Since we are interested in the qualitative frequency and temperature dependence of these quantities, we omitted the constant prefactors in Eq.~(1) of Ref.~\cite{Shi2023}.
Note that $\tau^{-1}(0) = \tau_0^{-1} = \rho(T)$, which is shown in Fig.~\ref{fig:optical}(c) of the main text. 
In a normal FL (without disorder) exhibiting a usual Drude peak in $\sigma(\omega)$, we expect $\tau^{-1}_0 \sim T^2$,  $\tau^{-1}(\omega) =  \tau^{-1}_0 + a \omega^2$, while $m^{\ast}(\omega) \simeq m^{\ast}(0) = m^{\ast}_0$ is expected to be a temperature-independent constant.

\begin{figure}[t!]
\includegraphics[width=\linewidth]{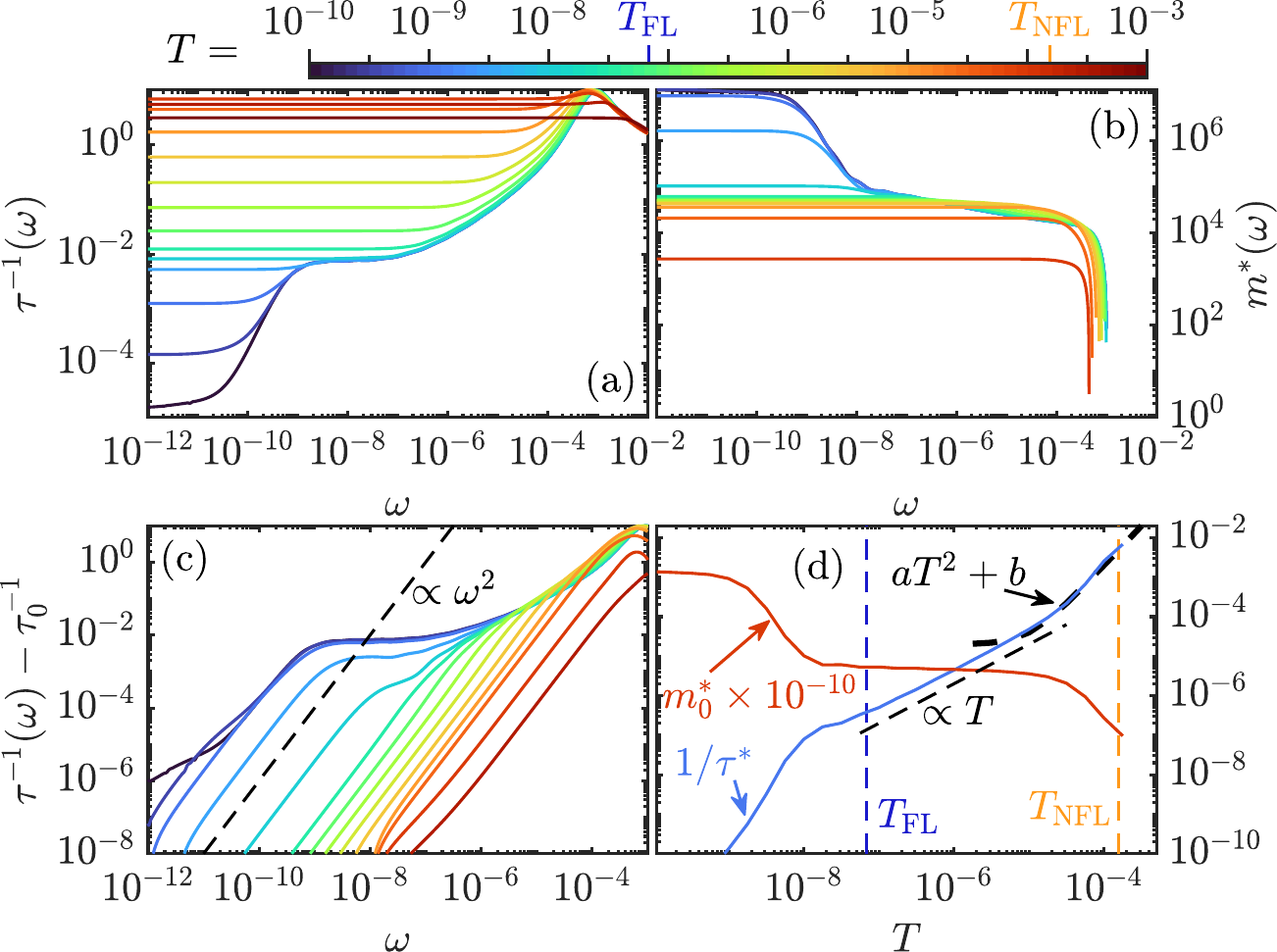}
\caption{%
Frequency dependence of (a) the transport scattering rate $\tau^{-1}(\omega)$, (b) the effective mass $m^{\ast}(\omega)$, and 
(c) $\tau^{-1}(\omega) -\tau^{-1}_0$, where $\tau^{-1}_0 = \tau^{-1}(0) = \rho$.
(d) Temperature dependence of $\tau^{\ast -1} = \tau_0^{-1}/m^{\ast}_0$ and $m^{\ast}_0 = m^{\ast}(0)$. 
}
\label{fig:tauw_mw}
\end{figure}

Figure~\ref{fig:tauw_mw}(a) shows our results for $\tau^{-1}(\omega)$. 
It shows a peak around $|\omega| = \TNF$ where the hybridization gap forms and then decreases towards $\omega=0$.
There, $\tau^{-1}(0) = \tau_0^{-1} = \rho(T) \propto T$ for $\TFL < T < \TNF$. 
At intermediate frequencies within the NFL region ($\max(\TFL,T) < |\omega| < \TNF$), $\tau^{-1}(\omega)$ has a non-trivial $\omega$- and $T$-dependence and does not 
seem to follow a simple power law with possible logarithmic corrections.
In this region, the optical conductivity does not fit to a usual Drude peak. 
This non-Drude behavior is most clearly visible from our data for $\sigma'(\omega,T)$ [Fig.~\ref{fig:optical}(a) in the main text],
which shows a $\omega^{-1}$ dependence in the NFL region, while a usual Drude peak would imply an $\omega^{2}$ dependence.
Similar non-Drude behavior of the optical conductivity has been observed in \YRS~\cite{Prochaska2020,Li2023}.

Remarkably, in the NFL region ($\TFL < T < \TNF$) at low frequencies $|\omega| \lesssim T$,
we also find
$\tau^{-1}(\omega) - \tau^{-1}_0 \sim \omega^2$ similar to a FL, cf.~Fig.~\ref{fig:tauw_mw}(c). 
An $\sim \omega^2$ dependence of $\tau^{-1}(\omega)$ has also been found in \CCI, cf.~Fig,~4(a,c) of Ref.~\cite{Shi2023} and its discussion.
However, an important difference to normal FL behavior lies in the temperature dependence
of the $\omega^2$ prefactor of $\tau^{-1}(\omega) - \tau^{-1}_0 = a(T) \, \omega^2$: in the NFL strange-metal region, $a(T)$ is temperature dependent, which is not the case in a normal FL phase. 
At low frequencies, $m^{\ast}(\omega) \simeq \mr{const.}$ [c.f. Fig.~\ref{fig:tauw_mw}(b)] and $\tau_0^{-1} \sim 1/T$, hence the $\omega/T$ scaling of $\sigma(\omega,T)$ we have shown in this work dictates $a(T) \sim 1/T$. This is in line with the data shown in Fig.~\ref{fig:tauw_mw}(c). 

\begin{figure}[t!]
\includegraphics[width=\linewidth]{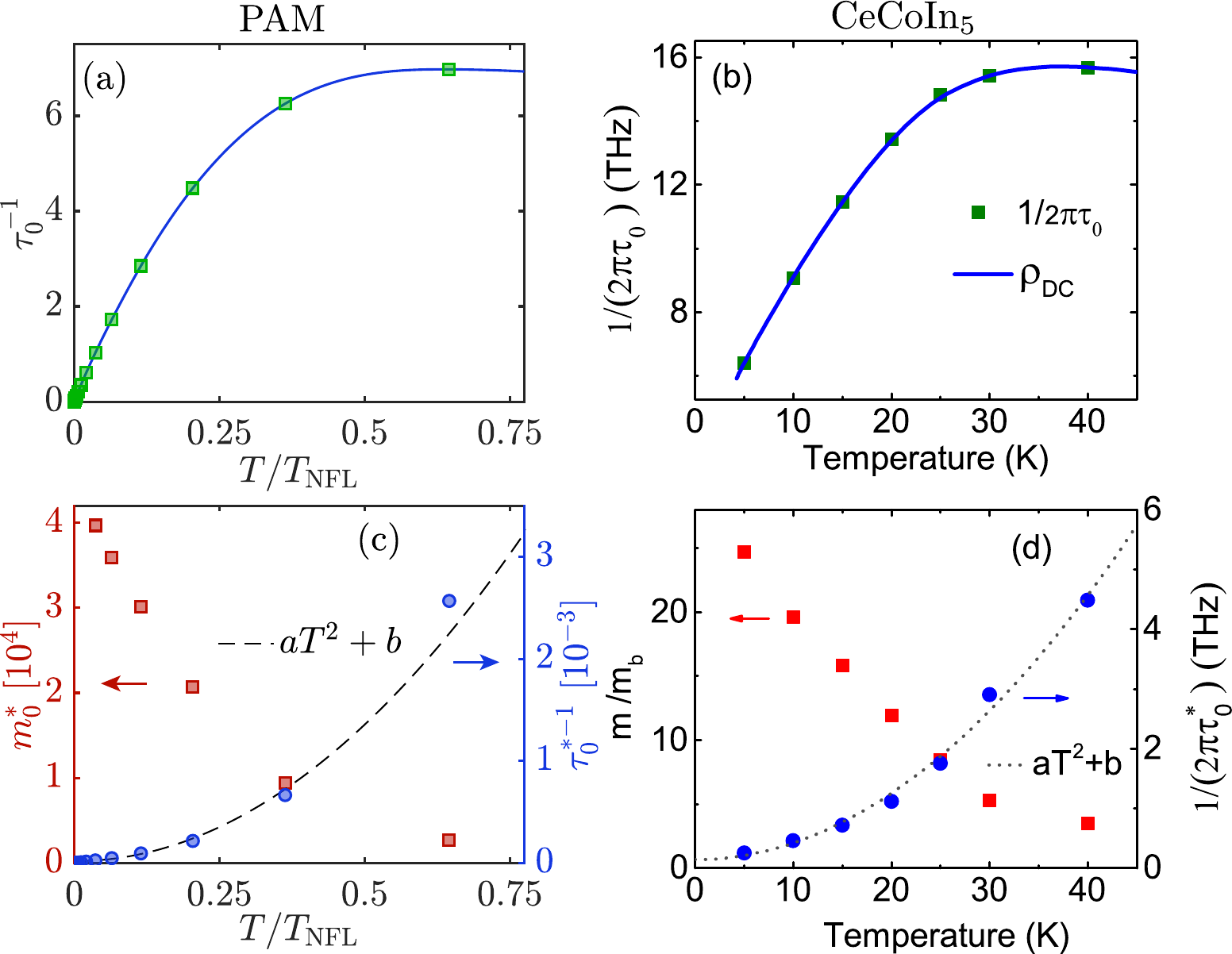}
\caption{%
(a) Scattering rate $\tau_0^{-1} = \rho(T)$ versus temperature for  $T \lesssim \TNF \simeq 1.5\cdot 10^{-4}$ for the PAM. Green squares are data points, and the blue line is a spline interpolation
that serves as a guide to the eye. (b) Scattering rate $\tau_0^{-1}$ (green squares) and rescaled resistivity (blue line) for \CCI\ close to its coherence temperature $T^{\ast} = 40 K$. The data in (b) is adapted from 
Fig.~4(b) of Ref.~\cite{Shi2023}. 
(c) Renormalized scattering rate (blue circles) and effective mass (red squares) versus temperature for the PAM. The black dashed line is a quadratic fit to the renormalized scattering rate in this temperature region.
(d) Renormalized scattering rate (blue circles) and effective mass (red squares) versus temperature for \CCI, adapted from Fig.~4(d) of Ref.~\cite{Shi2023}.
}
\label{fig:PAM_vs_CCI}
\end{figure}

We emphasize that in our results, $\tau^{-1}(\omega)$ is \emph{not} proportional to $-\im \, \Sigma(\omega)$ (\emph{without} vertex contributions, a proportionality
would be expected).
In our CDMFT+NRG approach to the PAM, $-\im \, \Sigma(\omega)$ has a logarithmic frequency and temperature dependence, cf.~Figs.~11 and 12 of Ref.~\cite{Gleis2023}.
The frequency and temperature dependence $\tau^{-1}(\omega)$ discussed above is different from that. This again directly illustrates the importance of vertex contributions.
In an MFL~\cite{Varma1989} as it appears for instance in the Yukawa--SYK approach~\cite{Patel2023} with interaction disorder, the strange-metal behavior arises due to a dominant bubble contribution
and therefore $\tau^{-1}(\omega) \sim -\im \, \Sigma(\omega) \sim \max (T,|\omega|)$ would be \emph{linear} in frequency. 

Figure~\ref{fig:tauw_mw}(b) shows $m^{\ast}(\omega)$. In the NFL region ($\TFL < T < \TNF$), $m^{\ast}(\omega)$ is strongly frequency dependent around the NFL scale, $\omega \simeq 10^{-3} \text{--} 10^{-4} \simeq \TNF$,
and then saturates to an almost frequency and temperature-independent value $m^{\ast}(\omega) \simeq m^{\ast}(0) = m^{\ast}_0$. 
The weak frequency and temperature dependence of $m^{\ast}(\omega)$ does not seem to follow a simple power law.
Interestingly, even though there are no well-defined QPs in the strange-metal region, there nevertheless seems to be a somewhat well-defined effective mass $m^{\ast}_0$. 
We emphasize though that in the NFL region, $m^{\ast}_0 \simeq 5\cdot 10^{4} \sim 10/\TNF$ is orders of magnitude smaller than in the FL region, 
where $m^{\ast}_0 \simeq 1.5\cdot 10^{7} \sim 1/\TFL$, cf.~Fig.~\ref{fig:tauw_mw}(d).
The effective mass in the NFL region is therefore decisively distinct from the QP mass in the low-temperature FL region. 

In Fig.~\ref{fig:tauw_mw}(d), we show the temperature dependence of the renormalized scattering rate $\tau^{\ast -1} = \tau^{-1}_0/m^{\ast}_0$ (blue), together with $m^{\ast}_0$ (red). 
Deep in the NFL region, we find $\tau^{\ast -1} \sim T$, since $\tau^{-1}_0 \sim T$ and $m^{\ast}_0 = \mr{const}$. 
Interestingly, in the crossover region between $T \simeq \TNF$ and $T \simeq 10^{-1} \TNF$, $\tau^{\ast -1}$ deviates from the linear-in-$T$ behavior and is consistent with FL-like $T^{2}$ behavior. 

A similar $T^2$ behavior was reported for {\CCI}  in Ref.~\cite{Shi2023}, where this behavior was interpreted as evidence for a hidden Fermi liquid. Our calculations suggest that 
the $T^{2}$ behavior is rather a crossover behavior and measurements at lower temperatures are necessary for a definite conclusion. Such measurements are presumably not possible in {\CCI} due to its 
relatively high $T_c$. A promising candidate material to clarify whether $\tau^{\ast -1} \sim T$ or $\sim T^{2}$ may be {\YRS}. 
To emphasize the similarity between the experimental data on \CCI\ and our results on the PAM more visually, we show the resistivity $\rho(T)$ of the PAM in Fig.~\ref{fig:PAM_vs_CCI}(a)
on a linear scale in the crossover region, next to the corresponding experimental data on \CCI\ [Figs.~\ref{fig:PAM_vs_CCI}(b)], adapted from Fig.~4(b) of Ref.~\cite{Shi2023}.
In Figs.~\ref{fig:PAM_vs_CCI}(c) and (d), we further show the data for the renormalized scattering rate and the effective mass for both the PAM and \CCI, respectively (adapted from Fig.~4(d) of Ref.~\cite{Shi2023}
for the latter). The experimental data on \CCI\ and our numerical data on the PAM show remarkable qualitative agreement in the crossover region: (i) the resistivity has a broad maximum and turns to linear-in-$T$;
(ii) the renormalized scattering rate $\tau^{\ast -1} \propto T^2$; and (iii) the effective mass $m^{\ast}_0$ increases with temperature in a remarkably similar fashion.

\end{document}